\documentclass[12pt]{article}
\usepackage{graphicx}
\usepackage{bm}

\usepackage{hhline}

\usepackage{amsmath,amssymb}
\usepackage{times}
\usepackage[varg]{txfonts}
\usepackage{xcolor}
\DeclareMathAlphabet{\mathbold}{OML}{txr}{b}{it}

\usepackage{array,multirow,dcolumn}
\usepackage[mathlines,displaymath]{lineno}
\usepackage{rotating}
\usepackage[english]{babel}
 
\usepackage[numbers,square,comma,sort&compress]{natbib}
\usepackage{hypernat}
\usepackage{textcomp}
\newcolumntype{.}{D{.}{.}{-1}}
\newcolumntype{-}{D{-}{-}{-1}}

\definecolor{rltred}{rgb}{0.75,0,0}
\definecolor{rltgreen}{rgb}{0,0.5,0}
\definecolor{rltblue}{rgb}{0,0,0.5}

\usepackage[hyperindex,bookmarks,bookmarksnumbered,breaklinks,a4paper,unicode]{hyperref}
\hypersetup{%
  pdftitle        = {Measurement of the inclusive ep Scattering Cross Section
 at low Q2 and x at HERA},
  urlcolor        = rltblue,       % \href{...}{...} external (URL)
  urlbordercolor  = 0 0 0.5,
  filecolor       = rltblue,       % \href{...} local file
  filebordercolor = 0 0 0.5,
  linkcolor       = rltred,        % \ref{...}
  linkbordercolor = 0.75 0 0,
  citecolor       = rltgreen,      % \cite{...}
  citebordercolor = 0 0.5 0,
  pagecolor       = rltgreen,      % \pageref{...}
  pagebordercolor = 0 0.5 0,
  menucolor       = rltgreen,      % Acrobat menu items
  menubordercolor = 0 0.5 0,
  colorlinks    = true,
  pdfauthor     = {H1 Collaboration},
  pdfsubject    = { },
  pdfkeywords   = {High-Energy Physics, Particle Physics, Proton Structure, DIS}
}

\newlength{\dinwidth}
\newlength{\dinmargin}
\newcommand{\ddif}[3]{\frac{d^{2}#1}{d#2 d#3}}

\begin{document}

% change natbib spacing between numbers in citations
\makeatletter \def\NAT@space{} \makeatother

\begin{titlepage}
 
\noindent
DESY 17-051
\vspace*{2.5cm}

\begin{center}
\begin{Large}

{\bfseries 
Investigation into the limits of perturbation theory at low  {\boldmath{$Q^2$}}\\
using HERA deep inelastic scattering data }

\vspace*{2cm}

I.~Abt$^a$,
A.M.~Cooper-Sarkar$^b$,
B.~Foster$^{b,c,d}$, 
V.~Myronenko$^d$,
K.~Wichmann$^d$, 
M.~Wing$^e$

\end{Large}
\end{center}

{\protect \hskip 0.cm} $^a$ Max-Planck-Institut f{\"u}r Physik, Werner-Heisenberg-Institut, M{\"u}nchen 80805, Germany 

\vspace*{-.12cm}
{\protect \hskip 0.cm} $^b$ Physics Department, University of Oxford, Oxford OX1 3RH, United Kingdom
 
\vspace*{-.12cm}
{\protect \hskip 0.cm} $^c$ Hamburg University, I. Institute of Experimental Physics, Hamburg 22607, Germany

\vspace*{-.12cm}
{\protect \hskip 0.cm} $^d$ Deutsches Elektronen Synchrotron DESY, Hamburg 22607, Germany

\vspace*{-.12cm}
{\protect \hskip 0.cm} $^e$ Department of Physics and Astronomy, University College London, London WC1E 6BT, United Kingdom

\vspace*{1cm}

\begin{abstract} \noindent

A phenomenological study of the final combined HERA data on 
inclusive deep inelastic scattering (DIS) has been performed.
The data are presented and investigated
for a kinematic range extending from values
of the four-momentum transfer, $Q^2$, above $10^4$\,GeV$^2$
down to the lowest values observable at HERA
% of $Q^2$, and Bjorken $x_{\rm Bj}$ 
of $Q^2 = 0.045$\,GeV$^2$ and Bjorken $x$, $x_{\rm Bj} = 6 \cdot 10^{-7}$.
The data are well described 
by fits based on perturbative quantum chromodynamics (QCD) 
using collinear factorisation and evolution of the parton densities
encompassed in the DGLAP formalism from the highest $Q^2$
down to $Q^2$ of a few GeV$^2$.
The Regge formalism with the soft Pomeron pole can describe the data
up to $Q^2 \approx 0.65$\,GeV$^2$.
The complete data set can be described by a new fit using
the Abramowicz--Levin--Levy--Maor (ALLM) parameterisation.
The region between the Regge and the perturbative
QCD regimes is of particular interest.

\end{abstract}

\vspace*{1.5cm}

\end{titlepage}

\newpage
~~
\newpage

\section{Introduction}
\label{sec:intro}

The HERA collider at DESY in Hamburg has provided a large amount
of data on electron--proton scattering over an extensive kinematic range.
Recently, the HERA collider experiments ZEUS and H1 published
a combination of all inclusive 
deep inelastic scattering (DIS) cross sections~\cite{HERAPDF20}.
 
In perturbative quantum chromodynamics (pQCD), 
scattering cross sections are computed by convoluting the 
partonic cross sections and parton density functions (PDFs),
which provide the probability that a parton, either gluon or quark, 
with a fraction $x$ of the proton's momentum takes part in the process.
The PDFs are scale dependent, i.e.\ they depend
on the four-momentum-transfer squared, $Q^2$, of the interaction.
A QCD fit to the combined HERA data resulted in 
the family of PDFs called HERAPDF2.0~\cite{HERAPDF20}.

One of the surprises thrown up by HERA data has been 
the apparent validity of pQCD down to values of $Q^2$ 
much smaller than had been thought likely~\cite{HERAIcombi,HERAPDF20}. 
The QCD analysis of the final combined data, however,   
indicated some tension between the standard 
Dokshitzer--Gribov--Lipatov--Altarelli--Parisi (DGLAP) 
evolution~\cite{Gribov:1972ri,Gribov:1972rt,Lipatov:1974qm,Dokshitzer:1977sg,Altarelli:1977zs} of PDFs and the cross-section data.

In recent papers~\cite{HHT,MMHT-HT-16}, the combined data were used 
to investigate the necessity of higher-twist corrections 
to the DGLAP evolution at low values of the observable Bjorken $x$, 
$x_{\rm Bj}$, which is equal to the $x$ of pQCD in the 
naive quark--parton model.  
The data suggest that such higher-twist corrections are needed
and the resulting  PDFs are uniquely suitable to investigate the 
validity of pQCD down to values of $Q^2 \approx 2$\,GeV$^2$. 

From the photoproduction regime, $Q^2 \approx 0$, to values of 
$Q^2 \approx 1$\,GeV$^2$,
pQCD can $a~priori$ not be applicable and models based on 
Regge theory~\cite{Regge1,Regge2,Regge3,Regge4} 
have been successfully used to describe the general features 
of early HERA cross-section data~\cite{ZEUSF2-98,ZEUSREGGE,Regge:H1}.
The transition from the photoproduction to the DIS regime
is phenomenologically interesting, especially at very low values
of $x_{\rm Bj}$.
While the present paper presents an exploration of this transition
region, it is not intended to be a comprehensive study of phenomenological
models.  Rather its purpose is to present the final HERA data in a variety
of forms that have been found useful in previous theoretical analyses.
These forms, both graphical and tabular, make use of a full knowledge of
the correlated errors and are important input for model building in the
low-$Q^2$ and low-$x_{\rm Bj}$ regime.

%%%% The exploration of this transition region 
%%%% is the purpose of the present paper. 

%Phenomenological models based both on non-perturbative QCD and
%Regge Theory are compared to the data in this transition region 
%in order to ascertain whether or not there are clear
%transition points, where the data
%change to follow descriptions by perturbative QCD, phenomenological models 
%or the Regge Theory 
%applicable to photoproduction.   
%In addition, we investigate whether there is any indication for 
%BFKL behaviour using a simple
%“saturation” model for the behaviour of the 
%total cross section.

\section{HERA Data, Cross Sections and Structure Functions}

All HERA data on neutral current (NC)  and charged current (CC)
$e^+p$ and $e^-p$ inclusive cross 
sections corrected to zero beam polarisation
were combined by the H1 and ZEUS collaborations
to provide a coherent set of data for further analysis.
The data were collected between 1994 and 2007 and 
represent a total integrated luminosity of $\approx 1$\,fb$^{-1}$.

The investigations presented in this paper focus on the $e^+p$ 
NC data, taken at centre-of-mass energies, $\sqrt{s}$, of 318\,GeV and
300\,GeV.
Their kinematic range spans six orders of magnitude in
$x_{\rm Bj}$ and $Q^2$, 
on a grid with $6.21\cdot10^{-7} \le x_{\rm Bj} \le 0.65$
and $0.045 \le Q^2 \le 30000$\,GeV$^2$.
The corresponding range of the energy available
at the photon--proton vertex, 
$W^2 = Q^2(1/x_{\rm Bj} - 1) + m_p^2 $,
where $m_p$ is the mass of the proton, 
is $ 10.7 \le W \le 301.2$\,GeV. 

\begin{figure}[tbp]
\vspace{-0.3cm} 
%%\vspace*{5pt}
\centerline{
\includegraphics[width=0.9\textwidth]{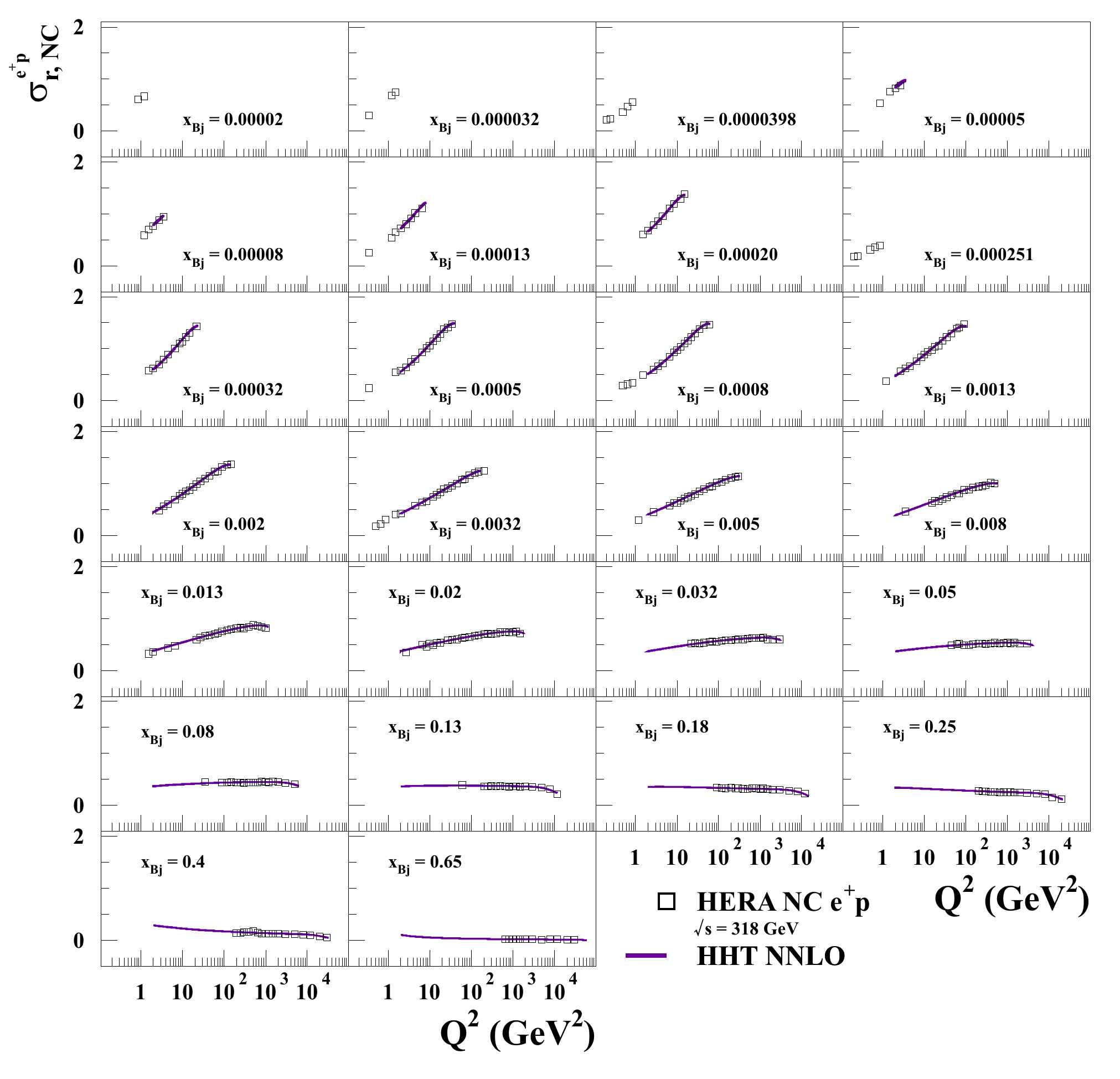}}
\vspace{0.5cm}
\caption {The combined HERA data on the 
          inclusive NC $e^+p$ reduced cross sections 
          with $\sqrt{s}= 318$\,GeV as a function of $Q^2$
          for 26 selected bins of $x_{\rm Bj}$. Also shown are 
          the predictions from the
          HHT NNLO~\cite{HHT} analysis down to $Q^2=2.0$\,GeV$^2$.
          The width of the bands represents the uncertainty 
          on the predictions. 
          The errors bars on the data are smaller than the symbols.
}
\label{fig:sred_318_data}
\end{figure}

The HERA NC combined data were published as 
reduced cross sections, $\sigma_{r, \rm NC}$, for $ep$
scattering. 
The cross-section data with $\sqrt{s} = 318$\,GeV
are shown in Fig.~\ref{fig:sred_318_data} as a function of $Q^2$ for
values of $x_{\rm Bj}$ for which more than one data point is available. 
For the regime of pQCD, predictions from the  HHT analysis~\cite{HHT} at
next-to-next-to-leading order (NNLO)
are also shown down to a $Q^2$ of 2.0\,GeV$^2$. 
At HERA, the lowest values of $x_{\rm Bj}$ are only reached at 
values of $Q^2$ sufficiently low that pQCD is not applicable. 
All predictions were made from sets of PDFs  
which were extracted from fits to data above a minimum $Q^2$, which,
for HHT NNLO, was $Q^2_{\rm min} = 2$\,GeV$^2$.
They describe the data well over their range of 
applicability.
The predictions from the HERAPDF2.0 analysis~\cite{HERAPDF20} at NNLO 
are very similar down to $Q^2 =3.5$\,GeV$^2$, 
which is the $Q^2_{\rm min}$ for HERAPDF2.0.

While the regime that can be treated by pQCD is clearly 
limited by theoretical considerations, 
the data themselves show no abrupt change in behaviour
in this regime. 
Scaling violations
are well established and well described by pQCD. The slope,
$d \sigma_{r,\rm{NC}} / d Q^2$, changes from negative to positive as 
$x_{\rm Bj}$ decreases.
Of particular interest are  $x_{\rm Bj}$
values with entries above and below $Q^2=1$\,GeV$^2$, such as
$x_{\rm Bj} = 0.0008$ and $x_{\rm Bj} = 0.0032$.
The $Q^2$ dependence does not change in any abrupt way around
$Q^2=1$\,GeV$^2$. It seems that nature does not know about
perturbation theory.

In order to describe $\sigma_{r,\rm NC}$, the
generalised structure functions,
$\tilde{F_{2}}$, $x \tilde{F_{3}}$ and $F_{\rm {L}}$,
are conventionally introduced. 
For $e^+p$,
\begin{linenomath*}\begin{equation}
  \sigma^{e^{+} p}_{r,{\rm NC}} 
  =
  \frac {x_{\rm Bj}Q^{4}} {2 \pi \alpha^{2}}
  \frac {1} {Y_{+}}
  \ddif{\sigma(e^{+}p)}{x_{\rm Bj}}{Q^{2}}
  =
  \tilde{F_{2}}(x_{\rm Bj},Q^{2}) - \frac {Y_{-}} {Y_{+}} \, x\tilde{F_{3}}
  (x_{\rm Bj},Q^{2})- \frac {y^2} {Y_{+}} \,F_{\rm L}(x_{\rm Bj},Q^{2}),
\label{eqn:red}
\end{equation}\end{linenomath*}
where $\alpha$ is the fine-structure constant and $Y_\pm=1 \pm (1-y)^2$,
with the inelasticity $y = Q^2/(s\,x_{\rm Bj})$.

The structure functions 
are $a~priori$ not limited to the perturbative regime.
This limitation only arises when they are expressed in terms
of parton distributions.
The structure-function $\tilde{F_{2}}$ 
has components due to photon exchange, due to $\gamma Z$
interference and due to $Z$ exchange. 
At $Q^2 \lesssim 5$\,GeV$^2$, only photon
exchange, described by $F_2$, has to be considered.
As $x\tilde{F_3}$
%, connected to valence quarks, 
does not have a
contribution from photon exchange, it can be 
neglected at low $Q^2$.

The structure-function $F_{\rm L}$ represents the exchange 
of longitudinally polarised photons, while
$F_2 = 2x_{\rm Bj}F_1 +F_{\rm L}$ 
is dominated by $F_1$, which describes 
the exchange of transversely polarised photons.
For low $Q^2$, $F_2$ is dominant and Eq.~\ref{eqn:red} 
can be expressed as

\begin{equation} 
  \sigma^{e^{+} p}_{r,{\rm NC}} 
        = F_2  - \frac{y^2}{Y_+} F_{\rm L}~
        = ~2x_{\rm Bj}F_1 + \frac {2(1-y)} {Y_{+}}~ F_{\rm L} ~.
\label{eqn:redlQ2}
\end{equation}

It has proven to be very difficult to extract $F_{\rm L}$
from HERA data. 
Special runs at the end of HERA operation provided
data that confirmed~\cite{H1FL2,ZEUSFL} the expectation~\cite{BKS,RT:FL:06}
that $F_{\rm L}$ is small compared to $F_2$ within the experimentally
accessible range of $Q^2 \ge 1.5$\,GeV$^2$.
In order to describe data with $Q^2$ as low as 1.5\,GeV$^2$,
a twist-four term was added to the description of $F_{\rm L}$
within the standard DGLAP formalism for the HHT analysis~\cite{HHT}. 
The resulting predictions describe the data quite well
even down to 1.2\,GeV$^2$, but at such low values
of $Q^2$, the resulting values for $F_{\rm L}$
are large and rapidly divergent and therefore unphysical.

At HERA, low $x_{\rm Bj}$ implies low $Q^2$ and vice versa. 
%As $Q^2$ approaches zero, 
%gauge invariance requires that $F_{\rm L}$ goes to zero.
Re-expressing the factor in front of $F_{\rm L}$ in Eq.~\ref{eqn:redlQ2}
in terms of $x_{\rm Bj}$ and $Q^2$ shows that it
goes to zero proportionally
to $Q^2$ and $x_{\rm Bj}$, while $F_1$ is only suppressed by $x_{\rm Bj}$.
Thus, at HERA, the exchange of transverse
photons dominates in the low-$Q^2$ region, independently
of the exact $Q^2$ dependence of $F_1$ and $F_{\rm L}$.

\section{Extraction of {\boldmath{$F_2$}} and {\boldmath{$\sigma^{\gamma^*p}$}} }

Traditionally, HERA physics at low $Q^2$ and $x_{\rm Bj}$
is discussed in terms of $F_2$ and $\sigma^{\gamma^*p}$, 
defined as the cross section for virtual photon 
exchange.
% associated with $F_2$. 
The values of $F_2$ 
have to be extracted from the reduced cross-section data.
This cannot be done in an unbiased way and it cannot be done
in the same way over the whole kinematic region. 
Indeed, on the contrary, very different models have to be used. 
However, in all cases, $F_2$ is extracted as

\begin{equation}\label{eq:f2extr}
 F_2^{{\rm extracted}} = F_2^{{\rm predicted}} \frac{\sigma_r^{{\rm measured}}}
                                        {\sigma_r^{{\rm predicted}}} ~~.
\end{equation}

Two different models were used to extract $F_2$ in two
overlapping $Q^2$ ranges for this paper.
The results of the HHT NNLO analysis~\cite{HHT}
and Eq.~\ref{eqn:red} were used 
for $Q^2 \ge 1.2$\,GeV$^2$.
The contributions from $\gamma Z$ interference
and $Z$ exchange, which become important
as $Q^2$ increases, were taken into account through Eq.~\ref{eqn:red}.
For $Q^2 \le 2.7$\,GeV$^2$,
Eq.~\ref{eqn:redlQ2} was used to extract $F_2$ using
estimates of $R=F_{\rm L}/(F_2-F_{\rm L})$ from the Badelek--Kwiecinski--Stasto (BKS) model~\cite{BKS}
for $F_{\rm L}$ at low $x_{\rm Bj}$ and low $Q^2$.
This model is based on the kinematic constraint 
that $F_{\rm L} \propto Q^4$ as $Q^2 \rightarrow 0$ and on the 
photon--gluon fusion mechanism. The contribution of
quarks having limited transverse momenta is treated 
phenomenologically, assuming 
the soft Pomeron exchange.
% Brian wants a Pomeron? 
The value of $R$ was predicted by extrapolating $F_2$
to the region of low $Q^2$.
In principle, $R$ depends not only on $Q^2$, but also on
$x_{\rm Bj}$. However, the dependence on $x_{\rm Bj}$ is small and
for the extraction of $F_2$, the average value of $R$ over 
the $x_{\rm Bj}$ range relevant for each $Q^2$ value was used.

\begin{figure}[tbp]
\vspace{-0.5cm} 
%\vspace*{5pt}
\centerline{
    \includegraphics[width=1.1\textwidth]{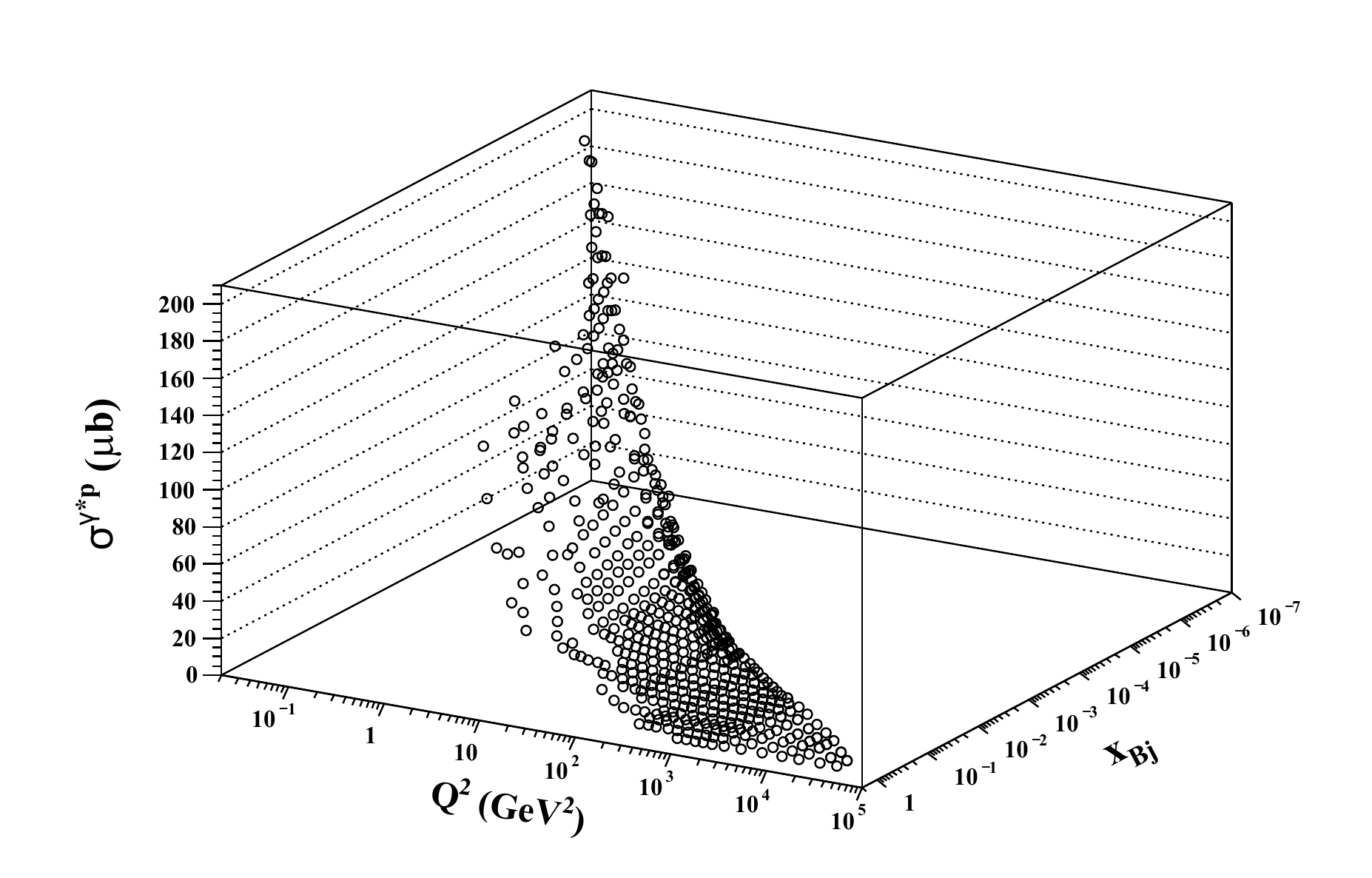}}
\caption {$\sigma^{\gamma^*p}(x_{\rm Bj},Q^2)$ 
         as extracted from the HERA
         $e^+p$ NC data at $\sqrt{s}=318$ and 300\,GeV. 
         For some values of $Q^2$ and $x_{\rm Bj}$, two data points
         corresponding to
         the two different centre-of-mass energies are shown.
}
\label{fig:sigma-xQ2}
\end{figure}

The cross section for the scattering of virtual
photons on protons, $\sigma^{\gamma^*p}$, was extracted 
from the structure-function $F_2$ 
by using the Hand convention~\cite{Hand63} to define the
photon flux, yielding the relation~\cite{Allen} 
\begin{equation}
\sigma^{\gamma^*p}(x_{\rm Bj},Q^2) =  
      \frac {4 \pi^2 \alpha (Q^{2} + (2 x_{\rm Bj}m_p)^2)}
            {Q^4 (1 - x_{\rm Bj})} 
                                  ~F_2(x_{\rm Bj},Q^2) ~.
\label{eqn:sigma}
\end{equation}
The extracted values of $\sigma^{\gamma^*p}(x_{\rm Bj},Q^2)$
for the complete HERA data set 
%from the structure function $F_2$ 
are tabulated in Tables~\ref{tab:sigma1}--\ref{tab:sigma7} 
and shown in Fig.~\ref{fig:sigma-xQ2}. 
The values of  $\sigma^{\gamma^*p}$ form a smooth plane, 
which again does not show any abrupt
features or transitions around $Q^2=1$\,GeV$^2$.
For small $x_{\rm Bj}$, Eq.~\ref{eqn:sigma} can be simplified to
\begin{equation}
\sigma^{\gamma^*p}(x_{\rm Bj},Q^2) =  \frac {4 \pi^2 \alpha}{Q^{2}} 
                                  ~F_2(x_{\rm Bj},Q^2) ~.
\label{eqn:sigma-lowx}
\end{equation}

\section{Features of {\boldmath{$\sigma^{\gamma^*p}$}}
         and  
         Fits to {\boldmath{$F_2$}} 
        }  

The structure-function $F_2$ was extracted on the 
$(x_{\rm Bj},Q^2)$-grid given by the published $ep$ cross-section data.
The cross-section $\sigma^{\gamma^*p}$
is related to the energy available at the $\gamma^* p$ vertex, 
$W$. For low enough $x_{\rm Bj}$,  $W^2 = Q^2/x_{\rm Bj}$.

\begin{figure}[tbp]
\vspace{-0.5cm} 
%\vspace*{5pt}
\centerline{
    \includegraphics[width=0.9\textwidth]{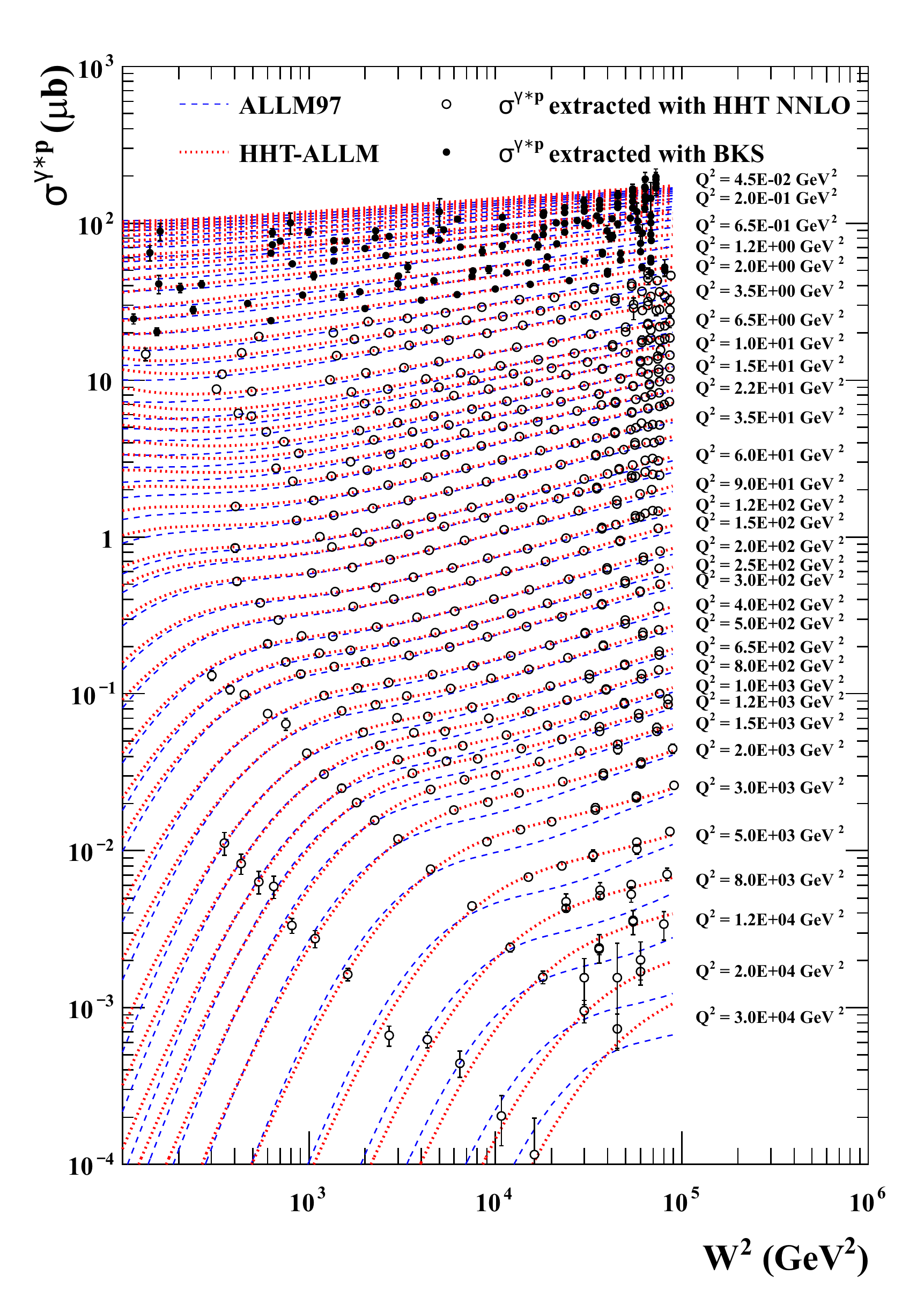}}
\caption{$\sigma^{\gamma^*p}(W^2)$ 
         as extracted from the HERA
         $e^+p$ NC data at $\sqrt{s}=318$ and 300\,GeV.
         For some values of $Q^2$ and $W$, two data points
         corresponding to
         the two different centre-of-mass energies are shown.
          Also shown are the predictions
          from the ALLM97 fit to early HERA data and the result
          of the new HHT-ALLM fit, see text for details about the fit.
          Below $Q^2=90$\,GeV$^2$, some $Q^2$ labels are omitted
          for reasons of legibility.}
\label{fig:sigma-W}
\end{figure}

Figure~\ref{fig:sigma-W} shows the complete set of extracted values
of $\sigma^{\gamma^*p}$ as a function of $W^2$ for fixed $Q^2$.
The BKS model was used to extract  $\sigma^{\gamma^*p}$
for $Q^2 \le 2.0$\,GeV$^2$ and HHT NNLO
was used for  $Q^2 > 2.0$\,GeV$^2$. 
As before, there is no indication of any abrupt change
in behaviour
around $Q^2 = 1$\,GeV$^2$.
The cross sections rise rapidly with $W$ for all $Q^2$.
For high enough $W$, i.e.\ low enough $x_{\rm Bj}$,
a smooth power rise as $W^{2 \lambda}$ is observed.  
As $Q^2$ decreases, $\lambda$ also decreases.            

The data can be described by fits to the
Abramowicz--Levin--Levy--Maor (ALLM) parameterisation~\cite{ALLM91,ALLM97}.
This $ansatz$ combines Regge phenomenology with some ideas from pQCD.
It describes $F_2$ as:
\begin{equation}
 F_2 = \frac {Q^2}{Q^2+m_0^2}
                  \cdot (F_2^{I\!P} + F_2^{I\!R} ) ~,
\end{equation}
where $m_0$ is an effective photon mass 
and $F_2^{I\!P}$ and $F_2^{I\!R}$
are the contributions of Pomeron and Reggeon exchange, respectively.
The parameterisation has 
23 parameters, including $m_0$, 
which are associated either with
Pomerons, Reggeons or mass scales. The complete set of formulae is
given in the Appendix. 
The predictions 
from the ALLM97 fit~\cite{ALLM97} to early ZEUS data 
and the results of a new fit to the combined HERA data set
called HHT-ALLM
are both shown in Fig.~\ref{fig:sigma-W}.

The ALLM97 fit and the HHT-ALLM fit differ mainly due
to the inclusion of high-$Q^2$ data, which have become available in the
later years of HERA operation. This is clearly visible in 
Fig.~\ref{fig:sigma-W}; the ALLM97 predictions do not describe the
high-$Q^2$ data at all. 
However, the description of the
low-$Q^2$ data is also improved by the HHT-ALLM fit.
The parameters of HHT-ALLM are listed and compared to the parameters
of ALLM97 in Table~\ref{tab:ALLM}. Also given are the parameters of 
a fit HHT-ALLM-FT, for which fixed-target 
data~\cite{Adams:1996gu,Arneodo:1996qe,Benvenuti:1989rh} 
were also included in the fit.

\begin{figure}[tbp]
\vspace{-0.5cm} 
%\vspace*{5pt}
\centerline{
    \includegraphics[width=0.9\textwidth]{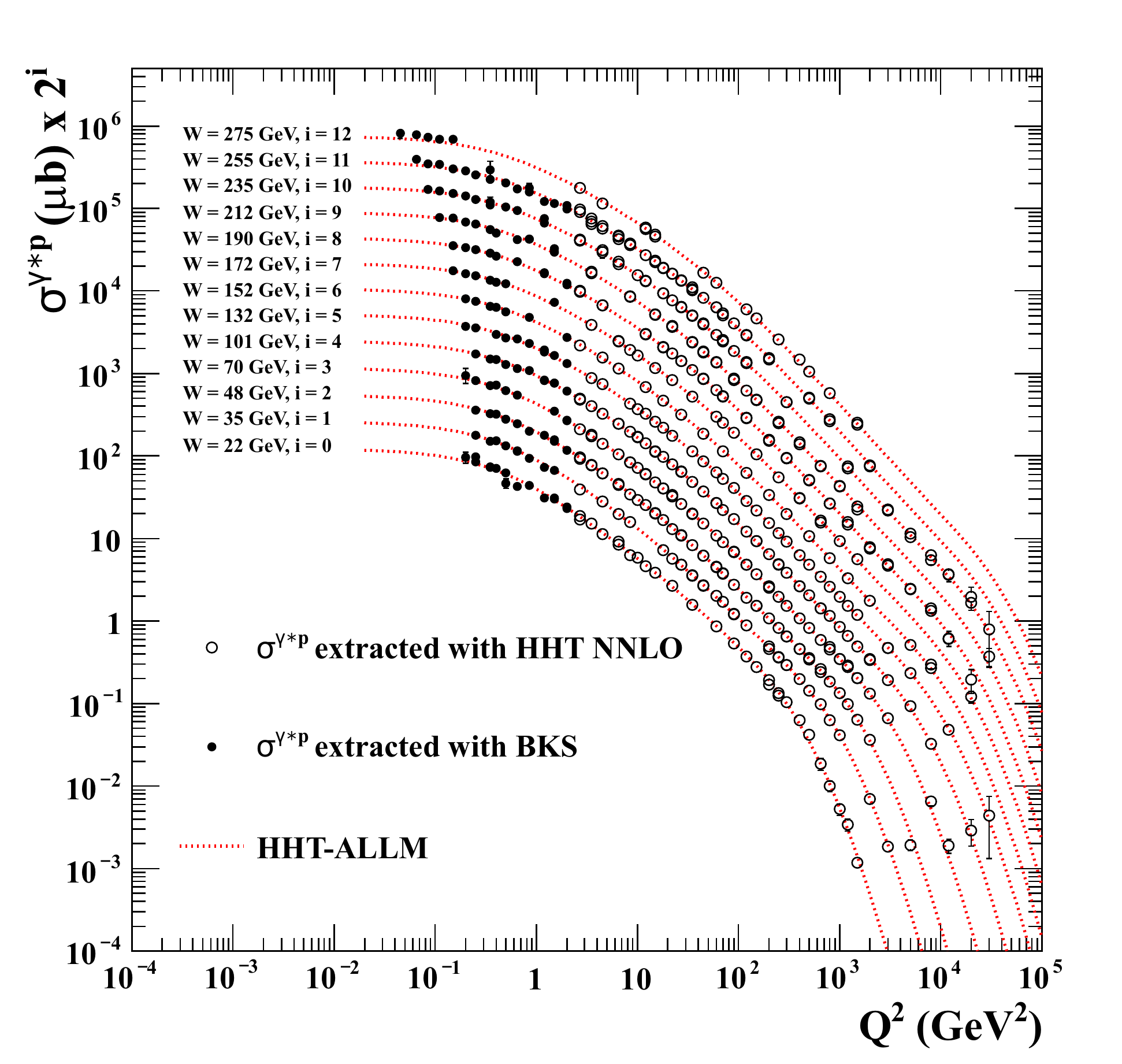}}
\caption{$\sigma^{\gamma^*p}(Q^2)$ 
         as extracted from the HERA
         $e^+p$ NC data at $\sqrt{s}=318$ and 300\,GeV for
         selected values of $W$, together with the predictions
         of the HHT-ALLM fit.
         For some values of $Q^2$ and $W$, two data points
         corresponding to
         the two different centre-of-mass energies are shown.
         The parameters from the HHT-ALLM fit were used
         to translate the data points to the $W$ values listed.
         The values of $\sigma^{\gamma^*p}(Q^2)$ are multiplied
         by $2^{\rm i}$ for different values of $W$  to make
         the curves individually visible.
         }
\label{fig:sigma-all}
\end{figure}

\begin{figure}[tbp]
\vspace{-0.5cm} 
%\vspace*{5pt}
\centerline{
    \includegraphics[width=0.9\textwidth]{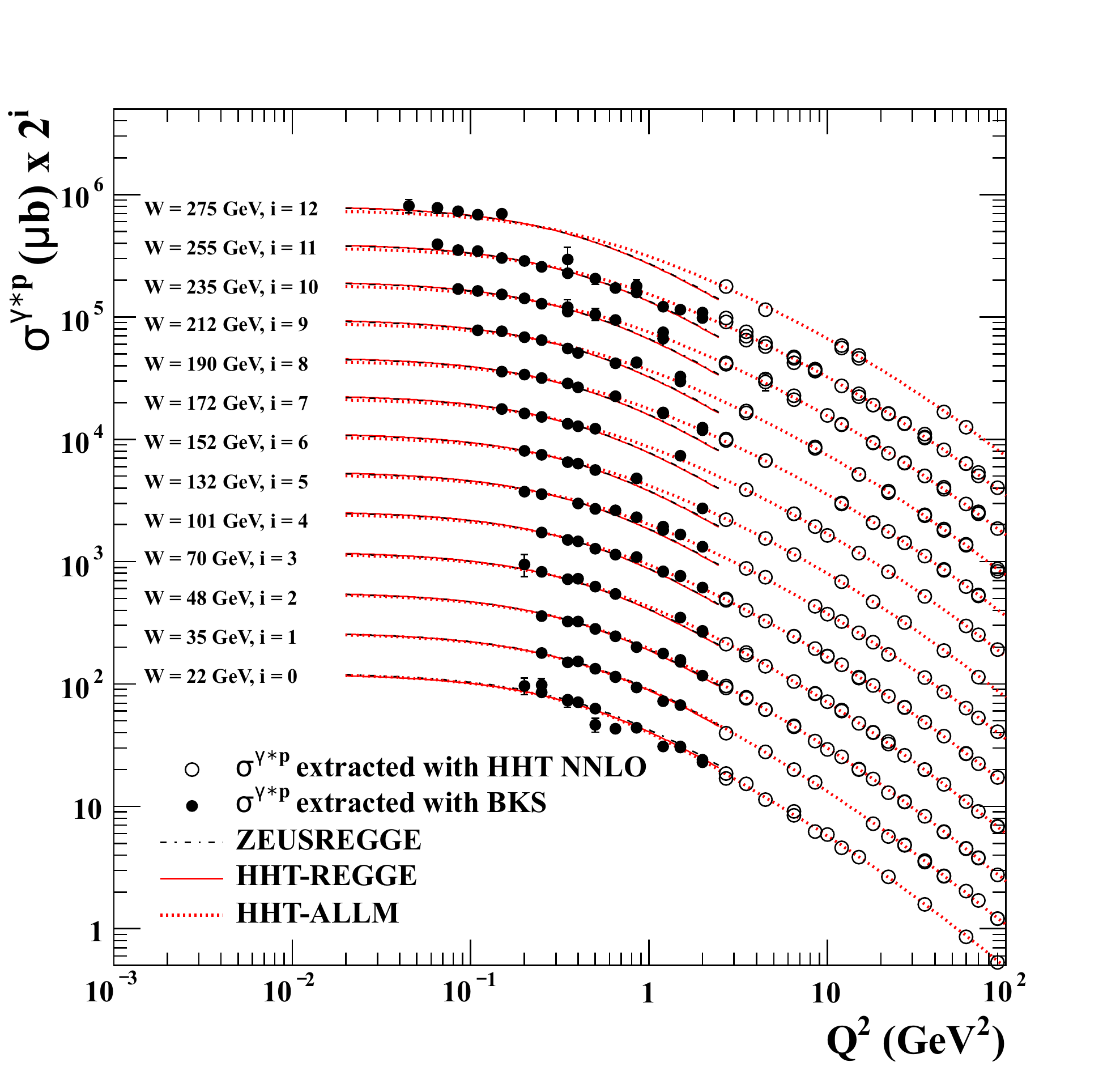}}
\caption{$\sigma^{\gamma^*p}(Q^2)$ in the low-$Q^2$ regime
         as extracted from the HERA
         $e^+p$ NC data at $\sqrt{s}=318$ and 300\,GeV for
         selected values of $W$, together with the predictions
         of the HHT-ALLM, HHT-REGGE and ZEUSREGGE fits, see text.
         The HHT-REGGE and ZEUSREGGE curves are almost identical.
         For some values of $Q^2$ and $W$, two data points
         corresponding to
         the two different centre-of-mass energies are shown.
         The parameters from the HHT-ALLM fit were used
         to translate the data points to the $W$ values listed.
         The values of $\sigma^{\gamma^*p}(Q^2)$ are multiplied
         by $2^{\rm i}$ for different values of $W$  to make
         the curves individually visible.}
\label{fig:sigma-low-Q2}
\end{figure}

The HHT-ALLM fit has a good 
$\chi^2/{\rm ndf} = 607 / 574 = 1.06$.
Thus, the HHT-ALLM parameters were used to move points 
close in $W$ to selected $W$ values by translating them, 
keeping $Q^2$ constant.
The result is shown in Fig.~\ref{fig:sigma-all}, which shows the
$Q^2$ dependence of  $\sigma^{\gamma^* p}$ for selected values of $W$.
The measured values of $\sigma^{\gamma^* p}$
extracted with the BKS model 
and with the results of the HHT NNLO analysis 
connect smoothly 
at the change-over value
of $Q^2=2.0$\,GeV$^2$ for all values of $W$.
The lack of a break in this $Q^2$ regime is striking.
However, the behaviour at high and low $Q^2$ differs.
At high $Q^2$, $F_2$ depends only weakly on $Q^2$
as QCD scaling violations depend on $\ln Q^2$. Thus,
$\sigma^{\gamma^* p}$ drops with $1/Q^2$ as indicated 
in Eq.~\ref{eqn:sigma} for fixed $W$.
As $Q^2 \rightarrow 0$, the values of $\sigma^{\gamma^* p}$ have 
to approach the finite limit of photoproduction at
$Q^2 = 0$ and $F_2$ has to be proportional to $Q^2$.
The challenge is to  model the
smooth transition from the high- to the low-$Q^2$ regime.
Figure~\ref{fig:sigma-low-Q2} focuses on 
this region.
Although generally providing a good fit,
the ALLM parameterisation 
predicts systematically lower $\sigma^{\gamma^*p}$ values
at the lowest $Q^2$ and highest $W$ values.

Regge phenomenology~\cite{Regge1}
has had considerable success in parameterising 
the data on soft hadron--hadron collisions.
In the low-$Q^2$ regime, the
photon can be considered a hadron.
Figure~\ref{fig:sigma-low-Q2} also shows predictions from
Regge fits to low-$Q^2$ data.
At sufficiently large values of $W$, the total hadronic cross section
is described in terms of the exchange in the $t$ channel
of a “Pomeron trajectory”,
$\alpha_{I\!P} (t) = \alpha_{I\!P}(0) + \alpha_{I\!P}' \cdot t$. 
This leads to the prediction that total hadron--hadron
cross sections depend on the intercept of the trajectory at $t = 0$, 
$\alpha_{I\!P} (0)$, as
\begin{equation}
 \sigma^{\gamma^* p} \propto  W^{2(\alpha_{I\!P}(0)-1)} ~,
\end{equation}
where for the soft Pomeron~\cite{pom:soft}  $\alpha_{I\!P}(0) \approx 1.08$.
%This corresponds to the observation of such a power-law 
%rise for low $Q^2$ as seen in 
Figure~\ref{fig:sigma-W} shows such a power-law rise for low $Q^2$.

At low $W$ values, 
%where the cross sections are falling with $W$, 
Reggeon-exchange terms also become important.
The full description of $F_2$ in the Regge 
formalism~\cite{ZEUSF2-98,alpha:don} is
\begin{equation}
F_2(x_{\rm Bj},Q^2) =
  \frac{Q^2}{4\pi^2\alpha} \cdot \frac{M_0^2}{M_0^2+Q^2}
  \cdot \Bigg(~ A_{I\!P} ~\Bigg(\frac{Q^2}{x_{\rm Bj}}\Bigg)^{\alpha_{I\!P}(0)-1} + 
      ~ A_{I\!R}~ \Bigg(\frac{Q^2}{x_{\rm Bj}} \Bigg)^{\alpha_{I\!R}(0)-1} ~ \Bigg)~, 
\label{eqn:regge}
\end{equation}
where $M_0$, $A_{I\!P}$, $A_{I\!R}$, $\alpha_{I\!P}(0)$  and $\alpha_{I\!R}(0)$
are the parameters to be determined.
The term ${M_0^2}/(M_0^2+Q^2)$ with an effective mass $M_0$
arises in the framework of the generalised vector-meson dominance (GVMD)
model~\cite{GVDM1,GVDM2}.
 
New Regge fits 
were performed to the $F_2$ values as extracted from the 
HERA inclusive NC $e^+p$ 
data with $Q^2 \le 0.65$\,GeV$^2$,
i.e.\ in the regime of the so-called soft Pomeron~\cite{pom:soft}.
As the $W$ range of the HERA data for this $Q^2$ regime does not
extend to low enough $W$ to require a Reggeon term, 
the Reggeon term was omitted for the default fit.
Thus, the default fit, HHT-REGGE, is a 3-parameter fit.
% fit with only the Pomeron term considered.

The predictions of the HHT-REGGE fit and of
a fit previously published by the ZEUS collaboration, 
ZEUSREGGE~\cite{ZEUSF2-98},
are shown in Fig.~\ref{fig:sigma-low-Q2}. 
For $Q^2 \lesssim 0.65$\,GeV$^2$,
the Regge predictions are unsurprisingly very similar to
the ALLM predictions, which are also based on Regge theory in this regime.
The Regge and ALLM fits describe the overall 
behaviour of the data reasonably well
for $Q^2 \lesssim 0.65$\,GeV$^2$.
However, for the highest $W$, both fits predict $\sigma^{\gamma^*p}$
to be systematically lower than is observed.
%for low $Q^2$ and high $W$ are slightly
%larger than the ALLM predictions, but are still below the data.
%In this regime, 
In addition, as $W$ increases from around 50\,GeV
and $Q^2$ approaches 1\,GeV$^2$, 
the Regge predictions diverge more and more 
from the ALLM fit.

\begin{figure}[tbp]
\vspace{-0.5cm} 
%\vspace*{5pt}
\centerline{
    \includegraphics[width=0.9\textwidth]{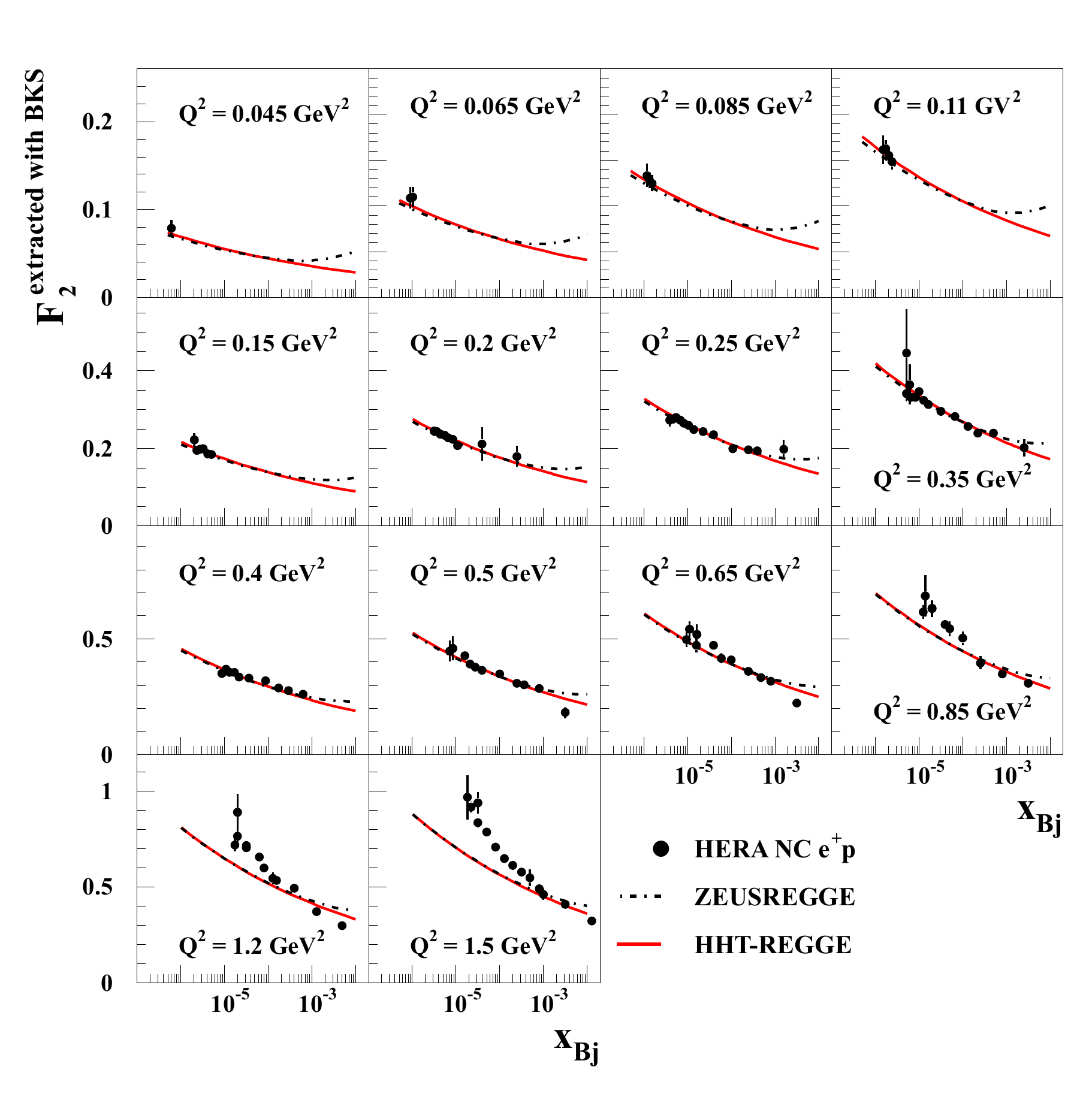}}
\caption {The structure-function $F_2$ as a function of 
          $x_{\rm Bj}$ for low $Q^2$ as extracted from the
          HERA NC $e^+p$ cross sections, 
          together with predictions from the HHT-REGGE 
          and ZEUSREGGE fits,
          see text for details.}
  \label{fig:regge-noft}
\end{figure}

A more detailed comparison is presented in 
Fig.~\ref{fig:regge-noft}, which shows the $F_2$ data at low $Q^2$
%as extracted from HERA NC $e^+p$ cross sections,
together with predictions from the HHT-REGGE 
and the old ZEUSREGGE fit.
The data are described  well by HHT-REGGE
for $Q^2 \le 0.65$\,GeV$^2$, 
i.e.\ the fitted range.
%The HHT-REGGE fit provides
%a somewhat better fit than ZEUSREGGE due to the 
%inclusion of more HERA data from 
%the combined HERA data set.

It is expected that the simple model of a single Pomeron trajectory
should start to break down beyond $Q^2 \approx 0.65$\,GeV$^2$~\cite{pom:soft}.
The HHT-REGGE fit has a $\chi^2/ {\rm ndf}$ of 0.83. 
Extending the fit to data with  $Q^2 \le 0.85$\,GeV$^2$, ZEUS-REGGE-3p-.85, 
leads to a $\chi^2/ {\rm ndf} = 1.13$, confirming this expectation.
Nevertheless, the data with $Q^2 \le 0.65$\,GeV$^2$ are still well described
by the ZEUS-REGGE-3p-.85 fit.
%The resulting parameters are, however, compatible to the 
%parameters of the default fit.
Table~\ref{tab:regge} summarises the results of all Regge fits.

If the Reggeon term is included in the HHT-REGGE fit with
$A_{I\!R}$ free and $\alpha_{I\!R}(0) = 0.5$ fixed~\cite{alpha:don},
the resulting HHT-REGGE-4p fit has a 
very good $\chi^2/ {\rm ndf} = 0.78$. 
However, $A_{I\!R}$ becomes negative with a large uncertainty.
This confirms that the range in $W$ of the HERA data alone
is not sufficient to constrain the Reggeon term.
The values for the Pomeron parameters are, however, consistent with the
default HHT-REGGE fit.
Adding the 
fixed-target data~\cite{Adams:1996gu,Arneodo:1996qe,Benvenuti:1989rh} 
does not significantly improve the constraints on the Reggeon
term, see Table~\ref{tab:regge} (HHT-REGGE-FT).

%======== 1st day

The ZEUSREGGE fit 
%was published by the ZEUS collaboration~\cite{ZEUSF2-98}.
%It 
was performed on low-$Q^2$ ($ 0.11 \le Q^2 \le 0.65$\,GeV$^2$) data 
on $F_2$ from early HERA running. 
This fit had  $A_{I\!R}$ as a free parameter with
$\alpha_{I\!R}(0) = 0.5$ fixed. This was possible because
selected photoproduction data~\cite{photo1,photo2}
with lower $W$ ($ 6 \lesssim W \lesssim 20$\,GeV)   
were included in the fit.

%--> Caption

\begin{figure}[tbp]
\vspace{-0.5cm} 
%\vspace*{5pt}
\centerline{
\includegraphics[width=0.95\textwidth]{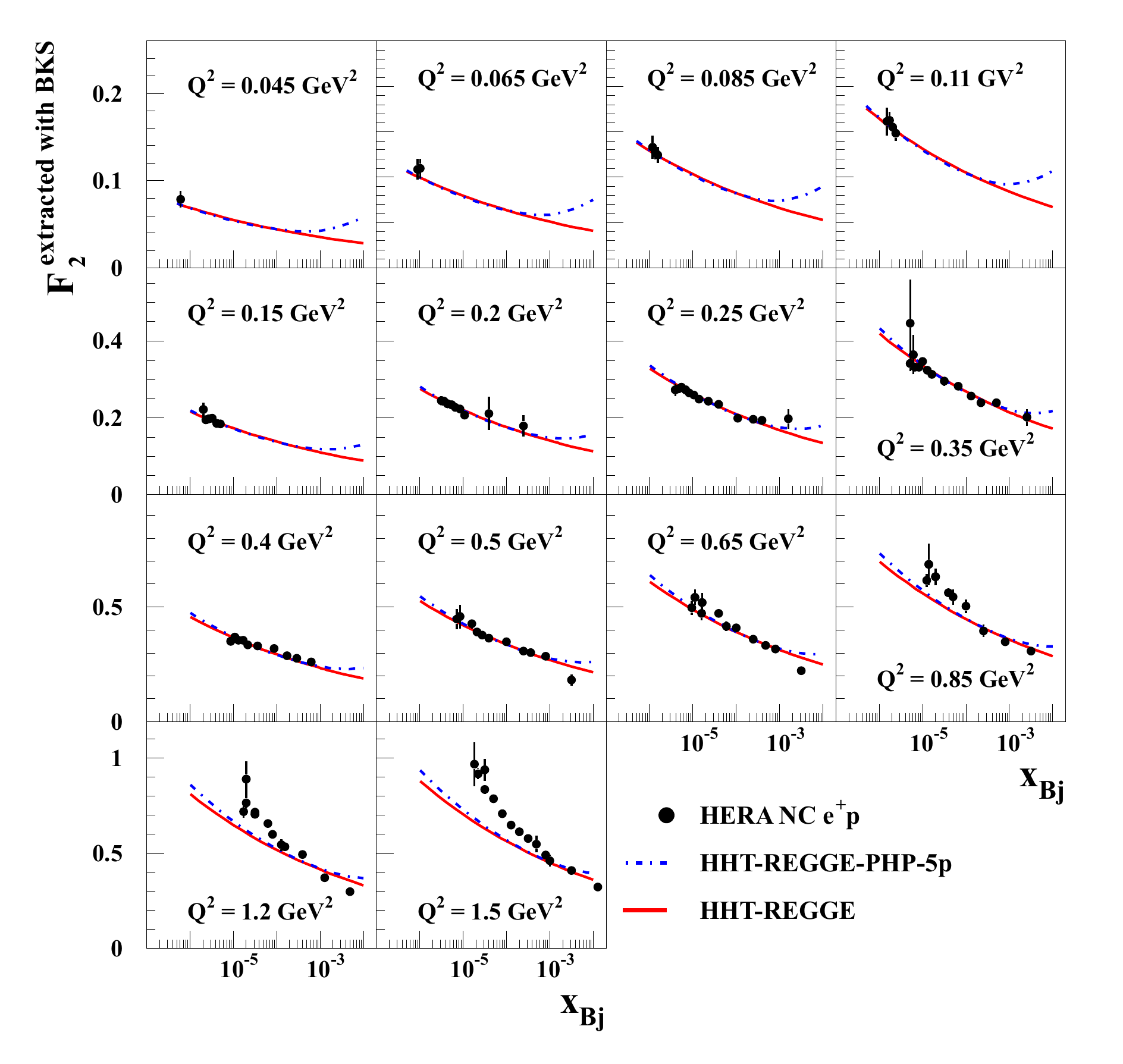}}
%  \epsfig{figure=figures/ncepa-all-regge-noft-withphp.eps,width=0.9\textwidth}
%}
\caption {The structure-function $F_2$ as a function of 
          $x_{\rm Bj}$ for low $Q^2$ as extracted from the
          HERA NC $e^+p$ cross sections, 
          together with predictions from the HHT-REGGE and
          HHT-REGGE-PHP-5p fits,
          see text for details.}
\label{fig:regge-php}
\end{figure}

Photoproduction data~\cite{photo2} were also included in
a fit HHT-REGGE-PHP-5p with both $A_{I\!R}$  
and $\alpha_{I\!R}(0)$ as free parameters. 
The inclusion of the low-$W$ photoproduction data
constrains the Reggeon term very effectively.
The parameters are listed 
in Table~\ref{tab:regge}.
Figure~\ref{fig:regge-php}  shows $F_2$ at low $Q^2$
as extracted from HERA $e^+p$ cross sections,               
together with the predictions from HHT-REGGE and HHT-REGGE-PHP-5p.
Within the kinematic range of the HERA data, the predictions
from the two fits are basically identical. 
The inclusion of the fixed-target data does not improve the 
fit, see Table~\ref{tab:regge}.

The value of the intercept $\alpha_{I\!P}(0)$ is of particular interest. 
The results are compatible for all HHT-REGGE fits.
The value of $1.097 \pm 0.004$\,(stat) from the default fit is
compatible with the old ZEUSREGGE results and with values
given in the literature~\cite{alpha:don,alpha:cud}.
%such as 1.081~\cite{alpha:don}
%and $1.093 \pm 0.003$~\cite{alpha:cud}. 

\section{Characteristics of {\boldmath${F_2}$} }

The DGLAP equations, on which the pQCD analyses are based, 
arise from the resummation of terms proportional to
$\alpha_s^n ({\rm ln} Q^2)^m$, where $\alpha_s$ is the 
strong coupling constant and
$n=m$ ($n=m-1$) at leading (next-to-leading) order.
These equations do not make a prediction on the shape of
the parton distributions themselves, but they
describe how the parton distributions evolve with $Q^2$.
At low $x$, 
the gluon PDF becomes dominant.
The steep rise of the gluon PDF with decreasing $x$ results in a
steep rise of $F_2$ with decreasing $x_{\rm Bj}$.

At fixed $Q^2$, $F_2$ can be parameterised as 
\begin{equation}
F_2 = C(Q^2) x_{\rm Bj}^{\,-\lambda(Q^2)} ~, 
\label{eqn:lambda}
\end{equation} 
where $C(Q^2)$ and $\lambda(Q^2)$ are parameters to be fit for each $Q^2$.
This parameterisation is inspired by QCD. 
At leading order (LO), the DGLAP evolution of the gluon PDF gives 
$x g(x,Q^2)$  proportional to $x^{-\lambda_g}$~\cite{Mandy2} with 
$\lambda_g= \sqrt { 12 \ln(t/t_0)/ (\beta_0 \ln(1/x)) }$, where $t$ and $t_0$ 
are $\ln(Q^2/\Lambda_{\rm QCD}^2)$ 
and $\ln(Q_0^2/\Lambda_{\rm QCD}^2)$, respectively, $Q_0^2$ is
the $Q^2$ at which the DGLAP evolution starts and 
$\Lambda_{\rm QCD}$ is the QCD scale parameter.
The parameter $\beta_0$  is the QCD beta function at leading order.
As $x g(x,Q^2)$ is dominant at low $x$, 
the evolution of $F_2$, $d F_2/d \ln(Q^2)$,
is dominated by the gluon and a dependence 
$F_2(x,Q^2) \propto x^{-\lambda_s}$ is expected, 
with $\lambda_s = \lambda_g -\epsilon$, where $\epsilon$ is a small
offset~\cite{Mandy2}. 
As a result, $\lambda_s$ has an approximately logarithmic dependence
on $Q^2$ via the $\ln(t)$ term, but it also depends on $\ln(1/x)$. 
Therefore, the parameterisation of Eq.~\ref{eqn:lambda} cannot be called
a QCD prediction, but rather an approximation of LO QCD.

Regge theory suggests that $\lambda = \alpha_{I\!P}(0)-1$ 
is approximately constant in the regime 
of soft Pomeron exchange, i.e.\ $Q^2 \lesssim 0.65$\,GeV$^2$. 
At higher $Q^2$, $\lambda$ can rise. This is also called
the regime of ``effective Pomeron'' exchange~\cite{Regge4}.
However, it is not included in the $ansatz$ of Eq.~\ref{eqn:regge}.
 
The values of $F_2$ in each $Q^2$ bin were fit according to
Eq.~\ref{eqn:lambda} for $x_{\rm Bj} < 0.01$, i.e.\ in a region
were valence quarks can be neglected. 
The four lowest-$Q^2$ bins were combined; 
as individual bins they have too few data points 
to produce a stable fit.
The combination was achieved 
by translating the points from their respective $Q^2$
values with fixed $x_{\rm Bj}$ to $Q^2 = 0.11$\,GeV$^2$, using the
predictions from the HHT-ALLM fit. The corrections to $F_2$
were typically around 25\,\%. 
%only for the lowest values of 
%$Q^2=0.045$\,GeV$^2$
%and $x_{\rm Bj}=6.21 \cdot 10^{-7}$, the correction was 140\,\%.

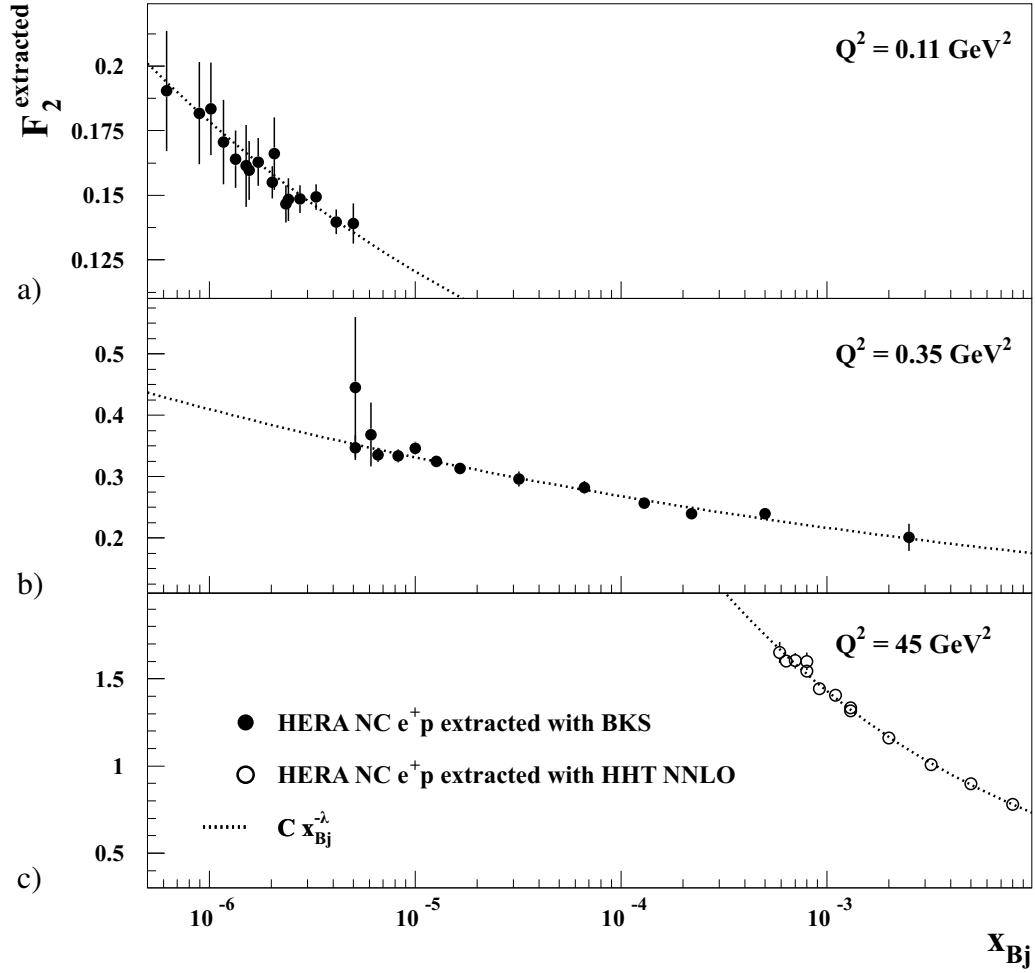
\begin{figure}[tbp]
\vspace{-5.5cm} 
 \centering
 \setlength{\unitlength}{0.1\textwidth}
 \begin{picture} (9,12)
  \put(0,0){\includegraphics[width=0.9\textwidth]{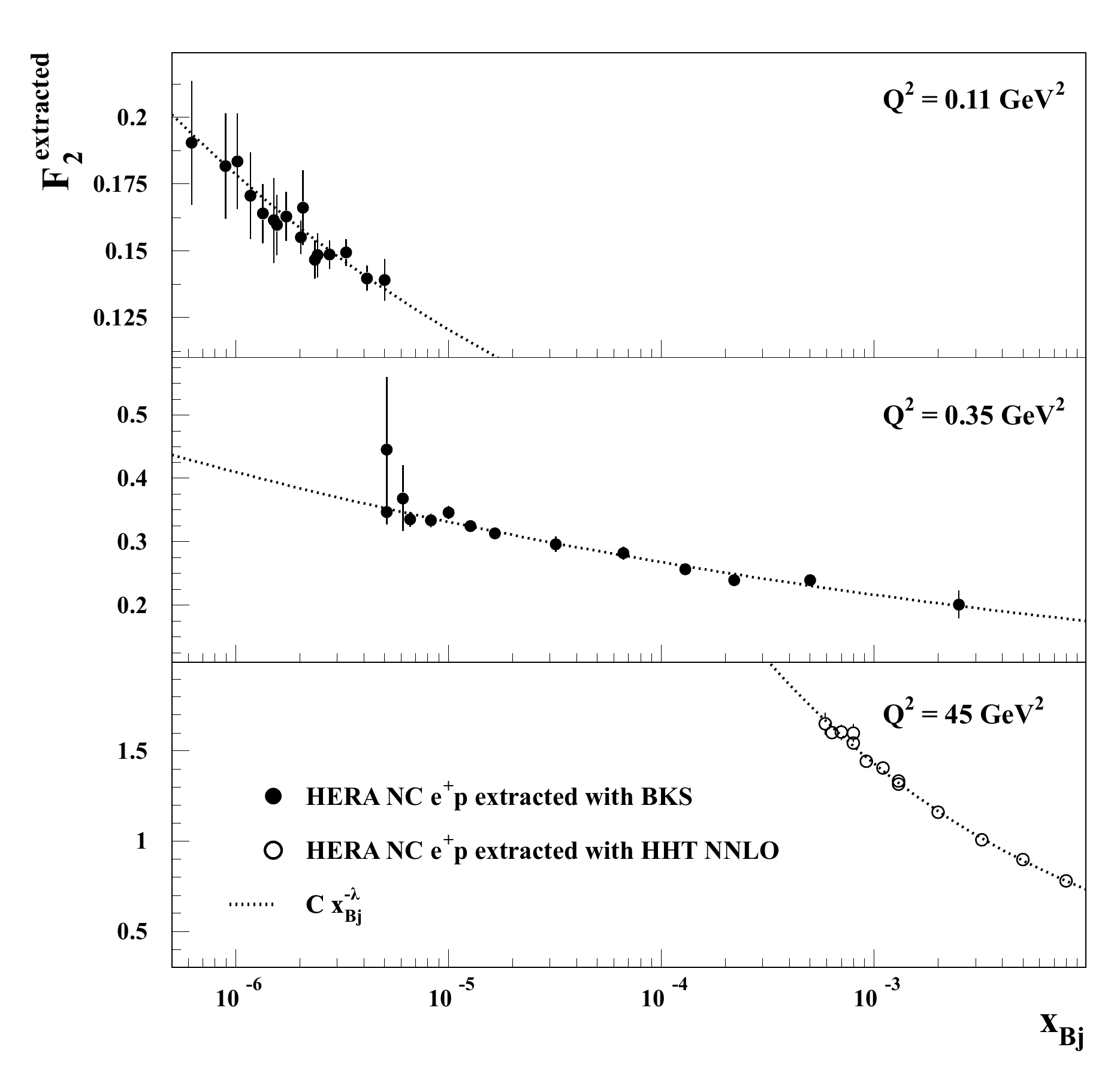}}
   \put (0.3,5.75) {a)} 
   \put (0.3,3.3) {b)}
   \put (0.3,0.85) {c)}
 \end{picture}
%\centerline{
%  \epsfig{figure=figures/lambda-fits.eps,width=0.9\textwidth}
%}
\caption {Values of $F_2$ extracted with the BKS model for 
          a) $Q^2=0.11$\,GeV$^2$, b) $Q^2=0.35$\,GeV$^2$, 
          and c) for $Q^2=45$\,GeV$^2$ extracted within 
          the framework of pQCD using the
          results of the HHT NNLO analysis.
          Also shown are fits according to
          $F_2 = C \,x_{\rm Bj}^{~-\lambda}$.
          }
\label{fig:lambda-bin-fits}
\end{figure}

The values of $F_2(x_{\rm Bj})$ and the fits are depicted 
in Fig.~\ref{fig:lambda-bin-fits} for $Q^2=0.11$, $ 0.35$ and
$45$\,GeV$^2$. 
The fits are very good for all $Q^2$ with $\chi^2/{\rm ndf} < 1$.
The values for $\lambda$ and $C$ are given in 
Tables~\ref{tab:lambda:BKS} and~\ref{tab:lambda:HHT} for fits
to BKS- and HHT-extracted $F_2$ values, respectively.
The $Q^2$ range is considerably extended
compared to values previously reported
by the H1 collaboration~\cite{LAMBDA:H1}; 
in the common range, the values of $\lambda$ and $C$ 
are compatible.

\begin{figure}[tbp]
\vspace{-5.5cm} 
 \centering
 \setlength{\unitlength}{0.1\textwidth}
 \begin{picture} (9,12)
  \put(0,4.3){\includegraphics[width=0.9\textwidth]{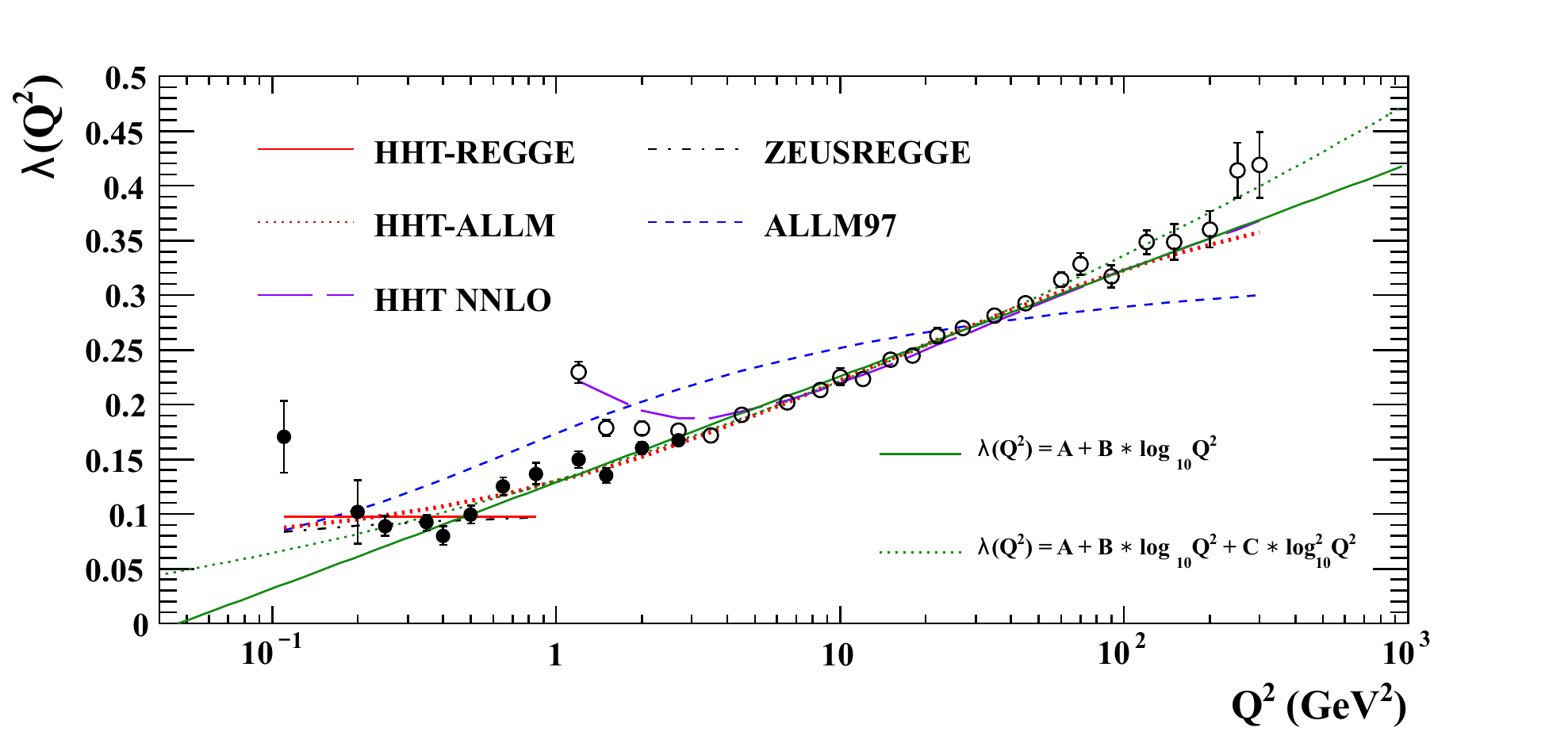}}
  \put(0,0){\includegraphics[width=0.9\textwidth]{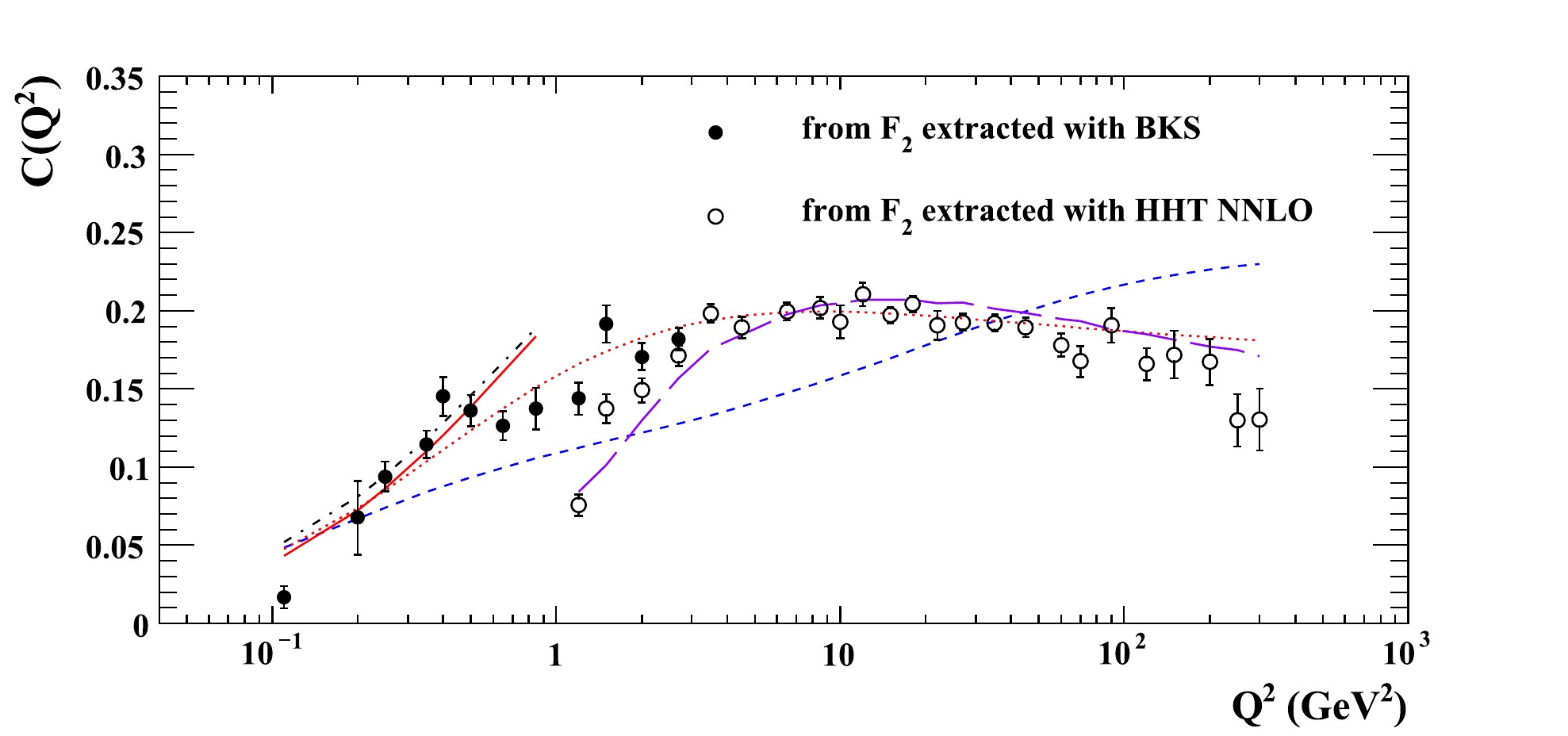}}
   \put (0.2,4.85) {a)} 
   \put (0.2,0.55)  {b)}
 \end{picture}
%\centerline{
%  \epsfig{figure=figures/lambda_fit.eps,width=0.9\textwidth}
%}
%\centerline{
%  \epsfig{figure=figures/C_fit.eps,width=0.9\textwidth}
%}
\caption {Values of a) $\lambda$ and b) $C$ determined in fits to
          $F_2(x_{\rm Bj},Q^2) = C(Q^2) \,x_{\rm Bj}^{~-\lambda(Q^2)}$,
          where $F_2$ was extracted either within the framework of pQCD using
          the results of the HHT NNLO analysis or with the BKS model,
          as appropriate. 
          Also shown are predictions from HHT NNLO, HHT-ALLM, ALLM97,
          HHT-REGGE and ZEUSREGGE, for details see text, and fits
          to $\lambda$
          with a straight line and a quadratic function. 
         }
\label{fig:lambda-fit}
\end{figure}

Figure~\ref{fig:lambda-fit}
depicts the dependence of the fit parameters 
$\lambda$ and $C$ on $Q^2$.
These parameters are shown down to $Q^2 = 1.2$\,GeV$^2$
for $F_2$ extracted using results from the HHT NNLO analysis, 
while they are shown for $Q^2$ up to 2.7\,GeV$^2$
for $F_2$ extracted  using the 
BKS model\,\footnote{The extraction is based on numerical
results provided by the authors~\cite{BKS}.}.   
%is not considered valid for higher $Q^2$. 
The values extracted with HHT NNLO and BKS differ markedly
in the overlap region.
The HHT NNLO analysis implies a strong rise of $\lambda$ for
$Q^2 < 2.0$\,GeV$^2$. 
This confirms earlier findings~\cite{HHT} that 
the HHT NNLO analysis becomes unphysical
below $Q^2 = 2.0$\,GeV$^2$, even though it can describe
cross-section data in this transition region.
By contrast, the extractions using the BKS model connect very 
smoothly to
the extractions and predictions of HHT NNLO around 2\,GeV$^2$.

Also shown in Fig.~\ref{fig:lambda-fit} are predictions
from HHT-ALLM, ALLM97, HHT-REGGE and ZEUSREGGE. 
The old ALLM97 fit is based on an early subset of HERA data and
cannot describe the full HERA data set. The predictions
of  HHT-ALLM
%the new fit to the combined HERA data, 
describe the
$Q^2$ dependence of $\lambda$ and $C$ quite well over 
the whole kinematic range. The large number of parameters
make the ALLM parameterisation very flexible.
The predictions of HHT-REGGE and ZEUSREGGE only describe the
extracted values of $\lambda$ and $C$ for values of 
$Q^2 \lesssim 0.5$\,GeV$^2$.
The almost constant $\lambda$ in this regime
is the Regge soft-Pomeron explanation 
for the behaviour of $\sigma^{\gamma^*p}$ for low $Q^2$,
as depicted in Fig.~\ref{fig:sigma-W}.
It should be noted that the BKS model 
and the ALLM parameterisation 
are both also based on the
Regge $ansatz$. 
Thus, the agreement 
between values of $\lambda$ from $F_2$ extracted with the BKS model
and the predictions from the HHT-ALLM parameterisation and the Regge
fits might be an artefact. % of the Regge relationship. 
The problem in this context is that no model-independent way
to extract $F_2$ exists.
The $\lambda$ values were also fitted with a 
first- or second-order polynomial in 
$\log_{10}Q^2$; %down to $Q^2=0.2$\,GeV$^2$;
for this, the HHT NNLO extraction was used for $Q^2 > 2.0$\,GeV$^2$
and the BKS extraction for lower $Q^2$.
The descriptions give  
$\chi^2/{\rm ndf} =2.5$ (1.5) for the first- (second-) order
polynomial.

The evolution of $F_2$, $d F_2/d \ln Q^2 $, is an interesting quantity
because, at sufficiently low values of $x_{\rm Bj}$, it 
is connected to the gluon content of the nucleon.
In pQCD, this is described by the gluon PDF.
In pure Regge phenomenology, there is no concept of a gluon.
However, Pomeron exchange is often interpreted as 
an exchange of a gluon ladder.  
In the alternative $ansatz$ of the
colour-dipole model~\cite{Allen}, the photon splits into a $q \bar{q}$ pair
which has time to evolve 
%into 
a complex hadronic structure
through gluon radiation over a coherence length proportional to
$1/Q^2$.
While this scenario seems conceptually different,
it is connected to the usual picture of the proton structure
via a Lorentz transformation.
It has been used~\cite{Allen} to investigate the 
data on $ \sigma^{\gamma^*p}$ and also describes the data well.

The values of $d F_2/d \ln Q^2 $ and their uncertainties were
derived from fits of quadratic functions in $\ln Q^2$
to the extracted values of $F_2$,
\begin{equation}
F_2 =  A(x_{\rm Bj}) + B(x_{\rm Bj}) \ln Q^2+  C(x_{\rm Bj}) \ln^2 Q^2 ~, 
\label{eqn:f2fit}
\end{equation} 
where $A$, $B$ and $C$ were free parameters. 
The values of $\chi^2/{\rm ndf}$ were $\lesssim 1$ for all fits.
As some values of the $x_{\rm Bj}$ grid had too few corresponding values
on the $Q^2$ grid, the ALLM predictions
were used to translate $F_2$ points along $Q^2$ trajectories 
to a reduced $x_{\rm Bj}$ grid.
The translation factors were on the percent level.

%=======

\begin{figure}[tbp]
\vspace{-0.5cm} 
%\vspace*{5pt}
\centerline{
  \includegraphics[width=0.9\textwidth]{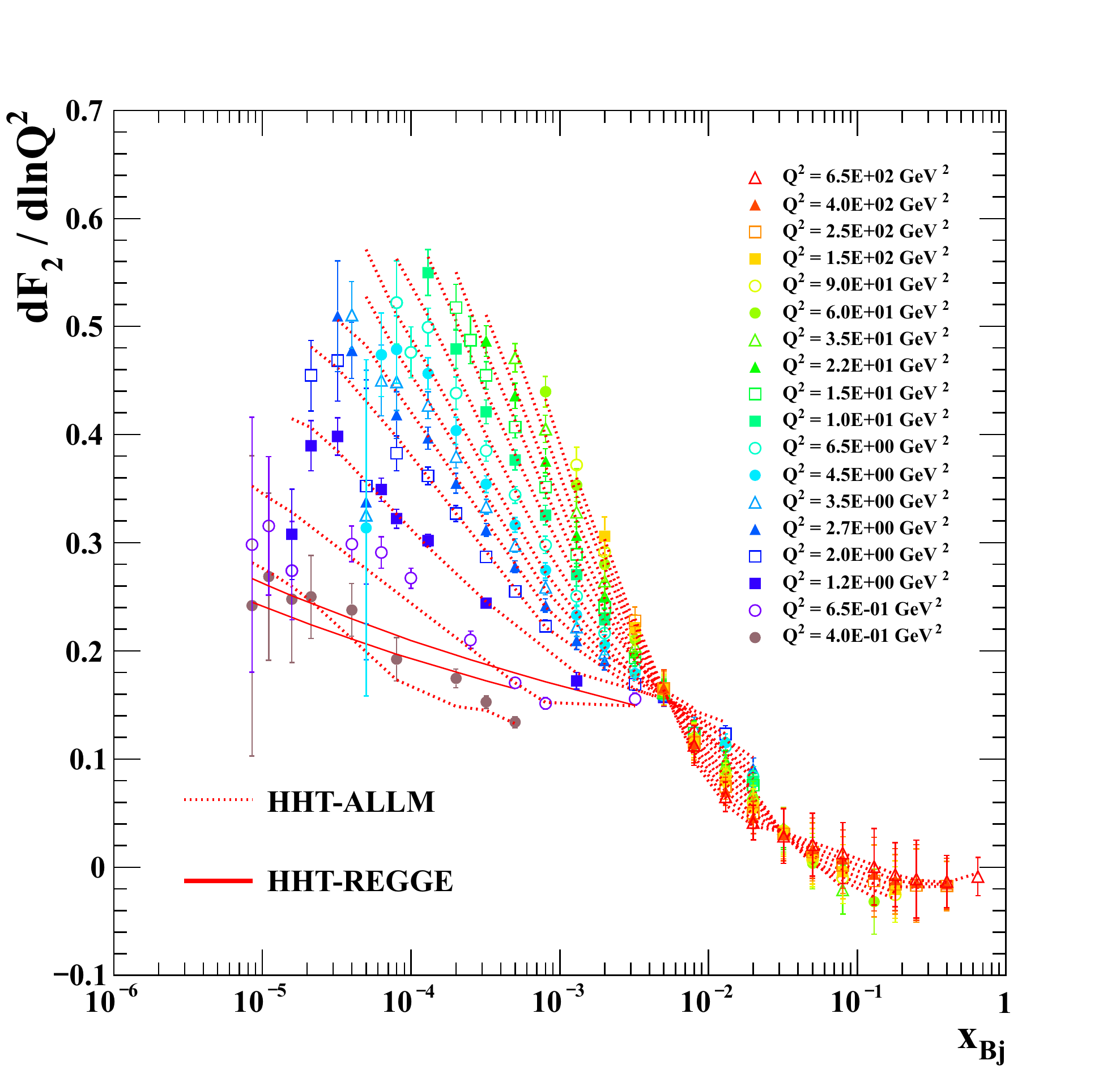}}
\caption {The derivative $d F_2/d \ln Q^2 $ as a function of 
          $x_{\rm Bj}$ for selected values of $Q^2$ over the full range
          in $x_{\rm Bj}$.
          Also shown are  HHT-ALLM predictions (dotted lines)
          and HHT-REGGE predictions for 
          $Q^2=0.4$ (lower solid line) and $Q^2=0.65$\,GeV$^2$
          (upper solid line).}
\label{fig:F2deriv}
\end{figure}

The results on $d F_2/d \ln Q^2$
are shown in Fig.~\ref{fig:F2deriv}
as a function of $x_{\rm Bj}$ 
for selected values of $Q^2$. 
Also shown are the predictions of the HHT-ALLM
and HHT-REGGE fits.
The HHT-ALLM predictions follow the data quite well
for $Q^2$ above 2\,GeV$^2$. 
At lower $Q^2$, neither the HHT-ALLM nor the 
HHT-REGGE fit agrees with the data for 
$x_{\rm Bj} \gtrsim 5 \cdot 10^{-5}$.

\begin{figure}[tbp]
\vspace{-0.5cm} 
%\vspace*{5pt}
\centerline{
  \includegraphics[width=0.9\textwidth]{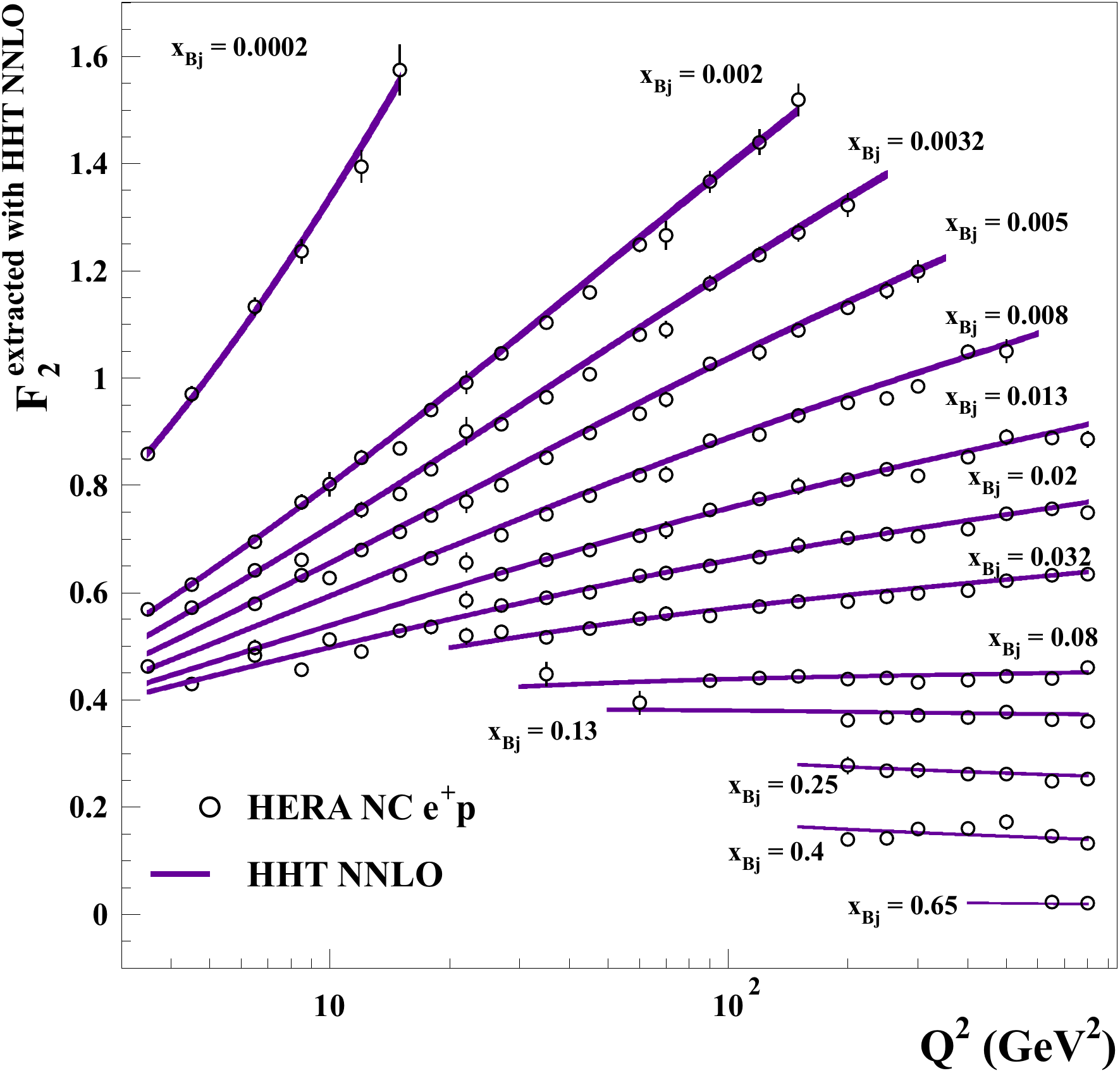}}
\caption {The structure-function $F_2(Q^2)$ for selected values 
          of $x_{\rm Bj}$ in the $Q^2$-range where  
          $F_2$ can be extracted within the framework of pQCD
          using the HHT NNLO analysis. Also shown are the predictions
          from HHT NNLO.
          The width of the bands represents the uncertainty 
          on the predictions. 
}
\label{fig:f2-h}
\end{figure}

For $ x_{\rm Bj} \ge 0.032$,
the HERA data show almost perfect scaling.
In this region, quarks and
gluons both contribute to the evolution of $F_2$ but with opposite signs.
The rate of scaling violation is determined by the 
value of $\alpha_s(Q^2)$, which is relatively small here. 
For $0.005 \le x_{\rm Bj} < 0.032$, 
the gluon contribution becomes increasingly important and results in
positive scaling violations.
The range of the data in $Q^2$ is sufficiently wide,
from $Q^2 = 2.0$ to $650$\,GeV$^2$, to demonstrate the 
$\alpha_s(Q^2)$ dependence of the scaling violations.
The scaling violations noticeably decrease as $\alpha_s(Q^2)$
decreases with $Q^2$.
This is also demonstrated in Fig.~\ref{fig:f2-h}, which shows 
$F_2$ depending on $Q^2$ for selected values of $x_{\rm Bj}$
in the regime of pQCD.

For $x_{\rm Bj} < 5 \cdot 10^{-3}$, 
where the $dF_2/d\ln Q^2$ values come together, 
the gluon contribution is dominant, 
leading to strong scaling violations, which increase as $Q^2$ 
increases.
In this region, the range of the data in $Q^2$ is no longer 
sufficient to see any damping of this due to the dependence
of $\alpha_s(Q^2)$ on $Q^2$.

\begin{figure}[tbp]
\vspace{-5.5cm} 
 \centering
 \setlength{\unitlength}{0.1\textwidth}
 \begin{picture} (9,12)
  \put(0,0){\includegraphics[width=0.9\textwidth]
                         {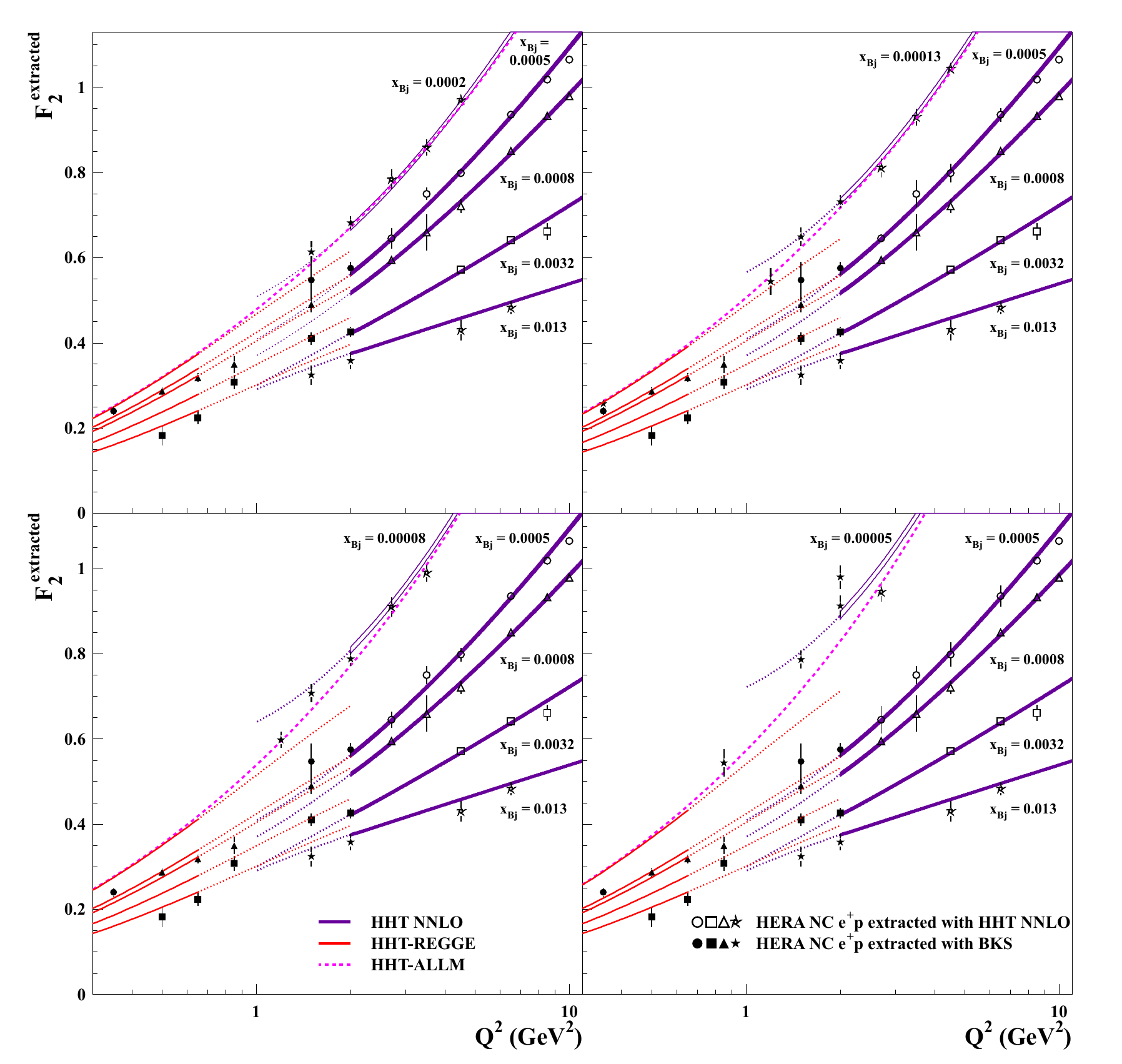}}
   \put (1.0,7.9) {a)} 
   \put (4.9,7.9) {b)}
   \put (1.0,4.0) {c)}
   \put (4.9,4.0) {d)}
 \end{picture}
%\centerline{
%  \epsfig{figure=figures/fig5mod-f2-b-all-allm.eps,width=0.9\textwidth}
%}
\caption {
          The structure-function $F_2(Q^2)$  
          for selected values of $x_{\rm Bj}$, extracted
          with the BKS model for $Q^2 \le 2$\,GeV$^2$ and 
          with results from the HHT NNLO
          analysis for $Q^2 > 2$\,GeV$^2$.
          Each of the four plots contains data for four $x_{\rm Bj}$ values 
          from $0.013$ to $5 \cdot 10^{-4}$ and one additional 
          $x_{\rm Bj}$ value, 
          ranging from $x_{\rm Bj} = 2 \cdot 10^{-4}$ in a) to 
          $x_{\rm Bj} = 5 \cdot 10^{-5}$ in d).
          Also shown are the predictions from the HHT-REGGE fit and
          the HHT NNLO analysis, together with the HHT-ALLM prediction.
          Dotted lines indicate extrapolations beyond the fitted regions.
          The width of the bands represents the uncertainty 
          on the HHT NNLO predictions. No uncertainties were computed
          for the HHT-REGGE and HHT-ALLM predictions. 
          For $x_{\rm Bj}=0.00005$ and $Q^2 = 2$\,GeV$^2$,
         two points are shown, extracted from data with
          $\sqrt{s}=318$ and 300\,GeV, respectively.
}
\label{fig:f2-l}
\end{figure}
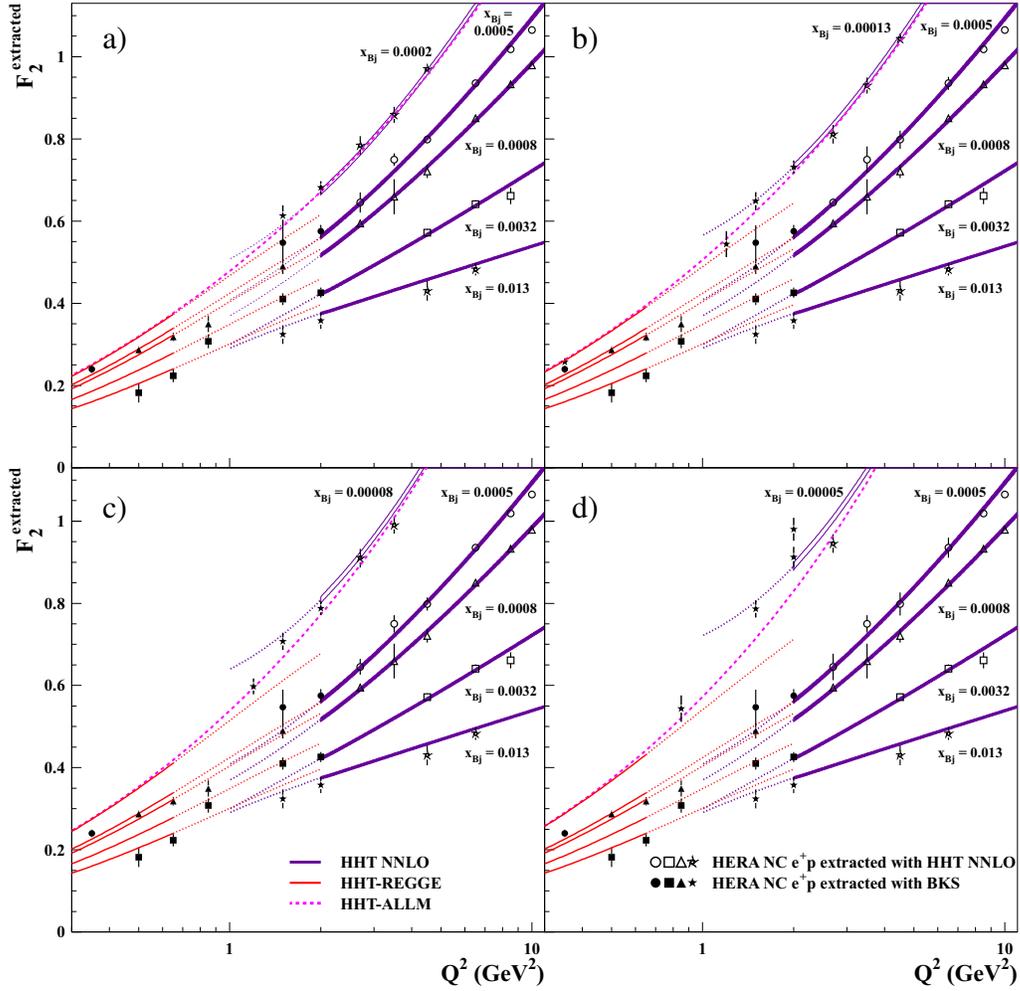

Figure~\ref{fig:f2-l} shows $F_2$ for different values of $x_{\rm Bj}$ 
as a function of $Q^2$, in order to explore in more detail the region 
at high $W$ (equivalent to low  $x_{\rm Bj}$) and low $Q^2$  
in which disagreement between the Regge-based models and 
the data was observed in Fig.~\ref{fig:sigma-low-Q2}.
Each plot contains data for the same four
relatively high $x_{\rm Bj}$ values with additional data for one lower
value of  $x_{\rm Bj}$. 
For the four high-$x_{\rm Bj}$ values repeated in each plot, 
the extrapolations of the HHT-REGGE and HHT-NNLO models 
are close and describe the data well.
For the highest  $x_{\rm Bj}$ value shown in Fig.~\ref{fig:f2-l}\,a),
$x_{\rm Bj} = 2 \cdot 10^{-4}$,  a gap between the extrapolations from
the HHT-REGGE and HHT-NNLO models begins to be visible. 
Figures~\ref{fig:f2-l}\,a)\,--\,d) demonstrate how  
this gap grows as $x_{\rm Bj}$ decreases, i.e\ as $W$ increases,
until in Fig.~\ref{fig:f2-l}\,d) it has become substantial.
Both models therefore clearly fail to describe the data in 
the transition region, 
whereas the data smoothly connect across it.
Even the ALLM fit has a tendency to undershoot 
the data for $x_{\rm Bj}=5 \cdot 10^{-5}$. 
The behaviour of the $F_2$ data, rising steeply in the transition region
to match onto the predicted pQCD behaviour at higher $Q^2$, 
is not an artefact of the method used to 
extract $F_2$, since the behaviour is observed 
for $F_2$ values all extracted using the BKS model; the BKS model might 
{\it a priori} have been expected to favour a much shallower 
Regge-like behaviour. 
Once again, 
the data indicate a smooth transition between the regions in 
which Regge theory and pQCD
give good descriptions of the data.

\begin{figure}[tbp]
\vspace{-0.5cm} 
%\vspace*{5pt}
\centerline{
  \includegraphics[width=0.9\textwidth]{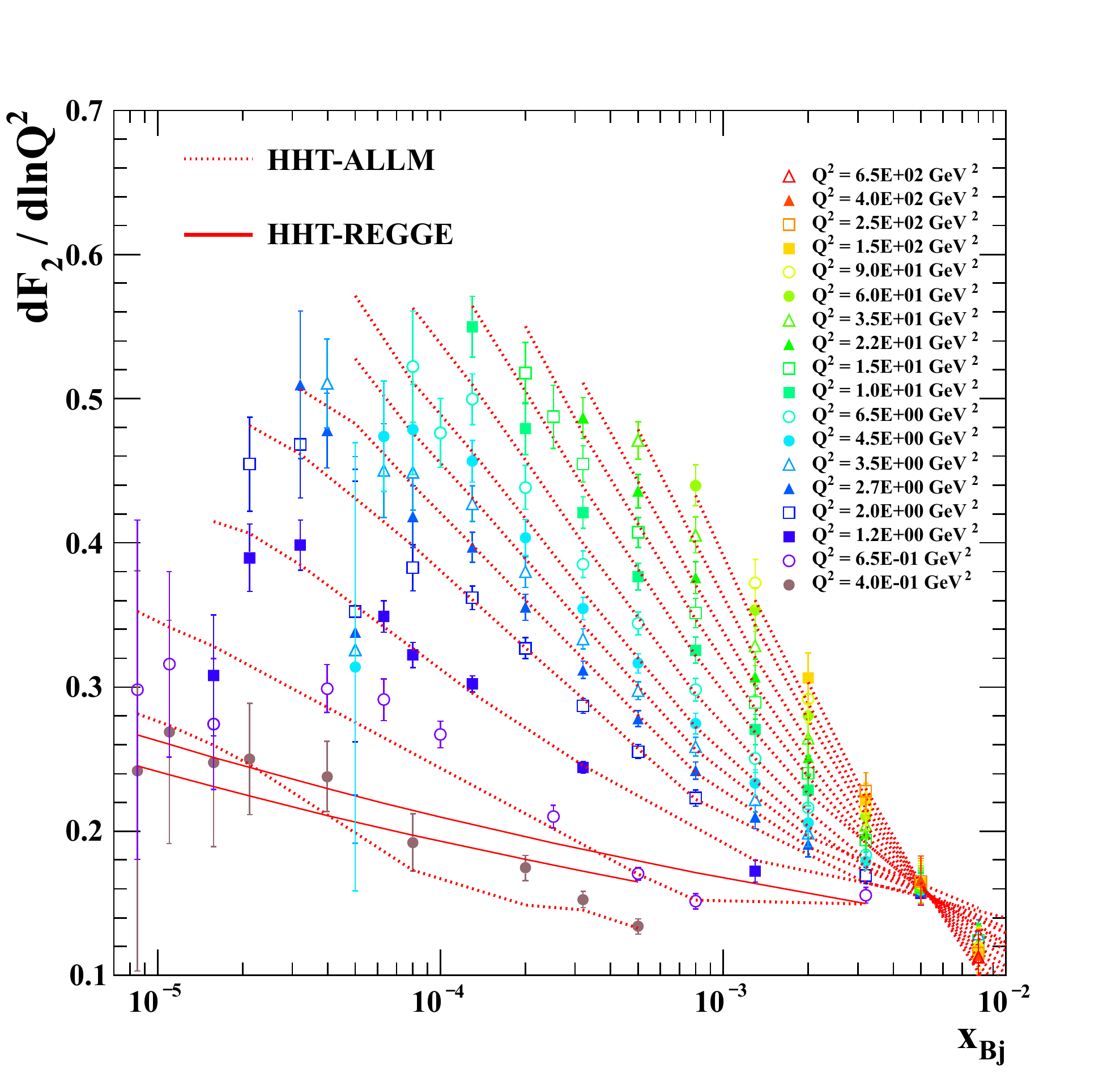}}
\caption {The derivative $d F_2/d \ln Q^2$ as a function of 
          $x_{\rm Bj}$ for selected values of $Q^2$ for
           $x_{\rm Bj} < 10^{-2}$.
          Also shown are  HHT-ALLM predictions (dotted lines)
          and HHT-REGGE predictions for 
          $Q^2=0.4$ (lower solid line) and $Q^2=0.65$\,GeV$^2$
          (upper solid line).
}
\label{fig:F2deriv-mag}
\end{figure}

The derivative $d F_2/d \ln Q^2$ is shown on an enlarged scale for  
$x_{\rm Bj} <  10^{-2}$ in Fig.~\ref{fig:F2deriv-mag}.
The data indicate that there is a turn-over which
moves to lower values of $x_{\rm Bj}$ as $Q^2$ 
%This follows the behaviour of the NC cross sections
%which also show such turn-overs.
decreases\,\footnote{At $x_{\rm Bj} = 5 \cdot 10^{-5}$, the values drop for all
available $Q^2$.  
This dip is associated with a feature of the
data, which is, however, of limited statistical
significance.}.
%The NC $e^+p$ cross sections show extra dips 
%at this  $x_{\rm Bj}$ for the respective $Q^2$ values.
%As the uncertainties are, however, large, this might
%be a coincidence.
A colour-dipole inspired
model introduced by Golec-Biernat and W{\"u}sthoff (GBW)~\cite{gb}
predicts that a ``critical line'' between regimes of ``soft'' and
``hard'' scattering associated with saturation would cross the
kinematic range of  Fig.~\ref{fig:F2deriv-mag}.
They predict that the critical value of 
$x_{\rm Bj}$ increases with decreasing $Q^2$ 
and would be
$x_{\rm Bj} \approx 2 \cdot 10^{-5}$ for $Q^2 = 2$\,GeV$^2$ and
$x_{\rm Bj} \approx 6 \cdot 10^{-4}$ at $Q^2 = 0.65$\,GeV$^2$.
However, there is no prediction on how the critical line
would manifest itself.
Thus, the observed lack of any transition in
the behaviour of $d F_2/d \ln Q^2$ in these regions cannot 
confirm or exclude  the GBW predictions.

\section{Summary and Outlook \label{sec:sum}}

The combined HERA $e^+p$ NC cross sections were studied over
their complete range in $x_{\rm Bj}$ and $Q^2$,
$6.21\cdot10^{-7} \le x_{\rm Bj} \le 0.65$
and $ 0.045 \le Q^2 \le 30000$\,GeV$^2$.
They were used to extract the structure-function $F_2$ and 
the photon--proton cross section $\sigma^{\gamma^*p}$. 
This was done by employing the pQCD analysis HHT NNLO
for $Q^2 > 2.0$\,GeV$^2$ and a low-$Q^2$ approximation
of the reduced cross-section together with the Regge-inspired BKS model for 
$Q^2 \le 2.0$\,GeV$^2$. 
The  extracted values of $F_2$
connect smoothly at this transition point.
A new fit, HHT-REGGE, can describe the $F_2$ data well
up to $Q^2 \le 0.65$\,GeV$^2$, i.e.\ in the regime of the soft pomeron.
The values of $\sigma^{\gamma^*p}$ derived from the
$F_2$ data are well described 
by a new HHT-ALLM fit over most of the phase space, with the exception
of data at the highest $W$, where
its failure can be ascribed to  the fast rise of the $F_2$ data
between $Q^2$ of 1 and 2\,GeV$^2$ for the lowest values of $x_{\rm Bj}$.

The characteristics of $F_2$ were also studied 
by extracting $\lambda$ parameters.
The dependence of $\lambda$ on $Q^2$ is reasonably well described 
by the HHT NNLO predictions for $Q^2 \ge 2.0$\,GeV$^2$
and predictions from the new HHT-REGGE fit for $Q^2 \le 0.65$\,GeV$^2$.
The absence of a clear transition between perturbative
and non-perturbative behaviour in the data is illustrated 
by the fact that the extracted values of $\lambda$ can also be fitted 
reasonably well with a second-order
polynomial in $\log_{10}Q^2$.

The derivative $d F_2/d \ln Q^2$ behaves as expected for higher $Q^2$.
For lower $Q^2$, it is observed that $d F_2/d \ln Q^2$
as a function of $x_{\rm Bj}$ shows turn-overs that move towards
lower $x_{\rm Bj}$ as $Q^2$ decreases.
It is unclear whether this can be ascribed to gluon saturation.

The construction of a new electron--proton or electron--ion
collider\,\cite{AM:newcolliders,Accardi:2012qut} would provide more data
to widen the kinematic range of future studies. In the meantime, it is
hoped that the data and studies presented in this paper will catalyse
theoretical work to shed further light on the transition region.

%%%% It can be hoped that a future electron--proton or electron--ion
%%%% collider\,\cite{AM:newcolliders,Accardi:2012qut}
%%%% will provide more data to widen the kinematic range
%%%% of future studies. 
%%%% The data as presented 
%%%% in this paper are important input for model building 
%%%% in the low-$Q^2$ and low-$x_{\rm Bj}$ regime.

\section{Acknowledgements}

We would like to thank Halina Abramowicz and Aharon Levy
for discussions about ALLM and
Barbara Bade{\l}ek and Anna Stasto for the discussions 
on the BKS model and their help in providing
their results numerically.
We would like to thank Paul Newman for discussions
on Regge phenomenology.
We thank our funding agencies, especially the
Humboldt foundation and the MPG, 
for financial support and DESY for the hospitality extended 
to the non-DESY authors.

\clearpage
\bibliography{new}

\begin{thebibliography}{39}
\expandafter\ifx\csname natexlab\endcsname\relax\def\natexlab#1{#1}\fi
\expandafter\ifx\csname bibnamefont\endcsname\relax
  \def\bibnamefont#1{#1}\fi
\expandafter\ifx\csname bibfnamefont\endcsname\relax
  \def\bibfnamefont#1{#1}\fi
\expandafter\ifx\csname citenamefont\endcsname\relax
  \def\citenamefont#1{#1}\fi
\expandafter\ifx\csname url\endcsname\relax
  \def\url#1{\texttt{#1}}\fi
\expandafter\ifx\csname urlprefix\endcsname\relax\def\urlprefix{URL }\fi
\providecommand{\bibinfo}[2]{#2}
\providecommand{\eprint}[2][]{#2}

\bibitem[{\citenamefont{Abramowicz \emph{et~al.}}(2015)}]{HERAPDF20}
\bibinfo{author}{\bibfnamefont{H.}~\bibnamefont{Abramowicz}}
  \bibnamefont{\emph{et~al.}} [\bibinfo{collaboration}{ZEUS and H1}
  Collaboration], \bibinfo{journal}{Eur.\ Phys.\ J.\ C}
  \textbf{\bibinfo{volume}{75}}, \bibinfo{pages}{580} (\bibinfo{year}{2015}),
  \eprint{[arXiv:1506.06042]}.

\bibitem[{\citenamefont{Aaron \emph{et~al.}}(2010)}]{HERAIcombi}
\bibinfo{author}{\bibfnamefont{F.}~\bibnamefont{Aaron}}
  \bibnamefont{\emph{et~al.}} [\bibinfo{collaboration}{{ZEUS and H1}}
  Collaboration], \bibinfo{journal}{JHEP} \textbf{\bibinfo{volume}{1001}},
  \bibinfo{pages}{109} (\bibinfo{year}{2010}), \eprint{[arXiv:0911.0884]}.

\bibitem[{\citenamefont{Gribov and
  Lipatov}(1972{\natexlab{a}})}]{Gribov:1972ri}
\bibinfo{author}{\bibfnamefont{V.}~\bibnamefont{Gribov}} \bibnamefont{and}
  \bibinfo{author}{\bibfnamefont{L.}~\bibnamefont{Lipatov}},
  \bibinfo{journal}{Sov.\ J.\ Nucl.\ Phys.} \textbf{\bibinfo{volume}{15}},
  \bibinfo{pages}{438} (\bibinfo{year}{1972}{\natexlab{a}}).

\bibitem[{\citenamefont{Gribov and
  Lipatov}(1972{\natexlab{b}})}]{Gribov:1972rt}
\bibinfo{author}{\bibfnamefont{V.}~\bibnamefont{Gribov}} \bibnamefont{and}
  \bibinfo{author}{\bibfnamefont{L.}~\bibnamefont{Lipatov}},
  \bibinfo{journal}{Sov.\ J.\ Nucl.\ Phys.} \textbf{\bibinfo{volume}{15}},
  \bibinfo{pages}{675} (\bibinfo{year}{1972}{\natexlab{b}}).

\bibitem[{\citenamefont{Lipatov}(1975)}]{Lipatov:1974qm}
\bibinfo{author}{\bibfnamefont{L.}~\bibnamefont{Lipatov}},
  \bibinfo{journal}{Sov.\ J.\ Nucl.\ Phys.} \textbf{\bibinfo{volume}{20}},
  \bibinfo{pages}{94} (\bibinfo{year}{1975}).

\bibitem[{\citenamefont{Dokshitzer}(1977)}]{Dokshitzer:1977sg}
\bibinfo{author}{\bibfnamefont{Y.}~\bibnamefont{Dokshitzer}},
  \bibinfo{journal}{Sov.\ Phys.\ JETP} \textbf{\bibinfo{volume}{46}},
  \bibinfo{pages}{641} (\bibinfo{year}{1977}).

\bibitem[{\citenamefont{Altarelli and Parisi}(1977)}]{Altarelli:1977zs}
\bibinfo{author}{\bibfnamefont{G.}~\bibnamefont{Altarelli}} \bibnamefont{and}
  \bibinfo{author}{\bibfnamefont{G.}~\bibnamefont{Parisi}},
  \bibinfo{journal}{Nucl.\ Phys.\ B} \textbf{\bibinfo{volume}{126}},
  \bibinfo{pages}{298} (\bibinfo{year}{1977}).

\bibitem[{\citenamefont{Abt \emph{et~al.}}(2016)}]{HHT}
\bibinfo{author}{\bibfnamefont{I.}~\bibnamefont{Abt}}
  \bibnamefont{\emph{et~al.}} [\bibinfo{collaboration}{HHT} Collaboration],
  \bibinfo{journal}{Phys.\ Rev.\ D} \textbf{\bibinfo{volume}{94}},
  \bibinfo{pages}{034032} (\bibinfo{year}{2016}), \eprint{[arXiv:1604.02299]}.

\bibitem[{\citenamefont{Harland-Lang
  \emph{et~al.}}(2016)\citenamefont{Harland-Lang, Martin, Motylinski, and
  Thorne}}]{MMHT-HT-16}
\bibinfo{author}{\bibfnamefont{L.~A.} \bibnamefont{Harland-Lang}},
  \bibinfo{author}{\bibfnamefont{A.~D.} \bibnamefont{Martin}},
  \bibinfo{author}{\bibfnamefont{P.}~\bibnamefont{Motylinski}},
  \bibnamefont{and} \bibinfo{author}{\bibfnamefont{R.~S.}
  \bibnamefont{Thorne}}, \bibinfo{journal}{Eur.\ Phys.\ J.\ C}
  \textbf{\bibinfo{volume}{76}}, \bibinfo{pages}{186} (\bibinfo{year}{2016}),
  \eprint{[arXiv:1601.03413]}.

\bibitem[{\citenamefont{Collins}({(1977)})}]{Regge1}
\bibinfo{author}{\bibfnamefont{P.}~\bibnamefont{Collins}},
  \emph{\bibinfo{title}{An Introduction to Regge Theory and High Energy
  Physics}} (\bibinfo{publisher}{Cambridge University Press, Cambridge, UK},
  \bibinfo{year}{{1977}}).

\bibitem[{\citenamefont{Barone and Predazzi}({(2002)})}]{Regge2}
\bibinfo{author}{\bibfnamefont{V.}~\bibnamefont{Barone}} \bibnamefont{and}
  \bibinfo{author}{\bibfnamefont{E.}~\bibnamefont{Predazzi}},
  \emph{\bibinfo{title}{High Energy Particle Diffraction}}
  (\bibinfo{publisher}{in Texts and Monographs in Physics. Springer, Berlin},
  \bibinfo{year}{{(2002)}}).

\bibitem[{\citenamefont{Donnachie \emph{et~al.}}({(2002)})}]{Regge3}
\bibinfo{author}{\bibfnamefont{A.}~\bibnamefont{Donnachie}}
  \bibnamefont{\emph{et~al.}}, \emph{\bibinfo{title}{Pomeron Physics and QCD}}
  (\bibinfo{publisher}{Cambridge University Press}, \bibinfo{year}{{(2002)}}).

\bibitem[{\citenamefont{Newman and Wing}(2014)}]{Regge4}
\bibinfo{author}{\bibfnamefont{P.}~\bibnamefont{Newman}} \bibnamefont{and}
  \bibinfo{author}{\bibfnamefont{M.}~\bibnamefont{Wing}},
  \bibinfo{journal}{Rev.\ Mod.\ Phys.} \textbf{\bibinfo{volume}{86}},
  \bibinfo{pages}{1037} (\bibinfo{year}{2014}), \eprint{[arXiv:1308.3368]}.

\bibitem[{\citenamefont{Breitweg \emph{et~al.}}(1999)}]{ZEUSF2-98}
\bibinfo{author}{\bibfnamefont{J.}~\bibnamefont{Breitweg}}
  \bibnamefont{\emph{et~al.}} [\bibinfo{collaboration}{ZEUS} Collaboration],
  \bibinfo{journal}{Eur.\ Phys.\ J.\ C} \textbf{\bibinfo{volume}{7}},
  \bibinfo{pages}{609} (\bibinfo{year}{1999}).

\bibitem[{\citenamefont{Breitweg \emph{et~al.}}(2000)}]{ZEUSREGGE}
\bibinfo{author}{\bibfnamefont{J.}~\bibnamefont{Breitweg}}
  \bibnamefont{\emph{et~al.}} [\bibinfo{collaboration}{ZEUS} Collaboration],
  \bibinfo{journal}{Phys. Lett. B} \textbf{\bibinfo{volume}{487}},
  \bibinfo{pages}{53} (\bibinfo{year}{2000}).

\bibitem[{\citenamefont{Adloff \emph{et~al.}}(1997)}]{Regge:H1}
\bibinfo{author}{\bibfnamefont{C.}~\bibnamefont{Adloff}}
  \bibnamefont{\emph{et~al.}} [\bibinfo{collaboration}{H1} Collaboration],
  \bibinfo{journal}{Nucl.\ Phys.\ B} \textbf{\bibinfo{volume}{497}},
  \bibinfo{pages}{3} (\bibinfo{year}{1997}).

\bibitem[{\citenamefont{Andreev \emph{et~al.}}(2014)}]{H1FL2}
\bibinfo{author}{\bibfnamefont{V.}~\bibnamefont{Andreev}}
  \bibnamefont{\emph{et~al.}} [\bibinfo{collaboration}{H1} Collaboration],
  \bibinfo{journal}{Eur.\ Phys.\ J.\ C} \textbf{\bibinfo{volume}{74}},
  \bibinfo{pages}{2814} (\bibinfo{year}{2014}), \eprint{[arXiv:1312.4821]}.

\bibitem[{\citenamefont{Abramowicz \emph{et~al.}}(2014)}]{ZEUSFL}
\bibinfo{author}{\bibfnamefont{H.}~\bibnamefont{Abramowicz}}
  \bibnamefont{\emph{et~al.}} [\bibinfo{collaboration}{ZEUS} Collaboration],
  \bibinfo{journal}{Phys.\ Rev.\ D} \textbf{\bibinfo{volume}{90}},
  \bibinfo{pages}{072002} (\bibinfo{year}{2014}), \eprint{[arXiv:1404.6376]}.

\bibitem[{\citenamefont{Badelek \emph{et~al.}}(1997)}]{BKS}
\bibinfo{author}{\bibfnamefont{B.}~\bibnamefont{Badelek}},
  \bibinfo{author}{\bibfnamefont{J.}~\bibnamefont{Kwiecinski}}
  \bibnamefont{and} \bibinfo{author}{\bibfnamefont{A.}
  \bibnamefont{Stasto}}, \bibinfo{journal}{Z.\ Phys.\ C}
  \textbf{\bibinfo{volume}{74}}, \bibinfo{pages}{297} (\bibinfo{year}{1997}).

\bibitem[{\citenamefont{Martin \emph{et~al.}}(2006)\citenamefont{Martin,
  Stirling, and Thorne}}]{RT:FL:06}
\bibinfo{author}{\bibfnamefont{A.~D.} \bibnamefont{Martin}},
  \bibinfo{author}{\bibfnamefont{W.~J.} \bibnamefont{Stirling}},
  \bibnamefont{and} \bibinfo{author}{\bibfnamefont{R.~S.}
  \bibnamefont{Thorne}}, \bibinfo{journal}{Phys.\ Lett.\ B}
  \textbf{\bibinfo{volume}{635}}, \bibinfo{pages}{305} (\bibinfo{year}{2006}),
  \eprint{[hep-ph/0601247]}.

\bibitem[{\citenamefont{Hand}(1963)}]{Hand63}
\bibinfo{author}{\bibfnamefont{L.}~\bibnamefont{Hand}},
  \bibinfo{journal}{Phys.\ Rev.} \textbf{\bibinfo{volume}{129}},
  \bibinfo{pages}{1834} (\bibinfo{year}{1963}).

\bibitem[{\citenamefont{Caldwell}(2016)}]{Allen}
\bibinfo{author}{\bibfnamefont{A.}~\bibnamefont{Caldwell}},
  \bibinfo{journal}{New. J. Phys.} \textbf{\bibinfo{volume}{18}},
  \bibinfo{pages}{073019} (\bibinfo{year}{2016}),
  \eprint{[arXiv:1601.04472v1]}.

\bibitem[{\citenamefont{Abramowicz \emph{et~al.}}(1991)}]{ALLM91}
\bibinfo{author}{\bibfnamefont{H.}~\bibnamefont{Abramowicz}}
  \bibnamefont{\emph{et~al.}}, \bibinfo{journal}{Phys.\ Lett.\ B}
  \textbf{\bibinfo{volume}{269}}, \bibinfo{pages}{465} (\bibinfo{year}{1991}).

\bibitem[{\citenamefont{Abramowicz and Levy}(1997)}]{ALLM97}
\bibinfo{author}{\bibfnamefont{H.}~\bibnamefont{Abramowicz}} \bibnamefont{and}
  \bibinfo{author}{\bibfnamefont{A.}~\bibnamefont{Levy}}
  (\bibinfo{year}{1997}), \eprint{[hep-ph/9712415]}.

\bibitem[{\citenamefont{Adams \emph{et~al.}}(1996)}]{Adams:1996gu}
\bibinfo{author}{\bibfnamefont{M.~R.} \bibnamefont{Adams}}
  \bibnamefont{\emph{et~al.}} [\bibinfo{collaboration}{E665} Collaboration],
  \bibinfo{journal}{Phys.\ Rev.\ D} \textbf{\bibinfo{volume}{54}},
  \bibinfo{pages}{3006} (\bibinfo{year}{1996}).

\bibitem[{\citenamefont{Arneodo \emph{et~al.}}(1997)}]{Arneodo:1996qe}
\bibinfo{author}{\bibfnamefont{M.}~\bibnamefont{Arneodo}}
  \bibnamefont{\emph{et~al.}} [\bibinfo{collaboration}{New Muon}
  Collaboration], \bibinfo{journal}{Nucl.\ Phys.\ B}
  \textbf{\bibinfo{volume}{483}}, \bibinfo{pages}{3} (\bibinfo{year}{1997}),
  \eprint{[hep-ph/9610231]}.

\bibitem[{\citenamefont{Benvenuti \emph{et~al.}}(1989)}]{Benvenuti:1989rh}
\bibinfo{author}{\bibfnamefont{A.~C.} \bibnamefont{Benvenuti}}
  \bibnamefont{\emph{et~al.}} [\bibinfo{collaboration}{BCDMS} Collaboration],
  \bibinfo{journal}{Phys.\ Lett.\ B} \textbf{\bibinfo{volume}{223}},
  \bibinfo{pages}{485} (\bibinfo{year}{1989}).

\bibitem[{\citenamefont{Cudell \emph{et~al.}}(1997)\citenamefont{Cudell, Kang,
  and Kim}}]{pom:soft}
\bibinfo{author}{\bibfnamefont{J.~R.} \bibnamefont{Cudell}},
  \bibinfo{author}{\bibfnamefont{K.}~\bibnamefont{Kang}}, \bibnamefont{and}
  \bibinfo{author}{\bibfnamefont{S.~K.} \bibnamefont{Kim}},
  \bibinfo{journal}{Phys.\ Lett.\ B} \textbf{\bibinfo{volume}{395}},
  \bibinfo{pages}{311} (\bibinfo{year}{1997}), \eprint{[hep-ph/9712235]}.

\bibitem[{\citenamefont{Donnachie and Landshoff}(1992)}]{alpha:don}
\bibinfo{author}{\bibfnamefont{A.}~\bibnamefont{Donnachie}} \bibnamefont{and}
  \bibinfo{author}{\bibfnamefont{P.}~\bibnamefont{Landshoff}},
  \bibinfo{journal}{Phys. Lett. B} \textbf{\bibinfo{volume}{296}},
  \bibinfo{pages}{227} (\bibinfo{year}{1992}).

\bibitem[{\citenamefont{Sakurai and Schildknecht}(1972)}]{GVDM1}
\bibinfo{author}{\bibfnamefont{J.~J.} \bibnamefont{Sakurai}} \bibnamefont{and}
  \bibinfo{author}{\bibfnamefont{D.}~\bibnamefont{Schildknecht}},
  \bibinfo{journal}{Phys.\ Lett.\ B} \textbf{\bibinfo{volume}{40}},
  \bibinfo{pages}{121} (\bibinfo{year}{1972}).

\bibitem[{\citenamefont{Schildknecht and Spiesberger}(1997)}]{GVDM2}
\bibinfo{author}{\bibfnamefont{D.}~\bibnamefont{Schildknecht}}
  \bibnamefont{and}
  \bibinfo{author}{\bibfnamefont{H.}~\bibnamefont{Spiesberger}}
  (\bibinfo{year}{1997}), \eprint{[hep-ph/9707447]}.

\bibitem[{\citenamefont{Alekhin \emph{et~al.}}(1987)}]{photo1}
\bibinfo{author}{\bibfnamefont{S.~I.} \bibnamefont{Alekhin}}
  \bibnamefont{\emph{et~al.}} (\bibinfo{year}{1987}), \bibinfo{note}{{CERN-HERA
  87-01}}.

\bibitem[{\citenamefont{Caldwell \emph{et~al.}}(1978)}]{photo2}
\bibinfo{author}{\bibfnamefont{D.}~\bibnamefont{Caldwell}}
  \bibnamefont{\emph{et~al.}}, \bibinfo{journal}{Phys.\ Rev.\ Lett.}
  \textbf{\bibinfo{volume}{40}}, \bibinfo{pages}{1222} (\bibinfo{year}{1978}).

\bibitem[{\citenamefont{Cudell \emph{et~al.}}(2000)}]{alpha:cud}
\bibinfo{author}{\bibfnamefont{J.~R.} \bibnamefont{Cudell}}
  \bibnamefont{\emph{et~al.}}, \bibinfo{journal}{Phys. Rev. D}
  \textbf{\bibinfo{volume}{61}}, \bibinfo{pages}{034019}
  (\bibinfo{year}{2000}), \eprint{[hep-ph/9908218]}.

\bibitem[{\citenamefont{Cooper-Sarkar
  \emph{et~al.}}(1998)\citenamefont{Cooper-Sarkar, Devenish, and
  De~Roeck}}]{Mandy2}
\bibinfo{author}{\bibfnamefont{A.~M.} \bibnamefont{Cooper-Sarkar}},
  \bibinfo{author}{\bibfnamefont{R.~C.~E.} \bibnamefont{Devenish}},
  \bibnamefont{and} \bibinfo{author}{\bibfnamefont{A.}~\bibnamefont{De~Roeck}},
  \bibinfo{journal}{Int.\ J.\ Mod.\ Phys.\ A} \textbf{\bibinfo{volume}{13}},
  \bibinfo{pages}{3385} (\bibinfo{year}{1998}),
  \eprint{[arXiv:hep-ph/97123019]}.

\bibitem[{\citenamefont{Adloff \emph{et~al.}}(2001)}]{LAMBDA:H1}
\bibinfo{author}{\bibfnamefont{C.}~\bibnamefont{Adloff}}
  \bibnamefont{\emph{et~al.}} [\bibinfo{collaboration}{H1} Collaboration],
  \bibinfo{journal}{Phys.\ Lett.\ B} \textbf{\bibinfo{volume}{520}},
  \bibinfo{pages}{183} (\bibinfo{year}{2001}), \eprint{[hep-ex:0108035]}.

\bibitem[{\citenamefont{Golec-Biernat and W{\"u}sthoff}(1998)}]{gb}
\bibinfo{author}{\bibfnamefont{K.}~\bibnamefont{Golec-Biernat}}
  \bibnamefont{and}
  \bibinfo{author}{\bibfnamefont{M.}~\bibnamefont{W{\"u}sthoff}},
  \bibinfo{journal}{Phys.\ Rev.\ D} \textbf{\bibinfo{volume}{59}},
  \bibinfo{pages}{014017} (\bibinfo{year}{1998}), \eprint{[hep-ph/9807513]}.

\bibitem[{\citenamefont{Caldwell and Wing}(2016)}]{AM:newcolliders}
\bibinfo{author}{\bibfnamefont{A.}~\bibnamefont{Caldwell}} \bibnamefont{and}
  \bibinfo{author}{\bibfnamefont{M.}~\bibnamefont{Wing}},
  \bibinfo{journal}{Eur.\ Phys.\ J.\ C} \textbf{\bibinfo{volume}{76}},
  \bibinfo{pages}{463} (\bibinfo{year}{2016}), \eprint{[arXiv:1606.00783]}.

\bibitem[{\citenamefont{Accardi \emph{et~al.}}(2016)}]{Accardi:2012qut}
\bibinfo{author}{\bibfnamefont{A.}~\bibnamefont{Accardi}}
  \bibnamefont{\emph{et~al.}}, \bibinfo{journal}{Eur.\ Phys.\ J.\ A}
  \textbf{\bibinfo{volume}{52}}, \bibinfo{pages}{268} (\bibinfo{year}{2016}),
  \eprint{[arXiv:1212.1701]}.

\end{thebibliography}
  
% \input{tables}
% tables ======================================================
%
\clearpage

\begin{flushleft}
\begin{table}
\renewcommand*{\arraystretch}{1.3}
\begin{center}
\begin{scriptsize}
\begin{tabular}{|c|c|r|c||c|c|r|c|}
\hline
$Q^2$ (GeV$^2$) & $x_{\rm{Bj}}$ & $W$ (GeV) & $\sigma^{\gamma^* p}$ $\pm$  $\delta$ $\sigma^{\gamma^* p}$ ($\mu$b) & $Q^2$ (GeV$^2$) & $x_{\rm{Bj}}$ & $W$ (GeV) & $\sigma^{\gamma^* p}$ $\pm$  $\delta$ $\sigma^{\gamma^* p}$\\

\hline

0.045  &  6.21$\times 10^{-7}$  &  269.2  &  197.3  $\pm$  24.1  &  0.35  &  6.62$\times 10^{-5}$  &  72.7  &  90.5  $\pm$  3.4 \\
0.065  &  8.97$\times 10^{-7}$  &  269.2  &  189.9  $\pm$  20.6  &  0.35  &  1.30$\times 10^{-4}$  &  51.9  &  82.5  $\pm$  2.9 \\
0.065  &  1.02$\times 10^{-6}$  &  252.4  &  191.2  $\pm$  18.7  &  0.35  &  2.20$\times 10^{-4}$  &  39.9  &  76.9  $\pm$  3.0 \\
0.085  &  1.17$\times 10^{-6}$  &  269.5  &  178.1  $\pm$  17.1  &  0.35  &  5.00$\times 10^{-4}$  &  26.5  &  76.9  $\pm$  2.9 \\
0.085  &  1.34$\times 10^{-6}$  &  251.9  &  170.6  $\pm$  11.6  &  0.35  &  2.51$\times 10^{-3}$  &  11.8  &  64.7  $\pm$  7.2 \\
0.085  &  1.56$\times 10^{-6}$  &  233.4  &  165.6  $\pm$  11.8  &  0.4  &   8.83$\times 10^{-6}$  &  212.8  &  98.9  $\pm$  5.1 \\
0.11  &  1.51$\times 10^{-6}$  &  269.9  &  167.2  $\pm$  16.4  &  0.4  &  1.10$\times 10^{-5}$  &  190.7  &  104.0  $\pm$  3.8 \\
0.11  &  1.73$\times 10^{-6}$  &  252.2  &  168.1  $\pm$  9.5  &  0.4  &  1.33$\times 10^{-5}$  &  173.4  &  100.1  $\pm$  3.4 \\
0.11  &  2.02$\times 10^{-6}$  &  233.4  &  159.4  $\pm$  6.4  &  0.4  &  1.70$\times 10^{-5}$  &  153.4  &  99.5  $\pm$  2.9 \\
0.11  &  2.43$\times 10^{-6}$  &  212.8  &  152.1  $\pm$  8.5  &  0.4  &  2.20$\times 10^{-5}$  &  134.8  &  94.0  $\pm$  2.7 \\
0.15  &  2.07$\times 10^{-6}$  &  269.2  &  168.9  $\pm$  14.2  &  0.4  &  3.68$\times 10^{-5}$  &  104.3  &  92.5  $\pm$  3.9 \\
0.15  &  2.36$\times 10^{-6}$  &  252.1  &  148.2  $\pm$  7.2  &  0.4  &  8.83$\times 10^{-5}$  &  67.3  &  89.8  $\pm$  3.3 \\
0.15  &  2.76$\times 10^{-6}$  &  233.1  &  149.4  $\pm$  5.4  &  0.4  &  1.76$\times 10^{-4}$  &  47.7  &  80.6  $\pm$  3.0 \\
0.15  &  3.31$\times 10^{-6}$  &  212.9  &  149.6  $\pm$  4.9  &  0.4  &  2.94$\times 10^{-4}$  &  36.9  &  77.6  $\pm$  3.0 \\
0.15  &  4.14$\times 10^{-6}$  &  190.3  &  139.5  $\pm$  4.7  &  0.4  &  6.31$\times 10^{-4}$  &  25.2  &  73.0  $\pm$  3.0 \\
0.15  &  5.02$\times 10^{-6}$  &  172.9  &  138.6  $\pm$  7.8  &  0.5  &  7.32$\times 10^{-6}$  &  261.4  &  100.6  $\pm$  10.1 \\
0.2  &  3.15$\times 10^{-6}$  &  252.0  &  139.4  $\pm$  7.2  &  0.5  &  8.60$\times 10^{-6}$  &  241.1  &  103.1  $\pm$  11.6 \\
0.2  &  3.68$\times 10^{-6}$  &  233.1  &  137.8  $\pm$  5.2  &  0.5  &  1.58$\times 10^{-5}$  &  177.9  &  96.2  $\pm$  4.6 \\
0.2  &  4.41$\times 10^{-6}$  &  213.0  &  134.0  $\pm$  4.2  &  0.5  &  2.12$\times 10^{-5}$  &  153.6  &  87.8  $\pm$  2.9 \\
0.2  &  5.52$\times 10^{-6}$  &  190.3  &  131.6  $\pm$  3.8  &  0.5  &  2.76$\times 10^{-5}$  &  134.6  &  84.8  $\pm$  2.7 \\
0.2  &  6.69$\times 10^{-6}$  &  172.9  &  127.5  $\pm$  3.4  &  0.5  &  3.98$\times 10^{-5}$  &  112.1  &  81.8  $\pm$  3.6 \\
0.2  &  8.49$\times 10^{-6}$  &  153.5  &  125.6  $\pm$  3.8  &  0.5  &  1.00$\times 10^{-4}$  &  70.7  &  78.2  $\pm$  2.3 \\
0.2  &  1.10$\times 10^{-5}$  &  134.8  &  116.7  $\pm$  5.1  &  0.5  &  2.51$\times 10^{-4}$  &  44.6  &  69.1  $\pm$  2.0 \\
0.2  &  3.98$\times 10^{-5}$  &  70.9  &  118.7  $\pm$  24.5  &  0.5  &  3.68$\times 10^{-4}$  &  36.9  &  67.4  $\pm$  2.2 \\
0.2  &  2.51$\times 10^{-4}$  &  28.2  &  100.8  $\pm$  15.5  &  0.5  &  8.00$\times 10^{-4}$  &  25.0  &  64.4  $\pm$  2.0 \\
0.25  &  3.94$\times 10^{-6}$  &  251.9  &  124.6  $\pm$  7.3  &  0.5  &  3.20$\times 10^{-3}$  &  12.5  &  41.0  $\pm$  5.4 \\
0.25  &  4.60$\times 10^{-6}$  &  233.1  &  125.3  $\pm$  5.1  &  0.65  &  9.52$\times 10^{-6}$  &  261.3  &  84.6  $\pm$  5.3 \\
0.25  &  5.52$\times 10^{-6}$  &  212.8  &  126.3  $\pm$  4.1  &  0.65  &  1.12$\times 10^{-5}$  &  240.9  &  92.4  $\pm$  5.9 \\
0.25  &  6.90$\times 10^{-6}$  &  190.3  &  123.3  $\pm$  3.6  &  0.65  &  1.58$\times 10^{-5}$  &  202.8  &  81.1  $\pm$  5.3 \\
0.25  &  8.36$\times 10^{-6}$  &  172.9  &  119.4  $\pm$  3.5  &  0.65  &  1.64$\times 10^{-5}$  &  199.1  &  89.2  $\pm$  7.5 \\
0.25  &  1.06$\times 10^{-5}$  &  153.6  &  116.7  $\pm$  3.1  &  0.65  &  3.98$\times 10^{-5}$  &  127.8  &  81.7  $\pm$  3.2 \\
0.25  &  1.38$\times 10^{-5}$  &  134.6  &  111.7  $\pm$  3.1  &  0.65  &  5.98$\times 10^{-5}$  &  104.3  &  71.9  $\pm$  3.7 \\
0.25  &  2.30$\times 10^{-5}$  &  104.3  &  109.1  $\pm$  4.3  &  0.65  &  1.00$\times 10^{-4}$  &  80.6  &  70.6  $\pm$  2.7 \\
0.25  &  3.98$\times 10^{-5}$  &  79.3  &  105.9  $\pm$  5.0  &  0.65  &  2.51$\times 10^{-4}$  &  50.9  &  62.3  $\pm$  2.1 \\
0.25  &  1.10$\times 10^{-4}$  &  47.7  &  89.4  $\pm$  4.5  &  0.65  &  4.78$\times 10^{-4}$  &  36.9  &  57.4  $\pm$  2.1 \\
0.25  &  2.51$\times 10^{-4}$  &  31.6  &  87.9  $\pm$  4.6  &  0.65  &  8.00$\times 10^{-4}$  &  28.5  &  55.0  $\pm$  1.7 \\
 0.25  &  3.94$\times 10^{-4}$  &  25.2  &  87.2  $\pm$  5.2  &  0.65  &  3.20$\times 10^{-3}$  &  14.3  &  38.7  $\pm$  2.6 \\
0.25  &  1.58$\times 10^{-3}$  &  12.6  &  88.8  $\pm$  11.2  &  0.85  &  1.24$\times 10^{-5}$  &  261.8  &  77.7  $\pm$  3.7 \\
0.35  &  5.12$\times 10^{-6}$  &  261.5  &  111.3  $\pm$  6.4  &  0.85  &  1.38$\times 10^{-5}$  &  248.2  &  86.6  $\pm$  11.3 \\
0.35  &  5.12$\times 10^{-6}$  &  261.5  &  144.7  $\pm$  37.2  &  0.85  &  2.00$\times 10^{-5}$  &  206.2  &  82.2  $\pm$  3.0 \\
0.35  &  6.10$\times 10^{-6}$  &  239.5  &  118.2  $\pm$  16.7  &  0.85  &  2.00$\times 10^{-5}$  &  206.2  &  82.3  $\pm$  4.9 \\
0.35  &  6.62$\times 10^{-6}$  &  229.9  &  107.6  $\pm$  3.7  &  0.85  &  3.98$\times 10^{-5}$  &  146.1  &  74.0  $\pm$  2.6 \\
0.35  &  8.28$\times 10^{-6}$  &  205.6  &  107.0  $\pm$  3.4  &  0.85  &  5.00$\times 10^{-5}$  &  130.4  &  71.6  $\pm$  4.2 \\
0.35  &  1.00$\times 10^{-5}$  &  187.1  &  111.4  $\pm$  3.2  &  0.85  &  1.00$\times 10^{-4}$  &  92.2  &  66.4  $\pm$  4.1 \\
0.35  &  1.27$\times 10^{-5}$  &  166.0  &  104.2  $\pm$  2.9  &  0.85  &  2.51$\times 10^{-4}$  &  58.2  &  52.4  $\pm$  3.8 \\
0.35  &  1.65$\times 10^{-5}$  &  145.6  &  100.6  $\pm$  2.8  &  0.85  &  8.00$\times 10^{-4}$  &  32.6  &  46.2  $\pm$  2.6 \\
0.35  &  3.20$\times 10^{-5}$  &  104.6  &  94.9  $\pm$  3.9  &  0.85  &  3.20$\times 10^{-3}$  &  16.3  &  40.7  $\pm$  2.2 \\

\hline

\hline                            
\end{tabular}
\end{scriptsize}
\end{center}
\caption{\label{tab:sigma1} $\sigma^{\gamma^*p}(x_{\rm Bj},Q^2)$ 
         as extracted from the HERA
         $e^+p$ NC data at $\sqrt{s}=318$ and 300\,GeV.
         For some values of $Q^2$ and $x_{\rm Bj}$, two values are listed for the
         two different centre-of-mass energies.}
\end{table}

\end{flushleft}

\begin{flushleft}
\begin{table}
\renewcommand*{\arraystretch}{1.3}
\begin{center}
\begin{scriptsize}
\begin{tabular}{|c|c|r|c||c|c|r|c|}
\hline
$Q^2$ (GeV$^2$) & $x_{\rm{Bj}}$ & $W$ (GeV) & $\sigma^{\gamma^* p}$ $\pm$  $\delta$ $\sigma^{\gamma^* p}$  ($\mu$b) & $Q^2$ (GeV$^2$) & $x_{\rm{Bj}}$ & $W$ (GeV) & $\sigma^{\gamma^* p}$ $\pm$  $\delta$ $\sigma^{\gamma^* p}$\\

\hline
1.2  & 	1.76$\times 10^{-5}$      & 261.1	 & 59.6	 $\pm$  2.7&   2.7   & 	8.00$\times 10^{-5}$      & 183.7	 & 37.9	 $\pm$  0.6 \\
1.2  & 	2.00$\times 10^{-5}$       & 244.9	 & 65.8	 $\pm$  2.6&   2.7   & 	8.00$\times 10^{-5}$      & 183.7	 & 38.9	 $\pm$  1.5 \\
1.2  & 	2.00$\times 10^{-5}$	    & 244.9   &    74.3	 $\pm$  7.8&   2.7   & 	1.30$\times 10^{-4}$    & 144.1	 & 33.7	 $\pm$  0.5 \\
1.2  & 	3.20$\times 10^{-5}$     & 193.6	 & 63.6	 $\pm$  2.2&   2.7   & 	2.00$\times 10^{-4}$     & 116.2	 & 32.6	 $\pm$  0.7 \\
1.2  & 	3.20$\times 10^{-5}$     & 193.6	 & 65.2	 $\pm$  2.3&   2.7   & 	3.20$\times 10^{-4}$    & 91.8 	 & 29.1	 $\pm$  0.6 \\
1.2  & 	6.31$\times 10^{-5}$     & 137.9	 & 60.9	 $\pm$  1.6&   2.7   & 	5.00$\times 10^{-4}$     & 73.5 	 & 26.8	 $\pm$  0.5 \\
1.2  & 	8.00$\times 10^{-5}$       & 122.5	 & 55.7	 $\pm$  1.8&   2.7   & 	8.00$\times 10^{-4}$     & 58.1 	 & 24.7	 $\pm$  0.6 \\
1.2  & 	1.30$\times 10^{-4}$     & 96.1 	 & 50.8	 $\pm$  2.9&   2.7   & 	1.30$\times 10^{-3}$     & 45.6 	 & 23.6	 $\pm$  0.5 \\
1.2  & 	1.58$\times 10^{-4}$      & 87.1 	 & 49.8	 $\pm$  1.4&   2.7   & 	2.00$\times 10^{-3}$      & 36.7 	 & 20.0	 $\pm$  0.6 \\
1.2  & 	3.98$\times 10^{-4}$      & 54.9 	 & 46.1	 $\pm$  2.0&   2.7   & 	5.00$\times 10^{-3}$      & 23.2 	 & 18.9	 $\pm$  0.5 \\
1.2  & 	1.30$\times 10^{-3}$      & 30.4 	 & 34.8	 $\pm$  1.6&   2.7   & 	2.00$\times 10^{-2}$       & 11.5 	 & 14.6	 $\pm$  1.3 \\
1.2  & 	5.00$\times 10^{-3}$       & 15.5    &    28.0	 $\pm$  1.5&   3.5   & 	4.06$\times 10^{-5}$   & 293.6	 & 32.3	 $\pm$  2.5 \\
1.5  & 	1.85$\times 10^{-5}$     & 284.5	 & 52.3	 $\pm$  6.2&   3.5   & 	4.32$\times 10^{-5}$  & 284.5	 & 34.2	 $\pm$  2.5 \\
1.5  & 	2.20$\times 10^{-5}$     & 261.4	 & 56.4	 $\pm$  1.8&   3.5   & 	4.60$\times 10^{-5}$    & 275.8	 & 36.4	 $\pm$  1.6 \\
1.5  & 	3.20$\times 10^{-5}$     & 216.5	 & 58.2	 $\pm$  1.8&   3.5   & 	5.12$\times 10^{-5}$  & 261.3	 & 34.1	 $\pm$  1.4 \\
1.5  & 	3.20$\times 10^{-5}$     & 216.5	 & 63.8	 $\pm$  3.8&   3.5   & 	5.31$\times 10^{-5}$   & 256.7	 & 31.6	 $\pm$  1.3 \\
1.5  & 	5.00$\times 10^{-5}$       & 173.2	 & 57.5	 $\pm$  1.5&   3.5   & 	5.73$\times 10^{-5}$   & 247.1	 & 36.7	 $\pm$  2.1 \\
1.5  & 	8.00$\times 10^{-5}$       & 136.9	 & 52.5	 $\pm$  1.5&   3.5   & 	8.00$\times 10^{-5}$      & 209.2	 & 31.8	 $\pm$  0.7 \\
1.5  & 	1.30$\times 10^{-4}$      & 107.4	 & 48.4	 $\pm$  1.6&   3.5   & 	8.00$\times 10^{-5}$      & 209.2	 & 33.2	 $\pm$  0.8 \\
1.5  & 	2.00$\times 10^{-4}$      & 86.6 	 & 45.8	 $\pm$  1.8&   3.5   & 	1.30$\times 10^{-4}$    & 164.1	 & 29.8	 $\pm$  0.4 \\
1.5  & 	3.20$\times 10^{-4}$      & 68.5 	 & 43.2	 $\pm$  1.4&   3.5   & 	2.00$\times 10^{-4}$     & 132.3	 & 27.5	 $\pm$  0.4 \\
1.5  & 	5.00$\times 10^{-4}$      & 54.8 	 & 41.0	 $\pm$  3.2&   3.5   & 	3.20$\times 10^{-4}$    & 104.6	 & 25.4	 $\pm$  0.4 \\
1.5  & 	8.00$\times 10^{-4}$      & 43.3 	 & 36.7	 $\pm$  1.4&   3.5   & 	5.00$\times 10^{-4}$     & 83.7 	 & 24.1	 $\pm$  0.4 \\
1.5  & 	1.00$\times 10^{-3}$       & 38.7 	 & 34.6	 $\pm$  2.3&   3.5   & 	8.00$\times 10^{-4}$     & 66.1 	 & 21.2	 $\pm$  0.4 \\
1.5  & 	3.20$\times 10^{-3}$      & 21.6 	 & 30.8	 $\pm$  1.1&   3.5   & 	1.30$\times 10^{-3}$     & 51.9 	 & 20.0	 $\pm$  0.4 \\
1.5  & 	1.30$\times 10^{-3}$	    & 10.7    &    24.6	 $\pm$  1.7&   3.5   & 	2.00$\times 10^{-3}$      & 41.8 	 & 18.3	 $\pm$  0.3 \\
2    & 	2.47$\times 10^{-5}$     & 284.6	 & 50.1	 $\pm$  4.2&   3.5   & 	8.00$\times 10^{-3}$      & 20.9 	 & 14.9	 $\pm$  0.3 \\
2    & 	2.93$\times 10^{-5}$     & 261.4	 & 48.3	 $\pm$  1.3&   4.5   & 	5.22$\times 10^{-5}$   & 293.6	 & 28.1	 $\pm$  2.2 \\
2    & 	3.27$\times 10^{-5}$     & 247.3	 & 52.5	 $\pm$  2.7&   4.5   & 	5.92$\times 10^{-5}$   & 275.7	 & 28.2	 $\pm$  1.2 \\
2    & 	5.00$\times 10^{-5}$       & 200.0	 & 47.3	 $\pm$  1.3&   4.5   & 	6.18$\times 10^{-5}$  & 269.9	 & 27.7	 $\pm$  1.9 \\
2    & 	5.00$\times 10^{-5}$       & 200.0	 & 49.4	 $\pm$  1.4&   4.5   & 	6.83$\times 10^{-5}$   & 256.7	 & 29.9	 $\pm$  0.9 \\
2    & 	8.00$\times 10^{-5}$       & 158.1	 & 43.1	 $\pm$  0.9&   4.5   & 	7.32$\times 10^{-5}$   & 247.9	 & 27.8	 $\pm$  0.9 \\
2    & 	1.30$\times 10^{-4}$       & 124.0	 & 40.7	 $\pm$  0.9&   4.5   & 	8.18$\times 10^{-5}$   & 234.5	 & 28.8	 $\pm$  4.5 \\
2    & 	2.00$\times 10^{-4}$        & 100.0	 & 38.1	 $\pm$  0.8&   4.5   & 	8.18$\times 10^{-5}$   & 234.5	 & 30.5	 $\pm$  1.0 \\
2    & 	3.20$\times 10^{-4}$       & 79.1 	 & 35.2	 $\pm$  0.9&   4.5   & 	1.30$\times 10^{-4}$    & 186.0	 & 26.0	 $\pm$  0.6 \\
2    & 	5.00$\times 10^{-4}$       & 63.2 	 & 32.3	 $\pm$  0.9&   4.5   & 	1.30$\times 10^{-4}$	   & 186.0    &    26.0	 $\pm$  0.5 \\
2    & 	1.00$\times 10^{-3}$       & 44.7 	 & 28.6	 $\pm$  0.7&   4.5   & 	2.00$\times 10^{-4}$     & 150.0	 & 24.2	 $\pm$  0.4 \\
2    & 	3.20$\times 10^{-3}$       & 25.0 	 & 24.0	 $\pm$  0.7&   4.5   & 	3.20$\times 10^{-4}$    & 118.6	 & 22.4	 $\pm$  0.3 \\
2    & 	1.30$\times 10^{-3}$       & 12.4 	 & 20.3	 $\pm$  1.1&   4.5   & 	5.00$\times 10^{-4}$     & 94.8 	 & 19.9	 $\pm$  0.3 \\
2.7  & 	3.09$\times 10^{-5}$    & 295.7	 & 46.5	 $\pm$  2.6&   4.5   & 	8.00$\times 10^{-4}$     & 75.0 	 & 18.0	 $\pm$  0.3 \\
2.7  & 	3.66$\times 10^{-5}$     & 271.6	 & 43.1	 $\pm$  1.5&   4.5   & 	1.30$\times 10^{-3}$     & 58.8 	 & 16.6	 $\pm$  0.3 \\
2.7  & 	4.09$\times 10^{-5}$     & 256.9	 & 44.6	 $\pm$  4.0&   4.5   & 	2.00$\times 10^{-3}$      & 47.4 	 & 15.4	 $\pm$  0.3 \\
2.7  & 	4.09$\times 10^{-5}$     & 256.9	 & 47.9	 $\pm$  3.7&   4.5   & 	3.20$\times 10^{-3}$     & 37.5 	 & 14.3	 $\pm$  0.3 \\
2.7  & 	5.00$\times 10^{-5}$       & 232.4	 & 39.8	 $\pm$  1.0&   4.5   & 	1.30$\times 10^{-2}$	   & 18.5        & 10.9	 $\pm$  0.3 \\
2.7  & 	5.00$\times 10^{-5}$       & 232.4	 & 41.0	 $\pm$  1.2&         &             &             &                  \\
                                                  
\hline

\hline                            
\end{tabular}
\end{scriptsize}
\end{center}
\caption{\label{tab:sigma2} Continuation of Table~\ref{tab:sigma1}}
\end{table}

\end{flushleft}

\begin{flushleft}
\begin{table}
\renewcommand*{\arraystretch}{1.3}
\begin{center}
\begin{scriptsize}
\begin{tabular}{|c|c|r|c||c|c|r|c|}
\hline
$Q^2$ (GeV$^2$) & $x_{\rm{Bj}}$ & $W$ (GeV) & $\sigma^{\gamma^* p}$ $\pm$  $\delta$ $\sigma^{\gamma^* p}$  ($\mu$b)  & $Q^2$ (GeV$^2$) & $x_{\rm{Bj}}$ & $W$ (GeV) & $\sigma^{\gamma^* p}$ $\pm$  $\delta$ $\sigma^{\gamma^* p}$\\

\hline

6.5  &  7.54$\times 10^{-5}$  & 293.6   & 23.4   $\pm$ 1.8 &     12  &  1.392$\times 10^{-4}$ &  293.6  &  14.4  $\pm$  0.5 \\    
6.5  &  8.03$\times 10^{-5}$  &284.5   & 22.1   $\pm$ 1.3 &      12  &  1.61$\times 10^{-4}$  &  273.0  &  13.6  $\pm$  0.8  \\   
6.5  &  8.55$\times 10^{-5}$  & 275.7   & 21.9   $\pm$ 0.9 &     12  &  1.61$\times 10^{-4}$  &  273.0  &  14.3  $\pm$  0.3  \\	 
6.5  &  9.52$\times 10^{-5}$  &261.4   & 20.9   $\pm$ 0.9 &      12  &  1.82$\times 10^{-4}$ &  256.7  &  13.3  $\pm$  0.4 \\	 
6.5  &  9.86$\times 10^{-5}$  & 256.7   & 22.5   $\pm$ 0.7 &     12  &  2.00$\times 10^{-4}$  &  244.9  &  13.0  $\pm$  0.3  \\   
6.5  &  9.86$\times 10^{-5}$  & 256.7   & 23.0   $\pm$ 1.1 &     12  &  2.00$\times 10^{-4}$  &  244.9  &  13.3  $\pm$  0.4  \\   
6.5  &  1.30$\times 10^{-4}$  & 223.6   & 20.2   $\pm$ 0.4 &     12  &  3.20$\times 10^{-4}$  &  193.6  &  11.7  $\pm$  0.2  \\   
6.5  &  1.30$\times 10^{-4}$  & 223.6   & 21.7   $\pm$ 0.5 &     12  &  3.20$\times 10^{-4}$  &  193.6  &  11.8  $\pm$  0.2   \\	 
6.5  &  2.00$\times 10^{-4}$  & 180.3   & 19.6   $\pm$ 0.3 &     12  &  5.00$\times 10^{-4}$  &  154.9  &  10.8  $\pm$  0.1  \\   
6.5  &  2.00$\times 10^{-4}$  & 180.3   & 19.7   $\pm$ 0.4 &     12  &  8.00$\times 10^{-4}$  &  122.4  &  9.8   $\pm$  0.1  \\	 
6.5  &  3.20$\times 10^{-4}$  & 142.5   & 17.4   $\pm$ 0.2 &     12  &  1.30$\times 10^{-3}$  &  96.0   &  8.8   $\pm$  0.1   \\	 
6.5  &  5.00$\times 10^{-4}$  & 114.0   & 16.2   $\pm$ 0.3 &     12  &  2.00$\times 10^{-3}$  &  77.4   &  8.0   $\pm$  0.1  \\   
6.5  &  8.00$\times 10^{-4}$  & 90.1    & 14.7   $\pm$ 0.2 &     12  &  3.20$\times 10^{-3}$  &  61.1   &  7.1   $\pm$  0.1  \\   
6.5  &  1.30$\times 10^{-3}$  & 70.7    & 13.1   $\pm$ 0.2 &     12  &  5.00$\times 10^{-3}$  &  48.9   &  6.4   $\pm$  0.1  \\   
6.5  &  2.00$\times 10^{-3}$  & 57.0    & 12.0   $\pm$ 0.2 &     12  &  2.00$\times 10^{-2}$  &  24.3   &  4.7   $\pm$  0.1  \\   
6.5  &  3.20$\times 10^{-3}$  & 45.0    & 11.1   $\pm$ 0.2 &     15  &  1.74$\times 10^{-4}$ &  293.5  &  12.0  $\pm$  0.5  \\   
6.5  &  5.00$\times 10^{-3}$  & 36.0    & 10.0   $\pm$ 0.2 &     15  &  2.00$\times 10^{-4}$  &  273.8  &  11.2  $\pm$  0.6  \\   
6.5  &  1.30$\times 10^{-2}$  & 22.2    & 8.4    $\pm$ 0.2 &     15  &  2.00$\times 10^{-4}$  &  273.8  &  11.8  $\pm$  0.4  \\   
6.5  &  2.00$\times 10^{-2}$  & 17.9    & 8.8    $\pm$ 0.3 &     15  &  2.28$\times 10^{-4}$ &  256.7  &  10.9  $\pm$  0.3 \\    
8.5  &  9.86$\times 10^{-5}$  & 293.6   & 18.5   $\pm$ 0.7 &     15  &  2.46$\times 10^{-4}$  &  246.9  &  11.3  $\pm$  0.3  \\   
8.5  &  1.05$\times 10^{-4}$& 284.5   & 18.4   $\pm$ 1.1 &     15    &  3.20$\times 10^{-4}$  &  216.5  &  10.2  $\pm$  0.2  \\   
8.5  &  1.12$\times 10^{-4}$ & 275.7   & 17.9   $\pm$ 0.7 &     15   &  3.20$\times 10^{-4}$  &  216.5  &  10.2  $\pm$  0.2  \\   
8.5  &  1.24$\times 10^{-4}$& 261.4   & 18.5   $\pm$ 0.6 &     15    &  5.00$\times 10^{-4}$  &  173.2  &  9.2   $\pm$  0.1  \\   
8.5  &  1.29$\times 10^{-4}$  & 256.7   & 17.9   $\pm$ 0.5 &     15  &  8.00$\times 10^{-4}$  &  136.9  &  8.3   $\pm$  0.1   \\  
8.5  &  1.39$\times 10^{-4}$  & 247.3   & 17.3   $\pm$ 0.9 &     15  &  1.30$\times 10^{-3}$  &  107.4  &  7.3   $\pm$  0.1  \\   
8.5  &  1.39$\times 10^{-4}$  & 247.3   & 18.0   $\pm$ 1.1 &     15  &  2.00$\times 10^{-3}$  &  86.5   &  6.5   $\pm$  0.1   \\  
8.5  &  2.00$\times 10^{-4}$  & 206.1   & 16.3   $\pm$ 0.3 &     15  &  3.20$\times 10^{-3}$  &  68.4   &  5.9   $\pm$  0.1   \\  
8.5  &  2.00$\times 10^{-4}$  & 206.1   & 16.7   $\pm$ 0.3 &     15  &  5.00$\times 10^{-3}$  &  54.6   &  5.4   $\pm$  0.1  \\   
8.5  &  3.20$\times 10^{-4}$  & 163.0   & 14.8   $\pm$ 0.2 &     15  &  8.00$\times 10^{-3}$  &  43.1   &  4.8   $\pm$  0.1  \\   
8.5  &  5.00$\times 10^{-4}$  & 130.4   & 13.4   $\pm$ 0.2 &     15  &  2.00$\times 10^{-2}$  &  27.1   &  4.0   $\pm$  0.1  \\   
8.5  &  8.00$\times 10^{-4}$  & 103.0   & 12.3   $\pm$ 0.2 &     18  &  2.09$\times 10^{-4}$  &  293.4  &  10.2  $\pm$  0.4  \\   
8.5  &  1.30$\times 10^{-3}$  & 80.8    & 11.1   $\pm$ 0.2 &     18  &  2.37$\times 10^{-4}$  &  275.6  &  9.8   $\pm$  0.3  \\	 
8.5  &  2.00$\times 10^{-3}$  & 65.1    & 10.2   $\pm$ 0.2 &     18  &  2.68$\times 10^{-4}$  &  259.1  &  9.3   $\pm$  0.5  \\   
8.5  &  3.20$\times 10^{-3}$  & 51.5    & 8.8    $\pm$ 0.2 &     18  &  2.68$\times 10^{-4}$  &  259.1  &  9.4   $\pm$  0.2   \\	 
8.5  &  5.00$\times 10^{-3}$  & 41.1    & 8.4    $\pm$ 0.2 & 	 18  &  3.28$\times 10^{-4}$  &  234.2  &  9.1   $\pm$  0.2  \\      
8.5  &  2.00$\times 10^{-2}$  & 20.4    & 6.1    $\pm$ 0.2 &     18  &  3.28$\times 10^{-4}$  &  234.2  &  9.2   $\pm$  0.2  \\   
10  &  1.30$\times 10^{-4}$  &  277.3   & 15.6   $\pm$ 1.0 & 	 18  &  5.00$\times 10^{-4}$  &  189.7  &  8.1   $\pm$  0.2  \\      
10  &  2.00$\times 10^{-4}$  &  223.6   & 14.9   $\pm$ 0.5 & 	 18  &  5.00$\times 10^{-4}$  &  189.7  &  8.2   $\pm$  0.1   \\	    
10  &  3.20$\times 10^{-4}$  &  176.8   & 13.0   $\pm$ 0.3 &     18  &  8.00$\times 10^{-4}$  &  149.9  &  7.3   $\pm$  0.1  \\   
10  &  5.00$\times 10^{-4}$  &  141.4  &  12.0  $\pm$  0.3 & 	 18  &  1.30$\times 10^{-3}$  &  117.6  &  6.5   $\pm$  0.1  \\      
10  &  8.00$\times 10^{-4}$  &  111.8  &  11.0  $\pm$  0.3 &     18  &  2.00$\times 10^{-3}$  &  94.8   &  5.9   $\pm$  0.1  \\    
10  &  1.30$\times 10^{-3}$  &  87.7   &  10.0  $\pm$  0.3 &     18  &  3.20$\times 10^{-3}$  &  74.9   &  5.2   $\pm$  0.1  \\   
10  &  2.00$\times 10^{-3}$  &  70.6   &  9.0   $\pm$  0.3 & 	 18  &  5.00$\times 10^{-3}$  &  59.9   &  4.7   $\pm$  0.1   \\     
10  &  5.00$\times 10^{-3}$  &  44.6   &  7.1   $\pm$  0.1 &     18  &  8.00$\times 10^{-3}$  &  47.3   &  4.2   $\pm$  0.1  \\   
10  &  2.00$\times 10^{-2}$  &  22.2   &  5.9   $\pm$  0.1 &     18  &  2.00$\times 10^{-2}$  &  29.7   &  3.4   $\pm$  0.1  \\   

\hline

\hline
\end{tabular}
\end{scriptsize}
\end{center}
\caption{\label{tab:sigma3} Continuation of Table~\ref{tab:sigma1}}
\end{table}

\end{flushleft}

\begin{flushleft}
\begin{table}
\renewcommand*{\arraystretch}{1.3}
\begin{center}
\begin{scriptsize}
\begin{tabular}{|c|c|r|c||c|c|r|c|}
\hline
$Q^2$ (GeV$^2$) & $x_{\rm{Bj}}$ & $W$ (GeV) & $\sigma^{\gamma^* p}$ $\pm$  $\delta$ $\sigma^{\gamma^* p}$  ($\mu$b)  & $Q^2$ (GeV$^2$) & $x_{\rm{Bj}}$ & $W$ (GeV) & $\sigma^{\gamma^* p}$ $\pm$  $\delta$ $\sigma^{\gamma^* p}$\\

\hline

22  &  2.90$\times 10^{-4}$  &  275.4  &8.32  $\pm$ 0.26   &   35  &  2.00$\times 10^{-3}$  &  132.2  & 3.54    $\pm$  0.04   \\    
22  &  3.20$\times 10^{-4}$  &  262.2  &8.07  $\pm$ 0.27   &   35  &  3.20$\times 10^{-3}$  &  104.4  & 3.10    $\pm$  0.04    \\   
22  &  3.45$\times 10^{-4}$  &  252.5  &7.82  $\pm$ 0.18   &   35  &  5.00$\times 10^{-3}$  &  83.5  &  2.74   $\pm$   0.03    \\	  
22  &  3.88$\times 10^{-4}$  &  238.1  &7.56  $\pm$ 0.15   &   35  &  8.00$\times 10^{-3}$  &  65.9  &  2.41   $\pm$   0.03   \\	  
22  &  5.00$\times 10^{-4}$  &  209.7  &7.12  $\pm$ 0.25   &   35  &  1.30$\times 10^{-2}$  &  51.6  &  2.15   $\pm$   0.03    \\   
22  &  5.00$\times 10^{-4}$  &  209.7  &7.29  $\pm$ 0.14   &   35  &  2.00$\times 10^{-2}$  &  41.4  &  1.93   $\pm$   0.03    \\   
22  &  5.92$\times 10^{-4}$  &  192.7  &6.92  $\pm$ 0.13   &   35  &  3.20$\times 10^{-2}$  &  32.6  &  1.71   $\pm$   0.03    \\   
22  &  8.00$\times 10^{-4}$  &  165.8  &6.34  $\pm$ 0.10   &   35  &  8.00$\times 10^{-2}$  &  20.1  &  1.56   $\pm$   0.08     \\  
22  &  1.30$\times 10^{-3}$  &  130.0  &5.42  $\pm$ 0.16   &   45  &  5.90$\times 10^{-4}$  &  276.1  & 4.12    $\pm$  0.15    \\   
22  &  2.00$\times 10^{-3}$  &  104.8  &5.07  $\pm$ 0.11   &   45  &  6.34$\times 10^{-4}$  &  266.3  & 4.00    $\pm$  0.09    \\	  
22  &  3.20$\times 10^{-3}$  &  82.8  & 4.61  $\pm$ 0.14  &    45  &  7.00$\times 10^{-4}$  &  253.5  & 4.00    $\pm$  0.11     \\  
22  &  5.00$\times 10^{-3}$  &  66.2  & 3.94  $\pm$ 0.11  &    45  &  8.00$\times 10^{-4}$  &  237.1  & 3.85    $\pm$  0.06    \\   
22  &  8.00$\times 10^{-3}$  &  52.2  & 3.37  $\pm$ 0.10  &    45  &  8.00$\times 10^{-4}$  &  237.1  & 3.99    $\pm$  0.13    \\   
22  &  1.30$\times 10^{-2}$  &  40.9  & 3.03  $\pm$ 0.09  &    45  &  9.20$\times 10^{-4}$  &  221.1  & 3.60    $\pm$  0.07    \\   
22  &  3.20$\times 10^{-2}$  &  25.8  & 2.74  $\pm$ 0.08  &    45  &  1.10$\times 10^{-3}$  &  202.2  & 3.51    $\pm$  0.06    \\   
27  &  3.14$\times 10^{-4}$  &  293.2  &7.30  $\pm$ 0.35   &   45  &  1.30$\times 10^{-3}$  &  185.9  & 3.29    $\pm$  0.04    \\   
27  &  3.35$\times 10^{-4}$  &  283.9  &7.00  $\pm$ 0.41   &   45  &  1.30$\times 10^{-3}$  &  185.9  & 3.33    $\pm$  0.05    \\   
27  &  3.55$\times 10^{-4}$  &  275.7  &6.79  $\pm$ 0.20   &   45  &  2.00$\times 10^{-3}$  &  149.9  & 2.90    $\pm$  0.03    \\   
27  &  4.10$\times 10^{-4}$  &  256.6  &6.56  $\pm$ 0.19   &   45  &  3.20$\times 10^{-3}$  &  118.4  & 2.52    $\pm$  0.03   \\    
27  &  4.10$\times 10^{-4}$  &  256.6  &6.63  $\pm$ 0.17   &   45  &  5.00$\times 10^{-3}$  &  94.6  &  2.25   $\pm$   0.03    \\   
27  &  5.00$\times 10^{-4}$  &  232.3  &6.19  $\pm$ 0.10   &   45  &  8.00$\times 10^{-3}$  &  74.7  &  1.96   $\pm$   0.02    \\   
27  &  5.00$\times 10^{-4}$  &  232.3  &6.28  $\pm$ 0.15   &   45  &  1.30$\times 10^{-2}$  &  58.5  &  1.72   $\pm$   0.02    \\   
27  &  8.00$\times 10^{-4}$  &  183.6  &5.45  $\pm$ 0.07   &   45  &  2.00$\times 10^{-2}$  &  47.0  &  1.53   $\pm$   0.02    \\   
27  &  8.00$\times 10^{-4}$  &  183.6  &5.47  $\pm$ 0.16   &   45  &  3.20$\times 10^{-2}$  &  36.9  &  1.37   $\pm$   0.02     \\  
27  &  1.30$\times 10^{-3}$  &  144.0  &4.83  $\pm$ 0.06   &   45  &  5.00$\times 10^{-2}$  &  29.3  &  1.27   $\pm$   0.03    \\   
27  &  2.00$\times 10^{-3}$  &  116.1  &4.36  $\pm$ 0.05   &   60  &  8.00$\times 10^{-4}$  &  273.8  & 3.07    $\pm$  0.06     \\  
27  &  3.20$\times 10^{-3}$  &  91.7  & 3.81  $\pm$ 0.05  &    60  &  8.60$\times 10^{-4}$  &  264.0  & 3.17    $\pm$  0.10     \\  
27  &  5.00$\times 10^{-3}$  &  73.3  & 3.34  $\pm$ 0.04  &    60  &  9.40$\times 10^{-4}$  &  252.5  & 3.07    $\pm$  0.10    \\   
27  &  8.00$\times 10^{-3}$  &  57.9  & 2.96  $\pm$ 0.04  &    60  &  1.10$\times 10^{-3}$  &  233.4  & 2.87    $\pm$  0.07    \\   
27  &  1.30$\times 10^{-2}$  &  45.3  & 2.67  $\pm$ 0.04  &    60  &  1.30$\times 10^{-3}$  &  214.7  & 2.69    $\pm$  0.04    \\   
27  &  2.00$\times 10^{-2}$  &  36.4  & 2.44  $\pm$ 0.04  &    60  &  1.30$\times 10^{-3}$  &  214.7  & 2.73    $\pm$  0.06    \\   
27  &  3.20$\times 10^{-2}$  &  28.6  & 2.26  $\pm$ 0.06  &    60  &  1.50$\times 10^{-3}$  &  199.9  & 2.52    $\pm$  0.05    \\	  
35  &  4.60$\times 10^{-4}$  &  275.7  &5.17  $\pm$ 0.13   &   60  &  2.00$\times 10^{-3}$  &  173.0  & 2.34    $\pm$  0.03    \\   
35  &  5.00$\times 10^{-4}$  &  264.5  &5.21  $\pm$ 0.14   &   60  &  3.20$\times 10^{-3}$  &  136.7  & 2.03    $\pm$  0.02     \\  
35  &  5.31$\times 10^{-4}$  &  256.6  &5.03  $\pm$ 0.09   &   60  &  5.00$\times 10^{-3}$  &  109.3  & 1.75    $\pm$  0.02    \\   
35  &  5.74$\times 10^{-4}$  &  246.9  &5.30  $\pm$ 0.15   &   60  &  8.00$\times 10^{-3}$  &  86.3  &  1.54   $\pm$   0.02    \\   
35  &  6.16$\times 10^{-4}$  &  238.3  &4.96  $\pm$ 0.10   &   60  &  1.30$\times 10^{-2}$  &  67.5  &  1.34   $\pm$   0.02    \\   
35  &  6.57$\times 10^{-4}$  &  230.7  &4.83  $\pm$ 0.09   &   60  &  2.00$\times 10^{-2}$  &  54.2  &  1.20   $\pm$   0.02     \\  
35  &  8.00$\times 10^{-4}$  &  209.1  &4.59  $\pm$ 0.06   &   60  &  3.20$\times 10^{-2}$  &  42.6  &  1.07   $\pm$   0.02    \\   
35  &  8.00$\times 10^{-4}$  &  209.1  &4.67  $\pm$ 0.09   &   60  &  5.00$\times 10^{-2}$  &  33.8  &  1.00   $\pm$   0.02    \\   
35  &  1.00$\times 10^{-3}$  &  187.0  &4.28  $\pm$ 0.07   &   60  &  1.30$\times 10^{-1}$  &  20.1  &  0.85   $\pm$   0.05    \\   
35  &  1.30$\times 10^{-3}$  &  164.0  &3.96  $\pm$ 0.05   &       &                        &        &                      \\   

\hline
\end{tabular}
\end{scriptsize}
\end{center}
\caption{\label{tab:sigma4} Continuation of Table~\ref{tab:sigma1}}
\end{table}

\end{flushleft}

\begin{flushleft}
\begin{table}
\renewcommand*{\arraystretch}{1.3}
\begin{center}
\begin{scriptsize}
\begin{tabular}{|c|c|r|c||c|c|r|c|}
\hline
$Q^2$ (GeV$^2$) & $x_{\rm{Bj}}$ & $W$ (GeV) & $\sigma^{\gamma^* p}$ $\pm$  $\delta$ $\sigma^{\gamma^* p}$  ($\mu$b)  & $Q^2$ (GeV$^2$) & $x_{\rm{Bj}}$ & $W$ (GeV) & $\sigma^{\gamma^* p}$ $\pm$  $\delta$ $\sigma^{\gamma^* p}$\\

\hline

70  &  9.22$\times 10^{-4}$   &  275.4  &2.481  $\pm$  0.158    &  120  &  5.00$\times 10^{-2}$  &  47.8  &  0.478 $\pm$ 0.009    \\    
70  &  1.00$\times 10^{-3}$   &  264.4  &2.516  $\pm$  0.091    &  120  &  8.00$\times 10^{-2}$  &  37.2  &  0.448 $\pm$ 0.009     \\   
70  &  1.10$\times 10^{-3}$   &  252.1  &2.593  $\pm$  0.081    &  120  &  1.80$\times 10^{-1}$  &  23.4  &  0.379 $\pm$ 0.015     \\	 
70  &  1.24$\times 10^{-3}$   &  237.4  &2.442  $\pm$  0.065    &  150  &  2.00$\times 10^{-3}$  &  273.6  & 1.138 $\pm$ 0.022   \\	 
70  &  1.30$\times 10^{-3}$  &  231.9  & 2.377 $\pm$ 0.056     & 150  &  3.20$\times 10^{-3}$  &  216.2  & 0.937	 $\pm$   0.023  \\   
70  &  1.30$\times 10^{-3}$  &  231.9  & 2.463 $\pm$ 0.061     & 150  &  3.20$\times 10^{-3}$  &  216.2  & 0.953	 $\pm$   0.013  \\   
70  &  2.00$\times 10^{-3}$  &  186.9  & 2.033 $\pm$ 0.044     & 150  &  5.00$\times 10^{-3}$  &  172.8  & 0.818	 $\pm$   0.010  \\   
70  &  2.00$\times 10^{-3}$  &  186.9  & 2.070 $\pm$ 0.043     & 150  &  8.00$\times 10^{-3}$  &  136.4  & 0.701	 $\pm$   0.011   \\  
70  &  2.50$\times 10^{-3}$  &  167.1  & 1.930 $\pm$ 0.035     & 150  &  1.30$\times 10^{-2}$  &  106.7  & 0.605	 $\pm$   0.012  \\   
70  &  3.20$\times 10^{-3}$  &  147.7  & 1.752 $\pm$ 0.027     & 150  &  2.00$\times 10^{-2}$  &  85.7  &  0.524	$\pm$    0.011  \\	 
70  &  5.00$\times 10^{-3}$  &  118.0  & 1.546 $\pm$ 0.023     & 150  &  3.20$\times 10^{-2}$  &  67.4  &  0.451	$\pm$    0.011   \\  
70  &  8.00$\times 10^{-3}$  &  93.2  &  1.325 $\pm$ 0.025    &  150  &  5.00$\times 10^{-2}$  &  53.4  &  0.401	$\pm$    0.009  \\   
70  &  1.30$\times 10^{-2}$  &  72.9  &  1.163 $\pm$ 0.026    &  150  &  8.00$\times 10^{-2}$  &  41.5  &  0.360	$\pm$    0.008  \\   
70  &  2.00$\times 10^{-2}$  &  58.6  &  1.041 $\pm$ 0.022    &  150  &  1.80$\times 10^{-1}$  &  26.2  &  0.296	$\pm$    0.011  \\   
70  &  3.20$\times 10^{-2}$  &  46.0  &  0.929 $\pm$ 0.023    &  200  &  2.60$\times 10^{-3}$  &  277.0  & 0.813	 $\pm$   0.017  \\   
70  &  5.00$\times 10^{-2}$  &  36.5  &  0.863 $\pm$ 0.020    &  200  &  3.20$\times 10^{-3}$  &  249.6  & 0.710	 $\pm$   0.034  \\   
90  &  1.30$\times 10^{-3}$  &  262.9  & 2.007 $\pm$ 0.038     & 200  &  3.20$\times 10^{-3}$  &  249.6  & 0.743	 $\pm$   0.013  \\   
90  &  1.50$\times 10^{-3}$  &  244.8  & 1.876 $\pm$ 0.048     & 200  &  5.00$\times 10^{-3}$  &  199.5  & 0.619	 $\pm$   0.015  \\   
90  &  2.00$\times 10^{-3}$  &  211.9  & 1.623 $\pm$ 0.041     & 200  &  5.00$\times 10^{-3}$  &  199.5  & 0.637	 $\pm$   0.007 \\    
90  &  2.00$\times 10^{-3}$  &  211.9  & 1.706 $\pm$ 0.027     & 200  &  8.00$\times 10^{-3}$  &  157.5  & 0.539	 $\pm$   0.006  \\   
90  &  3.20$\times 10^{-3}$  &  167.4  & 1.470 $\pm$ 0.019     & 200  &  1.30$\times 10^{-2}$  &  123.2  & 0.460	 $\pm$   0.005  \\   
90  &  5.00$\times 10^{-3}$  &  133.8  & 1.286 $\pm$ 0.016     & 200  &  2.00$\times 10^{-2}$  &  99.0  &  0.401	$\pm$    0.004  \\   
90  &  8.00$\times 10^{-3}$  &  105.6  & 1.110 $\pm$ 0.015     & 200  &  3.20$\times 10^{-2}$  &  77.8  &  0.337	$\pm$    0.004  \\   
90  &  1.30$\times 10^{-2}$  &  82.7  &  0.952 $\pm$ 0.014    &  200  &  5.00$\times 10^{-2}$  &  61.7  &  0.307	$\pm$    0.004   \\  
90  &  2.00$\times 10^{-2}$  &  66.4  &  0.826 $\pm$ 0.012    &  200  &  8.00$\times 10^{-2}$  &  48.0  &  0.267	$\pm$    0.003  \\   
90  &  3.20$\times 10^{-2}$  &  52.2  &  0.717 $\pm$ 0.011    &  200  &  1.30$\times 10^{-1}$  &  36.6  &  0.233	$\pm$    0.006   \\  
90  &  5.00$\times 10^{-2}$  &  41.4  &  0.641 $\pm$ 0.012    &  200  &  1.80$\times 10^{-1}$  &  30.2  &  0.234	$\pm$    0.004   \\  
90  &  8.00$\times 10^{-2}$  &  32.2  &  0.591 $\pm$ 0.011    &  200  &  2.50$\times 10^{-1}$  &  24.5  &  0.208	$\pm$    0.012  \\   
90  &  1.80$\times 10^{-1}$  &  20.3  &  0.522 $\pm$ 0.027    &  200  &  4.00$\times 10^{-1}$  &  17.3  &  0.131	$\pm$    0.010  \\   
120  &  1.60$\times 10^{-3}$ &  273.6  & 1.457 $\pm$ 0.027     & 250  &  3.30$\times 10^{-3}$  &  274.8  & 0.630	 $\pm$   0.014  \\   
120  &  1.72$\times 10^{-3}$ &  263.9  & 1.468 $\pm$ 0.062     & 250  &  5.00$\times 10^{-3}$  &  223.0  & 0.506	 $\pm$   0.016  \\   
120  &  1.88$\times 10^{-3}$ &  252.4  & 1.412 $\pm$ 0.052     & 250  &  5.00$\times 10^{-3}$  &  223.0  & 0.524	 $\pm$   0.007  \\	 
120  &  2.00$\times 10^{-3}$ &  244.7  & 1.348 $\pm$ 0.023     & 250  &  8.00$\times 10^{-3}$  &  176.1  & 0.434	 $\pm$   0.005  \\   
120  &  2.12$\times 10^{-3}$ &  237.7  & 1.322 $\pm$ 0.039     & 250  &  1.30$\times 10^{-2}$  &  137.8  & 0.377	 $\pm$   0.004   \\  
120  &  2.12$\times 10^{-3}$ &  237.7  & 1.362 $\pm$ 0.042     & 250  &  2.00$\times 10^{-2}$  &  110.7  & 0.324	 $\pm$   0.004  \\   
120  &  2.70$\times 10^{-3}$ &  210.5  & 1.199 $\pm$ 0.036     & 250  &  3.20$\times 10^{-2}$  &  87.0  &  0.274	$\pm$    0.003  \\   
120  &  3.20$\times 10^{-3}$  &  193.3  &1.130  $\pm$ 0.027      &250  &  5.00$\times 10^{-2}$  &  68.9  &  0.245	$\pm$    0.003  \\   
120  &  3.20$\times 10^{-3}$  &  193.3  &1.153  $\pm$ 0.014      &250  &  8.00$\times 10^{-2}$  &  53.6  &  0.215	$\pm$    0.003   \\  
120  &  5.00$\times 10^{-3}$  &  154.5  &0.984  $\pm$ 0.014      &250  &  1.30$\times 10^{-1}$  &  40.9  &  0.189	$\pm$    0.004  \\   
120  &  8.00$\times 10^{-3}$  &  122.0  &0.843  $\pm$ 0.012      &250  &  1.80$\times 10^{-1}$  &  33.8  &  0.181	$\pm$    0.003  \\   
120  &  1.30$\times 10^{-2}$  &  95.5  & 0.733 $\pm$ 0.012     & 250  &  2.50$\times 10^{-1}$  &  27.4  &  0.160	$\pm$    0.007  \\   
120  &  2.00$\times 10^{-2}$  &  76.7  & 0.635 $\pm$ 0.011     & 250  &  4.00$\times 10^{-1}$  &  19.4  &  0.106  $\pm$   0.007   \\   
120  &  3.20$\times 10^{-2}$  &  60.3  & 0.554 $\pm$ 0.010     &      &                        &        &                        \\

\hline
\end{tabular}
\end{scriptsize}
\end{center}
\caption{\label{tab:sigma5} Continuation of Table~\ref{tab:sigma1}}
\end{table}

\end{flushleft}

\begin{flushleft}
\begin{table}
\renewcommand*{\arraystretch}{1.3}
\begin{center}
\begin{scriptsize}
\begin{tabular}{|c|c|r|c||c|c|r|c|}
\hline
$Q^2$ (GeV$^2$) & $x_{\rm{Bj}}$ & $W$ (GeV) & $\sigma^{\gamma^* p}$ $\pm$  $\delta$ $\sigma^{\gamma^* p}$  ($\mu$b)  & $Q^2$ (GeV$^2$) & $x_{\rm{Bj}}$ & $W$ (GeV) & $\sigma^{\gamma^* p}$ $\pm$  $\delta$ $\sigma^{\gamma^* p}$\\

\hline

300  &  0.0039  &  276.8  &  0.500  $\pm$  0.011  &  650  &  0.032  &  140.2  &  0.112  $\pm$  0.002     \\   
300  &  0.005  &  244.3  &  0.450  $\pm$  0.008  &   650  &  0.05  &  111.1  &  0.094  $\pm$  0.001       \\  
300  &  0.005  &  244.3  &  0.451  $\pm$  0.019  &   650  &  0.08  &  86.5  &  0.082  $\pm$  0.001        \\  
300  &  0.008  &  192.9  &  0.370  $\pm$  0.004  &   650  &  0.13  &  66.0  &  0.072  $\pm$  0.001       \\   
300  &  0.008  &  192.9  &  0.375  $\pm$  0.011  &   650  &  0.18  &  54.4  &  0.070  $\pm$  0.002        \\  
300  &  0.013  &  150.9  &  0.309  $\pm$  0.003  &   650  &  0.25  &  44.2  &  0.057  $\pm$  0.001        \\  
300  &  0.02  &  121.2  &  0.268  $\pm$  0.003  &    650  &  0.4  &  31.2  &  0.042  $\pm$  0.002         \\  
300  &  0.032  &  95.3  &  0.231  $\pm$  0.003  &    650  &  0.65  &  18.7  &  0.011  $\pm$  0.002         \\ 
300  &  0.05  &  75.5  &  0.199  $\pm$  0.003  &     800  &  0.0105  &  274.6  &  0.142  $\pm$  0.004     \\  
300  &  0.08  &  58.7  &  0.176  $\pm$  0.002  &     800  &  0.013  &  246.5  &  0.125  $\pm$  0.002      \\  
300  &  0.13  &  44.8  &  0.159  $\pm$  0.003  &     800  &  0.013  &  246.5  &  0.133  $\pm$  0.005      \\  
300  &  0.18  &  37.0  &  0.149  $\pm$  0.002  &     800  &  0.02  &  198.0  &  0.106  $\pm$  0.002      \\   
300  &  0.25  &  30.0  &  0.134  $\pm$  0.007  &     800  &  0.02  &  198.0  &  0.107  $\pm$  0.005      \\   
300  &  0.4  &  21.2  &  0.099  $\pm$  0.004  &      800  &  0.032  &  155.6  &  0.091  $\pm$  0.001     \\   
400  &  0.0053  &  274.0  &  0.361  $\pm$  0.008  &  800  &  0.05  &  123.3  &  0.077  $\pm$  0.001      \\   
400  &  0.008  &  222.7  &  0.280  $\pm$  0.009  &   800  &  0.08  &  95.9  &  0.070  $\pm$  0.001        \\  
400  &  0.008  &  222.7  &  0.296  $\pm$  0.004  &   800  &  0.13  &  73.2  &  0.058  $\pm$  0.001        \\  
400  &  0.013  &  174.3  &  0.241  $\pm$  0.003  &   800  &  0.18  &  60.4  &  0.056  $\pm$  0.002        \\  
400  &  0.013  &  174.3  &  0.245  $\pm$  0.008  &   800  &  0.25  &  49.0  &  0.047  $\pm$  0.001       \\   
400  &  0.02  &  140.0  &  0.205  $\pm$  0.003  &    800  &  0.4  &  34.7  &  0.031  $\pm$  0.002         \\  
400  &  0.032  &  110.0  &  0.174  $\pm$  0.002  &   800  &  0.65  &  20.8  &  0.008  $\pm$  0.001        \\  
400  &  0.05  &  87.2  &  0.151  $\pm$  0.002  &     1000  &  0.013  &  275.5  &  0.100  $\pm$  0.003     \\  
400  &  0.08  &  67.8  &  0.133  $\pm$  0.002  &     1000  &  0.02  &  221.4  &  0.086  $\pm$  0.002      \\  
400  &  0.13  &  51.7  &  0.118  $\pm$  0.002  &     1000  &  0.02  &  221.4  &  0.087  $\pm$  0.006       \\ 
400  &  0.18  &  42.7  &  0.109  $\pm$  0.004  &     1000  &  0.032  &  173.9  &  0.073  $\pm$  0.002     \\  
400  &  0.25  &  34.7  &  0.098  $\pm$  0.002  &     1000  &  0.05  &  137.8  &  0.063  $\pm$  0.002       \\ 
400  &  0.4  &  24.5  &  0.075  $\pm$  0.003  &      1000  &  0.08  &  107.2  &  0.054  $\pm$  0.001      \\  
500  &  0.0066  &  274.3  &  0.256  $\pm$  0.006  &  1000  &  0.13  &  81.8  &  0.047  $\pm$  0.001      \\   
500  &  0.008  &  249.0  &  0.236  $\pm$  0.005  &   1000  &  0.18  &  67.5  &  0.043  $\pm$  0.001      \\   
500  &  0.008  &  249.0  &  0.246  $\pm$  0.016  &   1000  &  0.25  &  54.8  &  0.038  $\pm$  0.001      \\   
500  &  0.013  &  194.8  &  0.201  $\pm$  0.003  &   1000  &  0.4  &  38.7  &  0.025  $\pm$  0.001       \\   
500  &  0.013  &  194.8  &  0.205  $\pm$  0.013  &   1000  &  0.65  &  23.2  &  0.006  $\pm$  0.001      \\   
500  &  0.02  &  156.5  &  0.170  $\pm$  0.003  &    1200  &  0.014  &  290.7  &  0.086  $\pm$  0.002     \\  
500  &  0.032  &  123.0  &  0.144  $\pm$  0.002  &   1200  &  0.014  &  290.7  &  0.093  $\pm$  0.006      \\ 
500  &  0.05  &  97.5  &  0.126  $\pm$  0.002  &     1200  &  0.02  &  242.5  &  0.070  $\pm$  0.003      \\  
500  &  0.08  &  75.8  &  0.108  $\pm$  0.002  &     1200  &  0.02  &  242.5  &  0.075  $\pm$  0.001      \\  
500  &  0.13  &  57.9  &  0.097  $\pm$  0.002  &     1200  &  0.032  &  190.5  &  0.058  $\pm$  0.002     \\  
500  &  0.18  &  47.7  &  0.085  $\pm$  0.002  &     1200  &  0.032  &  190.5  &  0.062  $\pm$  0.001      \\ 
500  &  0.25  &  38.7  &  0.078  $\pm$  0.003  &     1200  &  0.05  &  151.0  &  0.051  $\pm$  0.001      \\  
500  &  0.4  &  27.4  &  0.064  $\pm$  0.006  &      1200  &  0.08  &  117.5  &  0.045  $\pm$  0.001      \\  
650  &  0.0085  &  275.4  &  0.177  $\pm$  0.010  &  1200  &  0.13  &  89.6  &  0.039  $\pm$  0.001       \\  
650  &  0.0085  &  275.4  &  0.187  $\pm$  0.005  &  1200  &  0.18  &  73.9  &  0.037  $\pm$  0.001      \\   
650  &  0.013  &  222.2  &  0.152  $\pm$  0.005  &   1200  &  0.25  &  60.0  &  0.031  $\pm$  0.001      \\
650  &  0.013  &  222.2  &  0.154  $\pm$  0.002  &   1200  &  0.4  &  42.4  &  0.020  $\pm$  0.001       \\
650  &  0.02  &  178.5  &  0.125  $\pm$  0.007  &    1200  &  0.65  &  25.4  &  0.006  $\pm$  0.001      \\
650  &  0.02  &  178.5  &  0.132  $\pm$  0.002  &          &        &        &                          \\

\hline
\end{tabular}
\end{scriptsize}
\end{center}
\caption{\label{tab:sigma6} Continuation of Table~\ref{tab:sigma1}}
\end{table}

\end{flushleft}

\begin{flushleft}
\begin{table}
\renewcommand*{\arraystretch}{1.3}
\begin{center}
\begin{scriptsize}
\begin{tabular}{|c|c|r|c||c|c|r|c|}
\hline
$Q^2$ (GeV$^2$) & $x_{\rm{Bj}}$ & $W$ (GeV) & $\sigma^{\gamma^* p}$ $\pm$  $\delta$ $\sigma^{\gamma^* p}$  ($\mu$b)  & $Q^2$ (GeV$^2$) & $x_{\rm{Bj}}$ & $W$ (GeV) & $\sigma^{\gamma^* p}$ $\pm$  $\delta$ $\sigma^{\gamma^* p}$\\

\hline

1500  &  0.02  &  271.1  &  0.0576  $\pm$  0.0016  &    5000  &  0.0547  &  294.0  &  0.0132  $\pm$  0.0008     \\   
1500  &  0.02  &  271.1  &  0.0608  $\pm$  0.0044  & 	5000  &  0.08  &  239.8  &  0.0102  $\pm$  0.0007        \\  
1500  &  0.032  &  213.0  &  0.0443  $\pm$  0.0031  & 	5000  &  0.08  &  239.8  &  0.0114  $\pm$  0.0003        \\  
1500  &  0.032  &  213.0  &  0.0476  $\pm$  0.0011  & 	5000  &  0.13  &  182.9  &  0.0093  $\pm$  0.0008       \\   
1500  &  0.05  &  168.8  &  0.0432  $\pm$  0.0008  & 	5000  &  0.13  &  182.9  &  0.0094  $\pm$  0.0003        \\  
1500  &  0.08  &  131.3  &  0.0369  $\pm$  0.0007  & 	5000  &  0.18  &  150.9  &  0.0080  $\pm$  0.0003        \\  
1500  &  0.13  &  100.2  &  0.0304  $\pm$  0.0007  & 	5000  &  0.25  &  122.5  &  0.0068  $\pm$  0.0003        \\  
1500  &  0.18  &  82.7  &  0.0284  $\pm$  0.0007  & 	5000  &  0.4  &  86.6  &  0.0044  $\pm$  0.0002           \\ 
1500  &  0.25  &  67.1  &  0.0246  $\pm$  0.0007  & 	5000  &  0.65  &  51.9  &  0.0007  $\pm$  0.0001         \\  
1500  &  0.4  &  47.4  &  0.0156  $\pm$  0.0006  & 	8000  &  0.0875  &  288.8  &  0.0071  $\pm$  0.0007      \\  
1500  &  0.65  &  28.4  &  0.0033  $\pm$  0.0004  & 	8000  &  0.13  &  231.4  &  0.0053  $\pm$  0.0006        \\  
2000  &  0.0219  &  298.9  &  0.0448  $\pm$  0.0029  & 	8000  &  0.13  &  231.4  &  0.0061  $\pm$  0.0002       \\   
2000  &  0.032  &  246.0  &  0.0358  $\pm$  0.0010  & 	8000  &  0.18  &  190.9  &  0.0052  $\pm$  0.0002       \\   
2000  &  0.032  &  246.0  &  0.0367  $\pm$  0.0024  & 	8000  &  0.18  &  190.9  &  0.0056  $\pm$  0.0006       \\   
2000  &  0.05  &  194.9  &  0.0301  $\pm$  0.0018  & 	8000  &  0.25  &  154.9  &  0.0043  $\pm$  0.0002       \\   
2000  &  0.05  &  194.9  &  0.0311  $\pm$  0.0008  & 	8000  &  0.25  &  154.9  &  0.0047  $\pm$  0.0006        \\  
2000  &  0.08  &  151.7  &  0.0276  $\pm$  0.0006  & 	8000  &  0.4  &  109.5  &  0.0024  $\pm$  0.0002         \\  
2000  &  0.13  &  115.7  &  0.0234  $\pm$  0.0006  & 	8000  &  0.65  &  65.6  &  0.0006  $\pm$  0.0001         \\  
2000  &  0.18  &  95.5  &  0.0204  $\pm$  0.0006  & 	12000  &  0.13  &  283.4  &  0.0034  $\pm$  0.0007      \\   
2000  &  0.25  &  77.5  &  0.0181  $\pm$  0.0006  & 	12000  &  0.18  &  233.8  &  0.0035  $\pm$  0.0006       \\  
2000  &  0.4  &  54.8  &  0.0119  $\pm$  0.0005  & 	12000  &  0.18  &  233.8  &  0.0036  $\pm$  0.0002       \\  
2000  &  0.65  &  32.8  &  0.0027  $\pm$  0.0003  & 	12000  &  0.25  &  189.7  &  0.0024  $\pm$  0.0002       \\  
3000  &  0.032  &  301.2  &  0.0260  $\pm$  0.0011  & 	12000  &  0.25  &  189.7  &  0.0024  $\pm$  0.0005       \\  
3000  &  0.05  &  238.7  &  0.0217  $\pm$  0.0005  & 	12000  &  0.4  &  134.2  &  0.0016  $\pm$  0.0001         \\ 
3000  &  0.05  &  238.7  &  0.0223  $\pm$  0.0014  & 	12000  &  0.65  &  80.4  &  0.0004  $\pm$  0.0001        \\  
3000  &  0.08  &  185.7  &  0.0182  $\pm$  0.0005  & 	20000  &  0.25  &  245.0  &  0.0017  $\pm$  0.0002        \\ 
3000  &  0.08  &  185.7  &  0.0188  $\pm$  0.0012  & 	20000  &  0.25  &  245.0  &  0.0020  $\pm$  0.0006       \\  
3000  &  0.13  &  141.7  &  0.0153  $\pm$  0.0004  & 	20000  &  0.4  &  173.2  &  0.0010  $\pm$  0.0002       \\   
3000  &  0.18  &  116.9  &  0.0136  $\pm$  0.0004  & 	20000  &  0.4  &  173.2  &  0.0015  $\pm$  0.0005       \\   
3000  &  0.25  &  94.9  &  0.0114  $\pm$  0.0004  & 	20000  &  0.65  &  103.8  &  0.0002  $\pm$  0.0001      \\   
3000  &  0.4  &  67.1  &  0.0076  $\pm$  0.0003  & 	30000  &  0.4  &  212.1  &  0.0007  $\pm$  0.0002       \\   
3000  &  0.65  &  40.2  &  0.0016  $\pm$  0.0002  & 	30000  &  0.4  &  212.1  &  0.0016  $\pm$  0.0010       \\   
      &	       &        &                         &	30000  &  0.65  &  127.1  &  0.0001  $\pm$  0.0001       \\  

\hline
\end{tabular}
\end{scriptsize}
\end{center}
\caption{\label{tab:sigma7} Continuation of Table~\ref{tab:sigma1}}
\end{table}

\end{flushleft}

\begin{flushleft}
\begin{table}
\renewcommand*{\arraystretch}{1.3}
\begin{center}
%\begin{scriptsize}
\begin{tabular}{|c|c|c|c|c|c|}

\hline
 parameter & \multicolumn{2}{|c|}{HHT-ALLM} & \multicolumn{2}{|c|}{HHT-ALLM-FT} &     ALLM-97  \\    
\hline

$  m_0 $    & \multicolumn{2}{|c|}{    0.446  $\pm$ 0.028} &\multicolumn{2}{|c|}{  0.388  $\pm$ 0.021   } &   0.320\\ 
  $p_1 $   & \multicolumn{2}{|c|}{    74.2  $\pm$ 17.5}  &\multicolumn{2}{|c|}{  50.8  $\pm$ 10.3   } &   49.5\\ 
  $p_2 $   & \multicolumn{2}{|c|}{    29.3  $\pm$ 13.6}   &\multicolumn{2}{|c|}{  0.838  $\pm$ 0.273   } &   0.151\\ 
  $p_3 $   & \multicolumn{2}{|c|}{    4.74$\times 10^{-5}$  $\pm \ 2.63\times 10^{-5}$} &\multicolumn{2}{|c|}{  $1.87\times 10^{-5}$  $\pm \ 5\times 10^{-7}$   } &   0.525\\ 
  $p_4 $   & \multicolumn{2}{|c|}{    2.2$\times 10^{-8}$  $\pm \ 1.7\times 10^{-8}$} &\multicolumn{2}{|c|}{  $4.4\times 10^{-9}$ $\pm \ 4\times 10^{-10}$  } &  0.065\\ 
  $p_5   $ & \multicolumn{2}{|c|}{    0.412  $\pm$ 0.018} &\multicolumn{2}{|c|}{  0.356  $\pm$ 0.013  } &   0.281\\ 
  $p_6  $  & \multicolumn{2}{|c|}{    0.164  $\pm$ 0.011} &\multicolumn{2}{|c|}{  0.171  $\pm$ 0.005 } &   0.223\\ 
  $p_7 $   & \multicolumn{2}{|c|}{    17.7  $\pm$ 1.1}  &\multicolumn{2}{|c|}{  18.6  $\pm$ 0.8   } &   2.20\\ 
  $p_8$    & \multicolumn{2}{|c|}{   $-$0.835  $\pm$ 0.007} &\multicolumn{2}{|c|}{ $-$0.075  $\pm$ 0.007 } &  $-$0.081\\
  $p_9 $   & \multicolumn{2}{|c|}{   $-$0.446  $\pm$ 0.022} &\multicolumn{2}{|c|}{ $-$0.470 $\pm$ 0.014 } &  $-$0.448\\ 
  $p_{10} $  & \multicolumn{2}{|c|}{  10.6  $\pm$ 1.2} &\multicolumn{2}{|c|}{  9.2 $\pm$ 0.5   } &   1.17\\ 
  $p_{11}$   & \multicolumn{2}{|c|}{ $-$45.8 $\pm$ 1.0} &\multicolumn{2}{|c|}{ $-$0.477 $\pm$ 0.329 } &   0.363\\ 
  $p_{12}  $ & \multicolumn{2}{|c|}{  55.7 $\pm$ 1.0} &\multicolumn{2}{|c|}{  54.0 $\pm$ 0.4 } &   1.89\\ 
  $p_{13} $  & \multicolumn{2}{|c|}{ $-$0.031 $\pm$ 0.152} &\multicolumn{2}{|c|}{ 0.073 $\pm$ 0.068 } &  1.84\\ 
  $p_{14}$   & \multicolumn{2}{|c|}{ $-$1.04  $\pm$ 0.13} &\multicolumn{2}{|c|}{ $-$0.636 $\pm$ 0.033 } &   0.801\\ 
  $p_{15} $  & \multicolumn{2}{|c|}{  2.97 $\pm$ 0.13} &\multicolumn{2}{|c|}{  3.37 $\pm$ 0.03 } &   0.97\\ 
  $p_{16}  $ & \multicolumn{2}{|c|}{  0.163  $\pm$ 0.044} &\multicolumn{2}{|c|}{ $-$0.660 $\pm$ 0.254} &   3.49\\ 
  $p_{17} $  & \multicolumn{2}{|c|}{  0.706 $\pm$ 0.081} &\multicolumn{2}{|c|}{ 0.882 $\pm$ 0.042  } &   0.584\\ 
  $p_{18}  $ & \multicolumn{2}{|c|}{  0.185 $\pm$ 0.085} &\multicolumn{2}{|c|}{ 0.082 $\pm$ 0.0279 } &   0.379\\ 
  $p_{19} $  & \multicolumn{2}{|c|}{ $-$16.4  $\pm$ 2.6} &\multicolumn{2}{|c|}{ $-$8.5  $\pm$ 1.6  } &   2.61\\ 
  $p_{20} $  & \multicolumn{2}{|c|}{ $-$1.29  $\pm$ 1.32} &\multicolumn{2}{|c|}{ 0.339  $\pm$ 0.021 } &   0.011\\ 
  $p_{21}$   & \multicolumn{2}{|c|}{  4.51  $\pm$ 1.30} &\multicolumn{2}{|c|}{ 3.38  $\pm$ 0.02 } &   3.76\\ 
  $p_{22}$   & \multicolumn{2}{|c|}{  1.16  $\pm$ 0.39} &\multicolumn{2}{|c|}{ 1.07  $\pm$ 0.10 } &   0.493\\
\hline 
  $\chi^2$/ndf & \multicolumn{2}{|c|} { 607/574=1.06 }  
           & \multicolumn{2}{|c|} {1014/1001=1.01 }   
%           & \multicolumn{2}{|c|} {1299/1357=0.97 }   \\
           & 1299/1357=0.97    \\
\hline
\end{tabular}
%\end{scriptsize}
\end{center}
\caption{\label{tab:ALLM} Parameters of the HHT-ALLM fit to HERA data only,
compared to parameters obtained when adding fixed-target 
data~\cite{Adams:1996gu,Arneodo:1996qe,Benvenuti:1989rh}  
to the fit, HHT-ALLM-FT.
The parameter values of the ALLM97 fit published 
previously~\cite{ALLM97}, using early HERA and other data, are also listed.
The formulae for the ALLM parameterisation are provided in the Appendix;
the units can be deduced from these formulae.}
\end{table}

\end{flushleft}

\clearpage

\begin{flushleft}
\begin{table}
\renewcommand*{\arraystretch}{1.3}
\begin{center}
%\begin{scriptsize}
\begin{tabular}{|l|c|c|c|r|c|c|}
\hline
Name   &\multicolumn{5}{c|}{Fit Parameters } &  $\chi^2$/ \\
\cline{2-6}

of Fit    & $M_0^2$ (GeV$^2$)& $A_{I\!P}$ ($\mu {\rm b}$) & $\alpha_{I\!P}(0)$ & $A_{I\!R}$ ($\mu b$) & $\alpha_{I\!R}(0)$ & ndf \\
\hline

HHT-REGGE & 0.50 $\pm 0.03$ &  66.3 $\pm$ 3.2 &1.097 $\pm$ 0.004 & fixed to 0 & -- & 0.83\\

\hline

~~~~~3p-.85 & 0.58 $\pm 0.03$ &  58.5 $\pm$ 2.5 &1.105 $\pm$ 0.003 &  fixed to 0 & --  &1.13\\

\hline

~~~~~4p & 0.49 $\pm 0.03$ &  78.5 $\pm$ 7.1 &1.082 $\pm$ 0.008 & $-$230 $\pm$105 & fixed to 0.5 & 0.78\\

\hline

~~~~~FT-4p & 0.50 $\pm 0.02$ &  77.4 $\pm$ 5.6 &1.083 $\pm$ 0.006 & $-$217 $\pm$ \,60 & fixed to 0.5 & 0.75\\

\hline

~~~~~PHP-5p & 0.52 $\pm 0.01$ &  57.0 $\pm$ 4.7 &1.110 $\pm$ 0.007 & 193 $\pm$ \,51 & 0.50 $\pm 0.11$ & 1.16\\

\hline
%HHT-REGGE-FT-5p & 0.50 $\pm 0.03$ &  71.0 $\pm$ 5.9 &1.091 $\pm$ 0.007 &  -1171 $\pm 2424$ & 0.10 $\pm 0.50$ &0.75\\

%\hline

~~~~~PHP-FT-5p & 0.48 $\pm 0.01$ &  58.9 $\pm$ 3.0 &1.110 $\pm$ 0.005 &  263 $\pm$ \,69 & 0.39 $\pm 0.09$ &1.35\\

\hline

ZEUSREGGE & fixed to 0.53 &  63.5 $\pm$ 0.9 &1.097 $\pm$ 0.002 &  145 $\pm$ ~~\,2 & fixed to 0.5 &1.12\\

\hline
~~~~~update & 0.52 $\pm$ 0.04 & 62.0 $\pm$ 2.3 &1.102 $\pm$ 0.007 &  148 $\pm$ ~~\,5 & fixed to 0.5 & --\\

\hline
\hline

\end{tabular}
%\end{scriptsize}
\end{center}
\caption{\label{tab:regge} Summary of the results of the HHT-REGGE fits, for
details see text. Also listed are results previously published as 
ZEUSREGGE~\cite{ZEUSF2-98}.
The HERA data at that time were not yet sufficient to constrain $M_0$
in the fit.
Therefore, $M_0$ was extracted from the data within the framework
of the GVMD model~\cite{GVDM1,GVDM2} 
to be
$M_0^2 = 0.53 \pm 0.04 (stat) \pm 0.09 (syst)$\,GeV$^2$
and fixed to 0.53 in the ZEUSREGGE fit.
An update~\cite{ZEUSREGGE} to the ZEUSREGGE fit was published 
including more early ZEUS data, allowing $M_0$ to be a free parameter.
This update provided compatible results but slightly larger
uncertainties and no $\chi^2$ was provided. Therefore, 
the parameters of the original ZEUSREGGE fit were
used for all comparisons throughout this paper.
}
\end{table}

\clearpage

\end{flushleft}

\begin{flushleft}
\begin{table}
\renewcommand*{\arraystretch}{1.3}
\begin{center}
%\begin{scriptsize}
\begin{tabular}{|c|c|c|}

\hline
 $Q^2$ (GeV$^2$) & $\lambda$ & $C$  \\    
\hline
0.11  & 0.171	 $\pm$  0.033 	 &  0.017  $\pm$	0.007   \\	
0.20  & 0.102	 $\pm$  0.029 	 &  0.068  $\pm$	0.024  	\\
0.25  & 0.089     $\pm$  0.009	 &  0.094  $\pm$	0.010 	\\
0.35  & 0.092	 $\pm$  0.007	 &  0.115  $\pm$	0.009 	\\
0.40  & 0.080     $\pm$  0.008	 &  0.145  $\pm$	0.013  	\\
0.50  & 0.100     $\pm$  0.008	 &  0.136  $\pm$	0.010  	\\
0.65  & 0.125	 $\pm$  0.008	 &  0.126  $\pm$	0.009 	\\
0.85  & 0.137     $\pm$  0.010	 &  0.137  $\pm$        0.013  	\\
1.20  & 0.150     $\pm$  0.008	 &  0.144  $\pm$	0.010  	\\
1.50  & 0.135	 $\pm$  0.007	 &  0.192  $\pm$	0.012  	\\
2.00  & 0.161	 $\pm$  0.006	 &  0.171  $\pm$	0.009 	\\
2.70  & 0.168	 $\pm$  0.005	 &  0.182  $\pm$	0.007 	\\

\hline
\end{tabular}
%\end{scriptsize}
\end{center}
\caption{\label{tab:lambda:BKS} 
 The fitted values of $\lambda$ and $C$ from 
 Eq.~\ref{eqn:lambda} for $F_2$ extracted with the BKS model.
}
\end{table}

\end{flushleft}

\begin{flushleft}
\begin{table}
\renewcommand*{\arraystretch}{1.3}
\begin{center}
%\begin{scriptsize}
\begin{tabular}{|c|c|c|}

\hline
 $Q^2$ (GeV$^2$) & $\lambda$ & $C$  \\    
\hline
1.2 & 0.230 $\pm$	0.010 	 &  0.076  $\pm$	0.007 	        \\
1.5 & 0.179 $\pm$	0.007  	 &  0.138  $\pm$	0.009 		\\
2.0   & 0.178 $\pm$	0.006 	 &  0.149  $\pm$	0.008 		\\
2.7 & 0.176 $\pm$	0.005 	 &  0.171  $\pm$	0.007 		\\
3.5 & 0.172 $\pm$	0.004 	 &  0.198  $\pm$	0.006 		\\
4.5 & 0.191 $\pm$	0.005 	 &  0.189  $\pm$	0.007 		\\
6.5 & 0.202 $\pm$	0.004  	 &  0.199  $\pm$	0.006   	\\
8.5 & 0.213 $\pm$	0.005 	 &  0.202  $\pm$	0.007 		\\
10  & 0.225 $\pm$	0.008  	 &  0.193  $\pm$	0.011  		\\
12  & 0.223 $\pm$	0.005 	 &  0.211  $\pm$	0.008  		\\
15  & 0.241 $\pm$	0.004 	 &  0.197  $\pm$	0.005 		\\
18  & 0.245 $\pm$	0.004 	 &  0.204  $\pm$	0.005 		\\
22  & 0.263 $\pm$	0.007 	 &  0.191  $\pm$	0.009   	\\
27  & 0.270 $\pm$	0.004 	 &  0.193  $\pm$	0.006 		\\
35  & 0.281 $\pm$	0.005 	 &  0.192  $\pm$	0.006  		\\
45  & 0.293 $\pm$	0.005  	 &  0.189  $\pm$	0.006  		\\
60  & 0.314 $\pm$	0.007 	 &  0.178  $\pm$	0.007 		\\
70  & 0.328 $\pm$	0.010  	 &  0.168  $\pm$	0.010  		\\
90  & 0.317 $\pm$	0.010  	 &  0.190  $\pm$	0.011  		\\
120 & 0.348 $\pm$	0.011  	 &  0.166  $\pm$	0.010  		\\
150 & 0.349 $\pm$	0.016  	 &  0.172  $\pm$	0.015  		\\
200 & 0.360 $\pm$	0.017  	 &  0.167  $\pm$	0.015  		\\
250 & 0.414 $\pm$	0.025  	 &  0.130  $\pm$        0.017  		\\
300 & 0.419 $\pm$	0.030  	 &  0.130  $\pm$	0.020  		\\

\hline
\end{tabular}
%\end{scriptsize}
\end{center}
\caption{\label{tab:lambda:HHT} 
 The fitted values of $\lambda$ and $C$ from 
 Eq.~\ref{eqn:lambda} for $F_2$ extracted with 
 results from the HHT NNLO analysis.
}
\end{table}

\end{flushleft}

%===============================================figures
% \input{figures}
\clearpage

\clearpage

\clearpage

\clearpage

\clearpage

\clearpage

\clearpage

\clearpage

\clearpage

\clearpage

\clearpage

\clearpage

\clearpage

\clearpage

{\bf{\Large Appendix}}

{\large \underline {The ALLM parameterisation}}

The ALLM parameterisation is based on Regge phenomenology, but tries to
incorporate ideas of pQCD.
The parameterisation has 23 parameters~\cite{ALLM97} defined as follows:
\begin{eqnarray}
  F_2 &=& \frac{Q^2}{Q^2+m_0} (F_2^{I\!P}+F_2^{I\!R})   \nonumber\\ [1em]
  F_2^{I\!P} &=& c_{I\!P}*x_{I\!P}^{a_{I\!P}} (1-x_{Bj})^{b_{I\!P}}  \nonumber\\[1em]
  F_2^{I\!R} &=& c_{I\!R}*x_{I\!R}^{a_{I\!R}} (1-x_{Bj})^{b_{I\!R}} \nonumber \\[1em]
  \frac{1}{x_{I\!P}} &=& 1 + \frac{W^2-m_p^2}{Q^2+p_1} ~~~~~{\rm where}~m_p~{\rm is~ the~ proton~ mass} \nonumber\\[1em]
  \frac{1}{x_{I\!R}} &=& 1 + \frac{W^2-m_p^2}{Q^2+p_2} \nonumber \\[1em]
  t &=& \ln \Bigg( \frac{ \ln \frac{Q^2+p_3}{p_4}} {\ln \frac{p_3}{p_4}} \Bigg) \nonumber \\[1em]
  c_{I\!P} &=& p_5 + (p_5 - p_6) ~ \Bigg[ \frac{1}{1 + t^{p_7}} - 1 \Bigg]   \nonumber\\[1em]
  a_{I\!P} &=& p_8 + (p_8 - p_9) ~\Bigg[ \frac{1}{1 + t^{p_{10}}} - 1 \Bigg]  \nonumber \\[1em]
  b_{I\!P} &=& p_{11} + p_{12} t^{p_{13}} \nonumber \\[1em]
  c_{I\!R} &=& p_{14} + p_{15} t^{p_{16}}  \nonumber\\[1em]
  a_{I\!R} &=& p_{17} + p_{18} t^{p_{19}}  \nonumber\\[1em]
  b_{I\!R} &=& p_{20} + p_{21} t^{p_{22}}  \nonumber ~.
\label{eqn:ALLMfull}
\end{eqnarray}

The parameters were determined in fits to the combined HERA
$e^+p$ NC cross sections, HHT-ALLM, and to the combined HERA data together
with fixed-target
data~\cite{Adams:1996gu,Arneodo:1996qe,Benvenuti:1989rh}, HHT-ALLM-FT.
They are listed in Table~\ref{tab:ALLM},
where they
are also compared to the parameters published previously~\cite{ALLM97}.

As the HHT-ALLM fit describes the data well with a 
$\chi^2/{\rm ndf} = 1.06$, the parameters were used to translate  
data points to selected $W$ or $Q^2$ values.

\end{document}